\documentclass[a4paper,11pt]{article}
\pdfoutput=1 % if your are submitting a pdflatex (i.e. if you have
\usepackage{jheppub} % for details on the use of the package, please
\usepackage[T1]{fontenc} % if needed
\usepackage{relsize}
\usepackage{slashed}
\usepackage{ytableau}
\usepackage{amsthm}
\usepackage[utf8]{inputenc}
\usepackage[force]{feynmp-auto} 

\usepackage{filecontents}

\def \12x12 {$\frac{1}{2}\otimes \frac{1}{2}$ }
\def \0x1 {$0\otimes1$ }

\newcommand{\fb}[1]{{\color{black} { #1}}}
\title{On the Double Copy for Spinning Matter}
\author[a,b]{Yilber Fabian Bautista}
\author[a,c,d]{and Alfredo Guevara}

\affiliation[a]{Perimeter Institute for Theoretical Physics, Waterloo, ON N2L 2Y5, Canada}
\affiliation[b]{Department of Physics   and  Astronomy, York University, Toronto, Ontario, M3J 1P3, Canada.}
\affiliation[c]{Department of Physics and Astronomy, University of Waterloo, Waterloo, ON N2L 3G1, Canada}
\affiliation[d]{CECs Valdivia and Departamento de F\'isica, Universidad de Concepci\'on, Casilla 160-C, Concepci\'on, Chile}

\emailAdd{ybautistachivata@perimeterinstitute.ca}
\emailAdd{aguevara@perimeterinstitute.ca}

\abstract{
We explore various tree-level double copy constructions for amplitudes including massive particles with spin. By working in general dimensions, we use that particles with spins $s\leq 2$ are fundamental to argue that the corresponding double copy relations partially follow from compactification of their massless counterparts. This massless origin fixes the coupling of gluons, dilatons and axions to matter in a characteristic way (for instance fixing the gyromagnetic ratio), whereas the graviton couples universally reflecting the equivalence principle. For spin-1 matter we conjecture all-order Lagrangians reproducing the interactions with up to two massive lines and we test them in a classical setup, where the massive lines represent spinning compact objects such as black holes. We also test the amplitudes via CHY formulae for both bosonic and fermionic integrands. At five points, we show that by applying generalized gauge transformations one can obtain a smooth transition from quantum to classical BCJ double copy relations for radiation, thereby providing a QFT derivation for the latter. As an application, we show how the theory arising in the classical double copy of Goldberger and Ridgway can be naturally identified with a certain compactification of $\mathcal{N}=4$ Supergravity.}

\begin{document} 
\maketitle
\flushbottom

\section{Introduction}
The Bern-Carrasco-Johansson double copy program \cite{Bern:2008qj} has demonstrated how certain gravitational
quantities can be obtained as a square of gauge-theory
ones. Originally introduced for QFT scattering amplitudes with the
aim of performing gravitational multiloop computations \cite{Bern:2010ue,Bern:2010yg}, the program
has seen many incarnations ranging among the construction of classical
space-times \cite{Luna:2016hge,Monteiro:2013rya,Luna:2015paa,Cardoso:2016amd,Cardoso:2016ngt,Carrillo-Gonzalez:2017iyj,Luna:2018dpt,CarrilloGonzalez:2019gof,Arkani-Hamed:2019ymq}, kinematic algebra realizations \cite{Monteiro:2013rya,Monteiro:2011pc,BjerrumBohr:2012mg,Cho:2019ype,Chen:2019ywi}, off-shell extensions \cite{Anastasiou:2014qba,Anastasiou:2018rdx,Mizera:2018jbh}, and more recently
applications to gravitational wave phenomena \cite{Luna:2017dtq,Luna:2016due,Goldberger:2016iau,Goldberger:2017frp,Goldberger:2017vcg,Chester:2017vcz,Goldberger:2017ogt,Li:2018qap,Shen:2018ebu,Plefka:2018dpa,Plefka:2019hmz,PV:2019uuv,Bern:2019crd,Chung:2019duq}.

To test the extent of the double copy, and also to study phenomenologically
relevant setups, it is desirable to introduce fundamental matter in
the construction. This has already been explored in the context of standard QCD \cite{Johansson:2014zca,Johansson:2015oia, delaCruz:2015dpa,delaCruz:2016wbr,Brown:2018wss,Plefka:2018zwm}. Also a number of other interesting cases
has been considered,\footnote{For matter-coupled YM theory the gravitational $D=4$ Lagrangians were first obtained from double copy in \cite{Chiodaroli:2015wal}, see also \cite{Anastasiou:2017nsz}.} including quiver theories with bifundamental matter \cite{Chiodaroli:2013upa,Huang:2012wr,Huang:2013kca} and theories with
spontaneously broken symmetries \cite{Chiodaroli:2015rdg,Chiodaroli:2017ehv}. On the other hand the classical double copy, in its many realizations, inherently contains massive matter and hence it is important to clarify the connection between the quantum and classical approaches.

One such step has been taken along refs. \cite{Luna:2016due,Luna:2017dtq,Shen:2018ebu} which studied gravitational radiation associated to accelerating black holes from the amplitudes point of view. In a recent work \cite{Bautista:2019tdr} we have outlined a direct connection of this phenomena with the spin-multipole expansion, soft theorems and a new operation defining double copy for massive amplitudes with spin. In this work we will thoroughly expand on this latter aspect and show how it arises in a purely QFT framework. We will consider tree-level double copy of massive particles with generic spins and explore several interesting cases.

One of the results of \cite{Bautista:2019tdr} was to obtain graviton-matter amplitudes from double copy at low
multiplicities but generic spin quantum number $s$. In order to summarize this in a schematic form, consider
a single massive particle of spin-$s$ propagating in a background
of photons. We denote the tree-level amplitude involving $n-2$ photons
and such a massive line as $A_{n}^{{\rm QED},s}$. Using a symmetric
product $\odot$ we then constructed a gravitational amplitude involving
one or two gravitons and a massive line, as

\begin{equation}
A_{n}^{{\rm {\rm GR}},s+\tilde{s}}\sim A_{n}^{{\rm QED},s}\odot A_{n}^{{\rm QED},\tilde{s}}\mbox{,}\quad n=3,4\label{eq:dcintro}
\end{equation}
where $s+\tilde{s}$ is the spin of the massive line in the graviton
amplitude. These amplitudes and their higher multiplicity extensions
are relevant for a number of reasons. First, they have been recently
pinpointed to control the classical limit where the massive lines correspond to compact objects \cite{Neill:2013wsa,Bern:2019nnu,Bautista:2019tdr}. Second,
they have been observed to have an exponential form in accord with their multipole expansion \cite{Guevara:2018wpp,Chung:2018kqs,Bautista:2019tdr,Guevara:2019fsj,Arkani-Hamed:2019ymq}. Third, they are dimension-independent and are
not polluted with additional states arising from the double copy \cite{Bautista:2019tdr}, the latter of which will be evident once we provide the corresponding Lagrangians. 

In this paper we will rederive and extend (\ref{eq:dcintro}), mainly focusing
on the simplest cases with $s,\tilde{s}\leq1$. These interactions are fundamental in the sense that they have a healthy high-energy behaviour \cite{Arkani-Hamed:2017jhn}. By promoting QED to
QCD\footnote{\fb{For the lower multiplicity cases $n=3,4$, one can choose QCD partial amplitudes to coincide with QED amplitudes.}}, studying higher multiplicity amplitudes and the relevant cases
for two massive lines, we will identify the gravitational
theories obtained by this construction, as promised in \cite{Bautista:2019tdr}. In order to do this we must observe
that formula (\ref{eq:dcintro}) has implicit a rather strong assumption,
namely the fact that the LHS only depends on the quantum number $s+\tilde{s}$
and not on $s,\tilde{s}$ individually. For instance, this means that
for gravitons coupled to a spin-1 field, it should hold that

\begin{equation}
A_{n}^{{\rm {\rm GR}},1}\sim A_{n}^{{\rm QCD},\frac{1}{2}}\odot A_{n}^{{\rm QCD},\frac{1}{2}}=A_{n}^{{\rm QCD},0}\odot A_{n}^{{\rm QCD},1}\, , \label{eq:dcintro-1}
\end{equation}
(we have changed QED to QCD in preparation for $n>4$). This means $A_{n}^{{\rm {\rm gr}},1}$ not only realizes the equivalence principle in the sense of Weinberg \cite{PhysRev.135.B1049} but extends it to deeper orders in the soft expansion \cite{Bautista:2019tdr,Guevara:2019fsj}. In the classical limit, the $A^{\rm gr,s}_n$ amplitudes so constructed will reproduce a well defined
compact object irrespective of its double copy factorization. In \cite{Bautista:2019tdr}
we exploited condition \eqref{eq:dcintro-1} at \textit{arbitrary} spin to argue that the 3-point
amplitude should indeed take an exponential structure, which has recently
been identified as a characteristic feature of the Kerr black hole
in the sense of \cite{Vines:2017hyw}. Here we will argue that despite having arbitrary spin, this 3-pt. amplitude can still be considered fundamental as it is essentially equal to its high-energy limit, which in fact implies \eqref{eq:dcintro}-\eqref{eq:dcintro-1}.

A simple instance of (\ref{eq:dcintro}) for gravitons was verified
explicitly by Holstein \cite{Holstein:2006pq,Holstein:2006wi} (see also \cite{Bjerrum-Bohr:2013bxa}) for $s=0\,,\tilde{s}\leq1$. He observed that as gravitational amplitudes have an intrinsic gravitomagnetic
ratio $g=2$, the double copy (\ref{eq:dcintro}) can only hold by
modifying $A_{3}^{{\rm QED},1}$ away from its ``minimal-coupling''
value of $g=1$. This modification yields the gyromagnetic ratio $g=2$
characteristic of the electroweak model and was proposed as natural by Weinberg \cite{osti_4073049}. As observed long ago by Ferrara, Porrati and Telegdi \cite{PhysRevD.46.3529} this modification precisely cancels all powers of $1/m^{2}$ in $A_{4}^{{\rm QED},1}$, which otherwise prevented the Compton amplitude to have a smooth high-energy limit.
This is a crucial feature, as it hints that the theories with a natural
value $g=2$ have a simple massless limit, and indeed can be obtained
conversely by compactifying pure massless amplitudes at any multiplicity. Furthermore,
it was pointed out in \cite{Cucchieri:1994tx} (and recently from a modern perspective \cite{Arkani-Hamed:2017jhn}) that the appearance of $1/m^{2}$ can
be avoided up to $s=2$ in the gravitational Compton amplitude $A_{4}^{{\rm GR},s}$
since it corresponds to fundamental interactions. By working on general dimensions, we will see that
indeed all such fundamental amplitudes follow from dimensional reduction of
massless amplitudes, and ultimately from a compactification of a pure graviton/gluon master amplitude. This is the underlying reason
they can be arranged to satisfy (\ref{eq:dcintro}), which in turn simplifies the multipole expansion as we exploited in \cite{Bautista:2019tdr}.

On a different front, 
it has long been known that the squaring relations in the massless sector
yield additional degrees of freedom corresponding to a dilaton $\phi$
and 2-form potential $B_{\mu\nu}$. Their classical counterparts also arise in classical solutions (e.g. string theory backgrounds \cite{CAMPBELL1992199,CAMPBELL1993137,PhysRevD.43.3140,Gibbons:1982ih}) and therefore emerge naturally (and perhaps inevitably) in the classical double copy \cite{Goldberger:2016iau,Luna:2015paa,Luna:2016hge,Luna:2017dtq}. It is therefore natural to ask
whether the condition (\ref{eq:dcintro-1}) also holds when the massless
states involve such fields. As we have explained this is a non-trivial
constraint, and in fact, it only holds for graviton states! To exhibit this phenomena we are led to identify two different gravitational theories, which we refer
to as \12x12 and \0x1 theories for brevity. The corresponding
tree amplitudes will be constructed as
\begin{equation}
A_{n}^{\frac{1}{2}\otimes\frac{1}{2}}\sim A_{n}^{{\rm QCD},\frac{1}{2}}\otimes A_{n}^{{\rm QCD},\frac{1}{2}}\,,\quad A_{n}^{0\otimes1}\sim A_{n}^{{\rm QCD},0}\otimes A_{n}^{{\rm QCD},1}\label{eq:2dcs}
\end{equation}
We conjecture that at all orders in $\kappa=\sqrt{32\pi G}$ such tree-level interactions
follow from the more general Lagrangians,
\begin{equation}
\frac{\mathcal{L}^{\frac{1}{2}\otimes\frac{1}{2}}}{\sqrt{g}}=-\frac{2}{\kappa^{2}}R+\frac{(d-2)}{2}(\partial\phi)^{2}-\frac{1}{4}e^{\fb{\frac{\kappa}{2}}(d-4)\phi}F^I_{\mu\nu} F_I^{\mu\nu}+\frac{m_I^{2}}{2}e^{\fb{\frac{\kappa}{2}}(d-2)\phi}A_{\mu}^I A^{\mu}_I\,,\label{eq:1212dc}
\end{equation}
and
\begin{align}
\frac{\mathcal{L}^{0\otimes1}}{\sqrt{g}}=&-\frac{2}{\kappa^{2}}R+\frac{(d-2)}{2}(\partial\phi)^{2}-\frac{e^{-2\kappa\phi}}{6}H_{\mu\nu\rho}(H^{\mu\nu\rho}+\frac{3\kappa}{2}A_I^{\mu}F_I^{\nu\rho}) \nonumber \\ 
&-\frac{1}{4}e^{-\kappa\phi}F^I_{\mu\nu}F_I^{\mu\nu}+\frac{m_I^{2}}{2}A^I_{\mu}A_I^{\mu}+\rm{quartic\,\, terms}\,,\label{eq:01dc}
\end{align}
where $H=dB$ is the field strength of a two-form $B$. Here a sum over $I=1,2$, the flavour index, is implicit and "quartic terms" denote contact interactions between two matter lines that we will identify. These actions will be constructed in general dimensions from simple considerations such as 1) classical behaviour and 2) massless limit/compactification in the
string frame. We will then cross-check them against the corresponding QFT amplitudes using modern tools such as massive versions of CHY \cite{Cachazo:2013hca,Cachazo:2013iea,Cachazo:2014nsa,Cachazo:2014xea} and the connected formalism \cite{Cachazo:2018hqa,Geyer:2018xgb,Schwarz:2019aat}. In the massless limit, the \12x12 Lagrangian
is known as the Brans-Dicke-Maxwell (BDM) model with unit coupling
\cite{Cai:1996pj}. This theory is simpler than \0x1 in
many features, for instance in that the $B$-field is not sourced by
the matter line and it does not feature quartic interactions. Not surprisingly, in $d=4$ and in the massless
limit the \0x1 theory reproduces the bosonic interactions of
$\mathcal{N}=4$ Supergravity \cite{CREMMER197861,PhysRevD.15.2805}, which is known to arise
as the double copy between $\mathcal{N}=4$ Super Yang-Mills (SYM)
and pure Yang-Mills (YM) theories  \cite{Bern:2010ue,Bern:2011rj,BoucherVeronneau:2011qv}. In general dimension we will see that the \0x1 theory is precisely the QFT version
of the worldline model constructed by Goldberger and Ridgway in \cite{Goldberger:2016iau,Goldberger:2017ogt} and later extended in \cite{Goldberger:2017vcg,Li:2018qap} to exhibit a classical double copy construction with spin. This explains their findings on the fact that the \textit{classical} double copy not only fixes $g=2$ on the YM side, but also precisely sets the dilaton/axion-matter coupling on the gravity side.

The long-range radiation of a two-body system, emerging in the classical double copy, has been directly linked to a five-point amplitude at leading order
\cite{Luna:2017dtq,Kosower:2018adc,Bautista:2019tdr,PV:2019uuv}. We show that
by implementing generalized gauge transformations \cite{Bern:2008qj} one can define
a BCJ gauge in which the $\hbar\to0$ limit is smooth, i.e. there are
no "superclassical" $\sim \frac{1}{\hbar}$ contributions to cancel \cite{Kosower:2018adc}. The result precisely takes the
form derived in \cite{Bautista:2019tdr} from different arguments, \fb{i.e. by using the factorization properties of the classical amplitude.} This then  allows us to translate between the QFT version of the double copy and
a classical version of it. We employ this formulae to test double copy in several cases, including the computation of dilaton-axion-graviton radiation with spin \cite{Goldberger:2017vcg,Li:2018qap}.

This paper is organized as follows. In Section \ref{sec:sec2} we introduce the double copy for one matter line by studying its massless origin, focusing on the \12x12 theory and later extending it in more generality. In section \ref{constructing the lagrangians} we construct the Lagrangians for both QCD and Gravity from simple arguments, which are then checked against the previous amplitudes. In section \ref{two matter lines} we extend both the amplitudes and the Lagrangian construction to two matter lines and define the classical limit to make contact with previous results. In the appendices we provide some further details on the constructions. We also perform checks such as tree-level unitarity and explicit evaluation via the CHY formalism.

\subsubsection*{Note Added. }

During the final stages of this project we were informed of the publication
of \cite{Johansson:2019dnu} which has considered the full spectrum of the $(\mbox{\ensuremath{\frac{1}{2}}},\mbox{\ensuremath{\frac{1}{2}}})$
theory in $d=4$ dimensions. In contrast, here we have proposed a consistent tree-level truncation
of the matter spectrum which holds in general dimensions. We are grateful to the authors for sharing their manuscript, which led us to include Appendix \ref{4 d double copy} to show that both constructions
are consistent in $d=4$ (see also Appendix \ref{residues at 4 points}). In addition, refs. \cite{Bern:2019nnu,Bjerrum-Bohr:2019nws} have appeared, which have focused on scalars and have also employed the compactification to construct the relevant gravitational amplitudes, $A_n^{\rm{GR},0}$.

\section{Double Copy from Dimensional Reduction}\label{sec:sec2}

In this section we will introduce the double copy construction by considering
a single massive line. In this case one should expect the double copy to hold for massive
scalars as their amplitudes can be obtained via compactification of
higher dimensional amplitudes \cite{Bjerrum-Bohr:2013bxa,Bern:2019crd,Bern:2019nnu}. Here we will explicitly demonstrate how
this holds even for the case of spinning matter as long as such particles
are \textit{elementary}. This means we consider particles of spin
$s\leq2$ coupled to GR and particles of spin $s\leq1$ coupled to
QCD, in accordance with the notion of \cite{Arkani-Hamed:2017jhn}, see also \cite{Cucchieri:1994tx,Deser:2001dt}. The
fact that these amplitudes can be chosen to have a smooth high-energy
limit can be used backwards to construct them directly from their
massless counterparts. On the other hand, once the double copy form of gravitational-matter amplitudes is achieved one may use it to manifest properties such as the multipole expansion \cite{Bautista:2019tdr}.\fb{ We will expand on this in Sec \ref{sec:multipoles}.}

\subsection{The \texorpdfstring{$\frac{1}{2}\otimes\frac{1}{2}$}{1212} construction}

Let us consider first the case $s=\tilde{s}=\frac{1}{2}$ in (\ref{eq:dcintro}) and
relegate the other configurations for the next section. For $D=4$
massless QCD, the double copy procedure was first studied by Johansson
and Ochirov \cite{Johansson:2014zca}. In particular they observed that Weyl-spinors
in QCD can be double copied according to the rule $2\otimes2=2\oplus1\oplus1$,
where the two new states correspond to a photon $\gamma^{\pm}$ and
the remaining ones to axion and dilaton scalars. This implies that
we can obtain amplitudes in a certain Einstein-Maxwell theory directly
from massless QCD. More precisely, for two massive particles we can
write

\begin{equation}
A_{n}^{\frac{1}{2}\otimes\frac{1}{2}}(\gamma_{1}^{-}H_{3}{\cdots}H_{n}\gamma_{2}^{+})=\sum_{\alpha\beta}K_{\alpha\beta}[2|A_{n,\alpha}^{{\rm QCD}}(g{\cdots}g)|1\rangle \langle 1|\bar{A}_{n,\beta}^{{\rm QCD}}(g{\cdots}g)|2],\label{eq:dconeline}
\end{equation}
($\bar{A}$ here denotes charge conjugation, which will be relevant
in the massive case). In the gravitational amplitude the two photon
states $\gamma_{1}^{+}$,$\gamma_{2}^{-}$ make a matter line while
interacting with the ``fat'' states $H_{i}$. The latter are obtained
from the double copy of the gluons $g_{i}$, and can be taken to be
either a Kalb-Ramond field\footnote{In $D=4$ this field can be dualized to an axion pseudoscalar. We
will indistinctly refer to the two-form $B_{\mu\nu}$ as axion or Kalb-Ramond
field.}, a dilaton or a graviton by projecting the product representation
into the respective irreps.,

\begin{equation}
  H_{i}^{\mu\nu}\to\epsilon_{i}^{\mu}\tilde{\epsilon}_{i}^{\nu}=\underbrace{\epsilon_{i}^{[\mu}\tilde{\epsilon}_{i}^{\nu]}}_{B^{\mu\nu}}+\underbrace{\frac{\eta^{\mu\nu}}{D-2}\epsilon_{i}\cdot\tilde{\epsilon}_{i}}_{\eta^{\mu\nu}\frac{\phi}{\sqrt{D-2}}}+\underbrace{\left(\epsilon_{i}^{(\mu}\tilde{\epsilon}_{i}^{\nu)}-\frac{\eta^{\mu\nu}}{D-2}\epsilon_{i}\cdot\tilde{\epsilon}_{i}\right)}_{h^{\mu\nu}}\,.
\end{equation}\label{eq:fatgraviton}
The sum over $\mbox{\ensuremath{\alpha}},\,\beta$ in (\ref{eq:dconeline})
ranges over $(n-3)!$ orderings, where $K_{\alpha,\beta}$ is the
standard KLT kernel \cite{Kawai:1985xq,Bern:1998sv,BjerrumBohr:2010ta}.\footnote{We define the KLT kernel with no coupling constants and  absorb the gauge theory coupling $\tilde{g}$ into the generators $\tilde{ T}^a$ as $\tilde{g}\tilde{T}^a\to T^a$. We also absorb the overall factors of $i$ in the definition of the amplitudes and  use the conventions for the metric to be in the mostly minus signature.} This construction can be implemented
because for a single matter line we can take the matter particles
to be either in the fundamental or in the adjoint representation and
the basis of partial amplitudes will be identical \cite{Johansson:2014zca}. In section \ref{two matter lines}  we will switch to a more natural prescription for the case of two
matter lines.

The RHS of (\ref{eq:dconeline}) exhibits explicitly the helicity
weight $\pm\frac{1}{2}$ associated to the Weyl spinors $v_{1}^{-}=|1\rangle$
and $\bar{u}_{2}^{+}=[2|$ of the (massless) matter particles. This
means the operators $A^{{\rm QCD}}$ and $\bar{A}^{{\rm QCD}}$ ,
defined as the amplitude with such spinors stripped, do not carry
helicity weight. They can be written as products of Pauli matrices
$\sigma^{\mu}$, $\bar{\sigma}^{\mu}$ where the free Lorentz index
is contracted with momenta $p_{i}^{\mu}$ or gluon polarizations $\epsilon_{i}^{\mu}$,
as we will see in the examples of the next section. We can alternatively
write them in terms of the corresponding spinor-helicity variables
as in \cite{Johansson:2015oia}.

Quite generally, the LHS of (\ref{eq:dconeline}) defines a gauge
invariant quantity due to the fact that it is constructed from partial
gauge-theory amplitudes. It also has the correct factorization properties
(see e.g. \cite{Bern:2010yg,Chiodaroli:2017ngp}). Furthermore, by providing the Lagrangian it
will become evident that when the states $H_{i}$ are chosen to be
gravitons the amplitude we get for a single matter-line is that of
\textit{pure} Einstein-Maxwell theory, where the dilatons and axions
simply decouple. This decoupling is one of the key properties of these
objects, which we have exploited in \cite{Bautista:2019tdr}. Similarly, the decoupling of further matter particles will be treated in Appendix \ref{residues at 4 points}.

In order to extend (\ref{eq:dconeline}) to the massive case we rewrite
it in a way in which it is not sensitive to the dimension, and then
use dimensional reduction. This can be done by introducing polarization
vectors for the photons $\gamma^{\pm}$. Recall that a photon polarization
vector can be taken to be $\epsilon_{\mu}^{+}\sigma^{\mu}=\sqrt{2}\frac{|\mu\rangle[p|}{\langle\mu p\rangle}$
where $[\mu|$ is a reference spinor carrying the gauge freedom, and
analogously $\epsilon_{\mu}^{-}\bar{\sigma}^{\mu}=\sqrt{2}\frac{|\mu]\langle p|}{[\mu p]}$.
We then have the identity

\begin{eqnarray}
[2|X|1\rangle \langle 1 |\bar{Y}|2] & = & \frac{{\rm Tr}(X|1\rangle [1\mu_1]  \langle 1|\bar{Y}|2]\langle 2\mu_{2}\rangle |2])}{[1\mu_{1}]\langle 2\mu_{2}\rangle}, \\
 & = & \frac{1}{2}{\rm Tr}\left(X\bar{p}_{1}\epsilon_{1}\bar{Y}p_{2}\bar{\epsilon}_{2}\right),\label{eq:massless}
\end{eqnarray}
where the bottom line now can be naturally extended to higher dimensions.\footnote{We represent the Dirac algebra in terms of the $2^{D/2}\times2^{D/2}$
matrices $\Gamma_{D}^{\mu}=\left(\begin{array}{cc}
0 & \sigma_{D}^{\mu}\\
\bar{\sigma}_{D}^{\mu} & 0
\end{array}\right)$ and define $X=X_{\mu}\sigma_{D}^{\mu},\,\bar{X}=X_{\mu}\bar{\sigma}_{D}^{\mu}$
etc. The extension of (\ref{eq:massless}) to general dimension simply
states that linear combinations $c_{ab}u_{i}^{a}\bar{v}_{i}^{b}$
of the Weyl spinors can be replaced as $c_{ab}v_{i}^{a}\bar{u}_{i}^{b}=p_{i}\bar{\epsilon}_{i}$
for some particular choice of $\epsilon_{i}^{\mu}$ depending on $c_{ab}$.
A formula for general dimension is of course obtained by replacing
$\sigma^{\mu},\bar{\sigma}^{\mu}\to\Gamma^{\mu}$, which in $D=4$
also reduces to (\ref{eq:massless}).} It is manifestly gauge invariant since the shift  $\epsilon_{i}\to \epsilon_{i}+p_{i}$
is projected out due to $p_{i}\bar{p}_{i}=0$. 

Using this identity, the double copy (\ref{eq:dconeline}) can be uplifted
to dimension $D=2m$ as
\begin{equation}
A_{n}^{\frac{1}{2}\otimes\frac{1}{2}}(\gamma_{1}H_{3}\cdots H_{n}\gamma_{2}^{*})=\frac{1}{2}\sum_{\alpha\beta}K_{\alpha\beta}{\rm Tr(}A_{n,\alpha}^{{\rm QCD}}(g{\cdots}g)\bar{p}_{1}\epsilon_{1}\bar{A}_{n,\beta}^{{\rm QCD}}(g{\cdots}g)p_{2}\bar{\epsilon}_{2}).\label{eq:kltmassless}
\end{equation}
Note that the operators $A_{n}^{{\rm QCD}}$, $\bar{A}_{n}^{{\rm QCD}}$
in (\ref{eq:dconeline}) are defined under the support of the Dirac
equation. This means that they can be shifted by operators proportional
to $p_{1}$ or $p_{2}$. The insertion of $p_{1},p_{2}$ in (\ref{eq:kltmassless})
certainly projects out these contributions by using the on-shell condition
$p\bar{p}=\bar{p}p=0$. For instance, if the matrix operator $A_{n}^{{\rm QCD}}$
is shifted by $p_{2\mu}\sigma^{\mu}$ the QCD amplitude $\bar{u}_{2}A_{n}^{{\rm QCD}}v_{1}$
is invariant, and consistently (\ref{eq:kltmassless}) picks
up no extra contribution, i.e.
\begin{equation}
    {\rm Tr(}p_{2}\bar{p}_{1}\epsilon_{1}\bar{A}_{n,\beta}^{{\rm QCD}}(g\cdots g)p_{2}\bar{\epsilon}_{2})=-{\rm Tr(}p_{2}p_{2}\bar{p}_{1}\epsilon_{1}\bar{A}_{n,\beta}^{{\rm QCD}}(g\cdots g)\epsilon_{2})=0,
\end{equation}
where we used $p_{2}\bar{\epsilon}_{2}={-}\epsilon_{2}\bar{p}_{2}$.
This kind of manipulations are usual when bringing the QCD amplitude
into multipole form \cite{Bautista:2019tdr}  to make explicit the corresponding form
factors. 

 We now proceed to dimensionally reduce our formulae in order to
obtain a KLT expression for massive spin-$\frac{1}{2}$ particles.
This follows from a standard KK compactification on a torus, as we
explain in the next section. In terms of momenta, we can define the
$d=D-1$ components $p_{1}$ and $p_{2}$ via
\begin{equation}\begin{split}
P_{1} & =  (m,p_{1}),\\
P_{2} & =  (-m,p_{2}),\\
P_{i} & =  (0,k_{i}),\qquad i\in\{3,\ldots,n\}    
\end{split}
\end{equation}
which trivially satisfies momentum conservation in the KK component, which we take with minus signature. We also take all momenta outgoing. In terms of Feynman diagrams,
the reduction induces the flow of KK momentum through the only path
that connects particles $p_{1}$ and $p_{2}$. The propagators in
this line are deformed to massive propagators as

\begin{equation}
    \frac{1}{P_{I}^{2}}=\frac{1}{p_{I}^{2}{-}m^{2}},
\end{equation}
where $P_{I}=(m,p_{I})$ is the internal momentum. The procedure works
straightforwardly when compactifying more particles as long as the
KK lines do not cross (i.e. we will not allow interactions between
massive particles), as we will explain in the case of two matter lines.\\
 By applying these rules to (\ref{eq:kltmassless}) the amplitudes
$A^{{\rm gr}}$, $A^{{\rm QCD}}$ now contain massive lines and lead
to a (gravitational) Proca theory and the massive QCD theory in $d=D-1$
dimensions, respectively. This can be observed easily by applying
the dimensional reduction to the Lagrangian as we do in Section \ref{constructing the lagrangians}.
In the case of the spin-1 theory we choose the polarization vectors
$\epsilon_{1}$,$\epsilon_{2}$ to be $d$-dimensional, i.e.
$\epsilon\to(0,\varepsilon)$, so that the transverse condition
$\epsilon{\cdot}P=0$ now imposes $\varepsilon{\cdot}p=0$. In
the QCD case we note that the Dirac equation now becomes
\begin{equation}\begin{split}
(p_{\mu}\Gamma_{d}^{\mu})u & =  m\,u,\\
(p_{\mu}\Gamma_{d}^{\mu})v & =  -m\,v,
\end{split}
\end{equation}
where we have used
\begin{equation}\label{sigmad}
\sigma_{D}=(\mathbb{I},\Gamma_{d})\,,\qquad \bar{\sigma}_{D}=(-\mathbb{I},\Gamma_{d})\,,
\end{equation}
in the chiral representation. Denoting by $W$ and $W^*$ the Proca fields obtained from the photons, the construction (\ref{eq:kltmassless})
now reads 
\begin{equation}
A_{n}^{\frac{1}{2}\otimes\frac{1}{2}}(W_{1}H_{3}{\cdots}H_{n}W_{2}^{*}){=}\sum_{\alpha\beta}\frac{K_{\alpha\beta}}{2^{\left\lfloor d/\text{2}\right\rfloor {-}1}} {\rm Tr(}A_{n,\alpha}^{{\rm QCD}}(g{\cdots}g)(\slashed p_{1}{-}m)\slashed\varepsilon_{1}\bar{A}_{n,\beta}^{{\rm QCD}}(g{\cdots}g)(\slashed p_{2}{-}m)\slashed\varepsilon_{2}),\label{eq:massiveklt}
\end{equation}
where the normalization factor follows from the Dirac trace ${\rm tr}(\mathbb{I})=2^{\left\lfloor D/\text{2}\right\rfloor }$. Even though our derivation used that $d=2m-1$ for the reduction procedure, our final result is written explicitly in terms of $d$-dimensional Dirac matrices so we assume it to be valid in generic dimensions. To confirm this we will indeed compute both sides of \eqref{eq:massiveklt} from generic-dimensional Lagrangians and find a precise agreement. 

From now on we refer to the double-copy theory as the \textit{$\frac{1}{2}\otimes \frac{1}{2}$ theory} because it is constructed from two (conjugated) copies of massive QCD. As in the massless case, the
role of the projectors $\slashed p_{i}\pm m$ is to put the QCD amplitudes
on the support of the massive Dirac equation. With a slight abuse
of notation, we have left here the symbol $K_{\alpha\beta}$ for the
massive KLT kernel, this simply corresponds to the inverse of the
biadjoint amplitude involving two massive scalars of the same species,
$K_{\alpha\beta}=m_{n}^{-1}(\alpha|\beta)$, see e.g. \cite{Naculich:2015zha}
for details on this theory.

We have thus derived an explicit KLT relation for massive amplitudes
of one matter line, (\ref{eq:massiveklt}) as a direct consequence
of the massless counterpart. The resulting theory will be extended to two matter lines in Section \ref{two matter lines}. The partial
amplitudes $A_{n,\alpha}^{{\rm QCD}}$ are associated to Dirac spinors
in general dimension, as opposed to Majorana ones, and hence the resulting
spin-1 field is a complex\footnote{We thank Henrik Johansson for emphasizing this. On the other hand, for the double-copy theories obtained in this work we will drop the distinction between real/complex bosonic matter fields. See footnote \ref{fn15}.}
Proca state coupled to gravity. Moreover, it follows from the massless
case that when all the gravitational states $H_{i}$ are chosen as
gravitons, the dilaton and axion field decouple and the theory simply
corresponds to Einstein-Hilbert gravity plus a covariantized (minimally
coupled) spin-1 Lagrangian. We will see that this holds quite generally
and is consistent with the observations made around eq. \eqref{eq:dcintro} for generic
spin.

In our formula the states $H_{i}$ denote the fat gravitons \eqref{eq:fatgraviton} characteristic of the double copy construction. However,
a particular feature arises in that amplitudes with an odd number
of axion fields vanish. This can be traced back to the symmetry in
the two QCD factors of the \12x12 construction.
To see this, let us slightly rewrite (\ref{eq:massiveklt}) as

\begin{equation}
A_{n}^{\frac{1}{2}\otimes\frac{1}{2}}(W_{1}H_{1}^{\mu_{1}\nu_{1}}{\cdots}H_{n{-}2}^{\mu_{n-2}\nu_{n-2}}W_{2}^{*})=\sum_{\alpha\beta}K_{\alpha\beta}(A_{n,\alpha}^{{\rm QCD}})^{\mu_{1}\cdots\mu_{n-2}}\otimes(A_{n,\beta}^{{\rm QCD}})^{\nu_{1}\cdots\nu_{n-2}}\,,\label{eq:abklt}
\end{equation}
where
\begin{equation}\label{tensor product }
X\otimes Y=\frac{1}{2^{\left\lfloor d/\text{2}\right\rfloor -1}}{\rm Tr(}X(\slashed p_{1}{-}m)\slashed\varepsilon_{1}\bar{Y}(\slashed p_{2}{-}m)\slashed\varepsilon_{2}).
\end{equation}
It is not hard to check that (see for instance the explicit form in
(\ref{eq:dconeline}))
\begin{equation}
    (A_{n,\alpha}^{{\rm QCD}})^{\mu_{1}\cdots\mu_{n-2}}\otimes(A_{n,\beta}^{{\rm QCD}})^{\nu_{1}\cdots\nu_{n-2}}=(A_{n,\beta}^{{\rm QCD}})^{\nu_{1}\cdots\nu_{n-2}}\otimes(A_{n,\alpha}^{{\rm QCD}})^{\mu_{1}\cdots\mu_{n-2}}\,.
\end{equation}
Now, since the Kernel $K_{\alpha\beta}$ in (\ref{eq:abklt}) can
be arranged to be symmetric in $\alpha\leftrightarrow\beta$, this
implies that the RHS of (\ref{eq:abklt}) is symmetric under the exchange
of \textit{all} $\mu_{i}\leftrightarrow\nu_{i}$ at the same time,
namely $(\mu_{1},\mu_{2}\ldots)\leftrightarrow(\nu_{1},\nu_{2}\ldots)$.
However, if we antisymmetrize an odd number of pairs $\{\mu_{k},\nu_{k}\}$,
i.e. compute the amplitude for an odd number of axions, and symmetrize
the rest of the pairs, we obtain an expression which is antisymmetric
under the full exchange $(\mu_{1},\mu_{2}\ldots)\leftrightarrow(\nu_{1},\nu_{2}\ldots)$.
Hence amplitudes with an odd number of axions must vanish.

The above considerations imply that the axion field is pair-produced
and cannot be sourced by the Proca field. This is also true for amplitudes with no matter (i.e. the massless double copy) and even for amplitudes with
more matter lines: For e.g. two matter lines, provided a double copy formula as in section \ref{two matter lines}, we can test axion propagation by examining all possible factorization channels. Since the factorization always contains amplitudes with either one or none matter lines we conclude that the axion will not emerge in the cut unless introduced also as an external state. The argument carries over for an arbitrary number of matter lines.

The previous fact is surprising from the gravitational perspective since it is known that the axion couples naturally to the spin of matter particles. We interpret this fact as an avatar of the spin-$\frac{1}{2}$ origin of the construction. In Appendices \ref{4 d double copy} and \ref{residues at 4 points} we will specialize the construction to $d=4$: In particular we will show that being a pseudoscalar,
the axion can only be sourced when the Proca field decays into a massive
pseudoscalar as well, as considered very recently in \cite{Johansson:2019dnu}. In the massless theory such field is obtained by selecting anticorrelated fermion helicities
in the RHS of (\ref{eq:dconeline}) which leads to massless (pseudo)scalars
instead of photons $\gamma^{\pm}$ \cite{Johansson:2014zca}. The analysis becomes
more involved in higher dimensions. For our purposes here we can neglect
these processes and simply keep the theory containing a Proca field,
a graviton and a dilaton as a consistent tree-level truncation of the spectrum
in arbitrary number of dimensions.

A further clarification is needed regarding the compactification and
the dilaton states. In the massless case these are obtained via the
replacement
\begin{equation}\label{eq:pr1}
   \epsilon_{i}^{\bar{\mu}}\tilde{\epsilon}_{i}^{\bar{\nu}}\to\frac{\eta^{\bar{\mu}\bar{\nu}}}{\sqrt{D-2}},
\end{equation}
where we have denoted the indices as $\bar{\mu},\mathbf{\bar{\nu}}$
to emphasize that the trace is taken in $D=d+1$ dimensions. However,
after dimensional reduction we have $\epsilon^{\bar{\mu}}\to\epsilon^{\mu}$,
and we extract the corresponding dilaton via
\begin{equation}\label{eq:pr2}
    \epsilon_{i}^{\mu}\tilde{\epsilon}_{i}^{\nu}\to\frac{\eta^{\mu\nu}}{\sqrt{d-2}}.
\end{equation}
This means that taking the dimensional reduction does not commute
with extracting dilaton states, as e.g. terms of the form $P_{1}\cdot\epsilon\,P_{2}\cdot\tilde{\epsilon}$
are projected to $P_{1}\cdot P_{2}=p_{1}\cdot p_{2}+m^{2}$ in the
first case and to $p_{1}\cdot p_{2}$ in the second case. In order to match certain results in the literature (e.g. \cite{Goldberger:2016iau}) we find that we need to adopt the second construction: first implement dimensional reduction
on the fat states, and then project onto either dilatons or gravitons. 

Let us close this subsection by providing some key examples of this procedure
for $n=3,4$. The 3-pt. dilaton amplitude from (\ref{eq:massiveklt}), using \eqref{eq:fatgraviton},
gives 
\begin{eqnarray}
A_{3}^{\frac{1}{2}\otimes\frac{1}{2}}(W{}_{1}\phi W_{2}^{*}) & = & \frac{2K_{3}}{2^{\left\lfloor d/2\right\rfloor }\sqrt{d-2}}{\rm Tr(}A_{3}^{{\rm \mu}}\slashed\varepsilon_{1}(\slashed p_{1}{-}m)\bar{A}_{3\mu}\slashed\varepsilon_{2}(\slashed p_{2}{-}m)), \nonumber \\
 & = & \frac{\kappa}{2\sqrt{d-2}}{\rm (}2m^{2}\varepsilon_{1}{\cdot}\varepsilon_{2}{+}(d-4)k_{3}{\cdot}\varepsilon_{1}k_{3}{\cdot}\varepsilon_{2}),\label{eq:dila3pt}
\end{eqnarray}
\textcolor{black}{where we have use the momentum conservation $p_{1}+p_{2}+k_{3}=0$,
and the dilaton projection $\epsilon_{3}^{\mu}\tilde{\epsilon}_{3}^{\nu}\to\frac{\eta^{\mu\nu}}{\sqrt{d-2}}$}.
This example will exhibit one of the main differences between the \12x12
construction and the other cases, namely that the dilaton (and the
axion) fields couple differently to matter in each case, as opposed
to gravitons which couple universally. 

%%%%%%%%%%%%%%%%%%%%%%%%%%%%%%%%%%%%%%%%%%%%%%%%%%%%%%%%%%%%%%%%%%%%%%%%%%%%%%%%%%%%%%%%%%%%%%%%%%%%%%%%%%%%%%%%%%%%%%%%%%%%%%%%%%%%%%%

Now we can move on to $n=4$. The only independent QCD amplitude reads 

\begin{equation}
A_{4,1324}^{\mu_{3}\mu_{4}}=-\frac{1}{4}\frac{\gamma^{\mu_{4}}(\slashed{p}_{1}{+}\slashed{k}_{3}{-}m)\gamma^{\mu_{3}}}{(p_1+k_3)^2-m^2}-\frac{1}{4}\frac{\gamma^{\mu_{3}}(\slashed{p}_{1}{+}\slashed{k}_{4}{-}m)\gamma^{\mu_{4}}}{(p_1+k_4)^2-m^2}\,. \label{eq:A41324}
\end{equation}
Analogously,
\begin{equation}
\bar{A}_{4,1324}^{\mu_{3}\mu_{4}}=-\frac{1}{4}\frac{\gamma^{\mu_{3}}(\slashed{p}_{1}{+}\slashed{k}_{3}{+}m)\gamma^{\mu_{4}}}{(p_1+k_3)^2-m^2}-\frac{1}{4}\frac{\gamma^{\mu_{4}}(\slashed{p}_{1}{+}\slashed{k}_{4}{+}m)\gamma^{\mu_{3}}}{(p_1+k_4)^2-m^2},\label{eq:A41324bar}
\end{equation}
where the conjugated amplitude is obtained by inverting the direction of the massive line. Note that this ordering corresponds to the QED amplitude.

The full Compton amplitude for fat gravitons can be computed from the double
copy $(\ref{eq:massiveklt})$,
\begin{equation}
A_{4}^{\frac{1}{2}\otimes\frac{1}{2}}(W_{1}H_{3}^{\mu_{3}\nu_{3}}H_{4}^{\mu_{4}\nu_{4}}W_{2}^{*})=\frac{1}{2^{\left\lfloor d/2\right\rfloor -1}}K_{1324,1324}\,\text{tr} \left[A_{4}^{\mu_{3}\mu_{4}}\slashed{\varepsilon}_{1}(\slashed{p}_{1}{+}m)\bar{{A}}_{4}^{\nu_{3}\nu_{4}}\slashed{\varepsilon}_{2}(\slashed{p}_{2}{+}m)\right],
\end{equation}\label{eq:4pt1212}
where the massive KLT kernel takes the compact form
\begin{equation}
K_{1324,1324}=\frac{2p_{1}{\cdot}k_{3}\,p_{1}{\cdot}k_{4}}{k_{3}{\cdot}k_{4}}.\label{eq:klt 1324}
\end{equation}
For instance, the two-dilaton emission amplitude reads

\begin{equation}\label{eq:ABphiphiB-2}
\begin{split}
A_{4}^{\frac{1}{2}\otimes\frac{1}{2}}(W_{1}\phi_{3}\phi_{4}W_{2}^{*}) & {=}\frac{\kappa^{2}\varepsilon_{1,\alpha}\varepsilon_{2,\beta}^{*}}{32(d-2)p_{1}{\cdot}k_{3}\,p_{1}{\cdot}k_{4}k_{3}{\cdot}k_{4}}\left\{ \left[(d{-}4)^{2}s_{34}^{2}{-}16(d{-}2)p_{1}{\cdot}k_{3}p_{2}{\cdot}k_{3}\right]\times\right.\\
&
\left[p_{1}{\cdot}k_{3}k_{4}^{\alpha}k_{3}^{\beta}{+}p_{2}{\cdot}k_{3}(k_{3}^{\alpha}k_{4}^{\beta}{+}p_{1}{\cdot}k_{3}\eta^{\alpha\beta})\right] 
 {+}2m^{2}s_{34}\left[4p_{1}{\cdot}k_{3}\left(k_{4}^{\alpha}k_{3}^{\beta}{-}k_{3}^{\alpha}k_{4}^{\beta}\right.\right.
\\& \left.\left.\left.{+}2p_{2}{\cdot}k_{3}\eta^{\alpha\beta}\right)
 {+}s_{34}\left((d{-}4)(k_{3}^{\alpha}k_{3}^{\beta}{+}k_{4}^{\alpha}k_{4}^{\beta})-2(k_{3}^{\alpha}k_{4}^{\beta}{+}m^{2}\eta^{\alpha\beta})\right)\right]\right\},
 \end{split}
\end{equation}
which again exhibits explicit mass dependence in accord with our discussion. On the other hand, extracting the pure graviton emission from \eqref{eq:4pt1212} gives 

\begin{equation}\label{eq:ABhhB-1}
\begin{split}
A_{4}^{\frac{1}{2}\otimes\frac{1}{2}}(W_{1}h_{3}h_{4}W_{2}^{*}) & =\frac{\kappa^{2}\varepsilon_{1,\alpha}\varepsilon_{2,\beta}^{*}}{2p_{1}{\cdot}k_{3}\,p_{1}{\cdot}k_{4}\,k_{3}{\cdot}k_{4}}p_{1}{\cdot}F_{3}{\cdot}F_{4}{\cdot}p_{1}\big[p_{1}{\cdot}p_{3}F_{4}^{\mu\alpha}F_{3,\mu}^{\beta}{+}\\
 &\quad p_{1}{\cdot}k_{4}F_{3}^{\mu\alpha}F_{4,\mu}^{\beta}{+}F_{3}^{\alpha\beta}p_{1}{\cdot}F_{4}{\cdot}p_{2}{+}F_{4}^{\alpha\beta}p_{1}{\cdot}F_{3}{\cdot}p_{2}{+}p_{1}{\cdot}F_{3}{\cdot}F_{4}{\cdot}p_{1}\eta^{\alpha\beta}\big],
\end{split}
\end{equation}
 where $F_{i}^{\mu\nu}=2k_{i}^{[\mu}\epsilon_{i}^{\nu]}$. Quite non-trivially, we find that the Dirac trace leads to a factorized formula. The underlying reason is of course that the graviton amplitudes are universal as announced. This means these results can also be obtained via the $0 \otimes 1$ factorization that we introduce in the next subsection.

%%%%%%%%%%%%%%%%%%%%%%%%%%%%%%%%%%%%%%%%%%%%%%%%%%%%%%%%%%%%%%%%%%%%%%%%%%%%%%%%%%%%%%%%%%%%%%%%%%%%%%%%%%%%%%%%%%%%%%%%%%%%%%%%%%%%%%
\subsubsection{Exempli Gratia: The Multipole Expansion}\label{sec:multipoles}

We have introduced the operation (\ref{eq:massiveklt}) with a slight
modification in \cite{Bautista:2019tdr}. This is because the main utility of this
construction is \textit{not} the fact that we can build gravitational
amplitudes by squaring those of QCD (we have just seen that the
former follow from a dimensional reduction of the Einsten-Maxwell
system), but the fact that by rearranging the massive QCD amplitudes
in a multipole form we obtain a multipole expansion on the gravitational
side \cite{Porto:2005ac,Porto:2006bt,Levi:2015msa,Levi:2014gsa,Levi:2018nxp,Porto:2008tb,Porto:2008jj}. To our knowledge there is no systematic way of performing such
expansion in general (however, see \cite{Cotogno:2019xcl} for a recent discussion).

For spin $\frac{1}{2}$ the multipole expansion is obtained by writing
the operator $A_{n}^{{\rm QCD}}$ in powers of the intrinsic angular-momentum
operator $J^{\mu\nu}=\frac{\gamma^{\mu\nu}}{2}=\frac{1}{4}\gamma^{[\mu}\gamma^{\nu]}$.
This is usually achieved by employing the Dirac equation. For instance,
at $n=3$ it is easy to derive the textbook identity
\begin{equation}
\bar{u}_{2}A_{3}^{{\rm QCD}}v_{1}\propto m\,\epsilon_{\mu}\bar{u}_{2}\gamma^{\mu}v_{1}=\epsilon_{3}\cdot p_{1}\bar{u}_{2}v_{1}-\frac{g}{4}\,k_{3\mu}\epsilon_{3\nu}\bar{u}_{2}\gamma^{\mu\nu}v_{1}\label{eq:3ptqcd}
\end{equation}
 which also holds for the operators in (\ref{eq:massiveklt})
as they are under the support of the Dirac equation. The first term
we call the scalar piece while the second we associate to a dipole \cite{Holstein:2006pq,Holstein:2006wi}. Here we interpret $g=2$ as the corresponding form
factor and its (tree-level) value is fixed for a Dirac spinor coupled
to a photon/gluon. In the following sections we shall see that this is not true for higher spins and in fact it
is the double copy criteria above what fixes $g=2$ in general \cite{Bautista:2019tdr,Holstein:2006pq,Holstein:2006wi}.

Now consider two such multipole operators $X,Y$ of order $p,q$ respectively, namely $X\sim(\gamma^{\mu\nu})^{p}$ and $Y\sim(\gamma^{\mu\nu})^{q}$ acting on Dirac spinors.
As they involve an even number of gamma matrices, and the Dirac trace
vanishes for an odd number of such, we have
\begin{equation}
{\rm Tr(}X\fb{(g{\cdots}g)}(\slashed p_{1}{-}m)\slashed\varepsilon_{1}\bar{Y}(g{\cdots}g)(\slashed p_{2}{-}m)\slashed\varepsilon_{2})={\rm Tr}(Xp_{1}\varepsilon_{1}\bar{Y}p_{2}\varepsilon_{2})+m^{2}{\rm Tr}(X\varepsilon_{1}\bar{Y}\varepsilon_{2}),\label{eq:expdc}
\end{equation}
where the conjugated operator $\bar{Y}$ is obtained by $\gamma^{\mu\nu}\to-\gamma^{\mu\nu}$.
In the cases studied in \cite{Bautista:2019tdr} (for $n=3,4$) both terms in the
RHS coincide and hence we defined the double copy product simply as
\begin{equation}
X\odot Y=\frac{1}{2^{\left\lfloor D/\text{2}\right\rfloor }}{\rm Tr}(X\varepsilon_{1}\bar{Y}\varepsilon_{2}),
\end{equation}
i.e. using twice the second term. At $s=\frac{1}{2}$ we explicitly tested
this definition for operators up to the quadratic order in $\gamma^{\mu\nu}.$
Let us here just recall the example of $A_{3}$, which exhibits an
explicit exponential form. Combining the Dirac algebra with 3-pt.
kinematics we find $(k_{\mu}\epsilon_{\nu}\gamma^{\mu\nu})^{2}=0$,
which we use to rewrite (\ref{eq:3ptqcd}) as
\begin{equation}\label{eq:3ptextr}
\bar{u}_{2}A_{3}^{{\rm QCD}}v_{1}\propto\epsilon\cdot p_{1}\times\bar{u}_{2}e^{J}v_{1}\,,
\end{equation}
where $J$ is a Lorentz generator that reads
\begin{equation}
J=-\frac{k_{3\mu}\epsilon_{3\nu}}{\epsilon_{3}\cdot p_{1}}J^{\mu\nu}=-\frac{k_{3\mu}\epsilon_{3\nu}}{\epsilon_{3}\cdot p_{1}}\frac{\gamma^{\mu\nu}}{2}.\label{eq:gens12}
\end{equation}
The exponential form for $s=\frac{1}{2}$ generators is only linear in this case since higher multipoles vanish. Note now that while the second equality holds for $s=\frac{1}{2}$, the generator $J$
itself makes sense in any representation  \cite{Bautista:2019tdr}. In the representation
$(J^{\mu\nu})_{\beta}^{\alpha}=\eta^{\alpha[\mu}\delta_{\beta}^{\nu]}$
we can check that $(e^{J})_{\alpha}^{\beta}p_{1}^{\alpha}=(p_{1}+k)^{\beta}=-p_{2}^{\beta}$
and hence the generator acts as a boost $p_{1}\to-p_{2}$. Now we
can plug the operator \eqref{eq:3ptextr} and its conjugate in (\ref{eq:expdc}) and check that in fact
both terms yield the same contribution:

\begin{eqnarray}
A_{3}^{{\rm QCD}}\otimes A_{3}^{{\rm QCD}} & \propto & {\rm Tr(}e^{J}(\slashed p_{1}{-}m)\slashed\varepsilon_{1}e^{-J}(\slashed p_{2}{-}m)\slashed\varepsilon_{2}),\nonumber \\
 & = & {\rm Tr}(e^{J}p_{1}e^{-J}e^{J}\varepsilon_{1}e^{-J}p_{2}\varepsilon_{2})+m^{2}{\rm Tr}(e^{J}\varepsilon_{1}e^{-J}\varepsilon_{2}), \nonumber\\
 & = &- {\rm Tr}(p_{2}\tilde{\varepsilon}_{2}p_{2}\varepsilon_{2})+m^{2}{\rm Tr}(\tilde{\varepsilon}_{2}\varepsilon_{2})=2m^{2}{\rm {\rm Tr(\mathbb{I})}}\tilde{\varepsilon}_{2}\cdot\varepsilon_{2}, \label{eq:expdc-1}
\end{eqnarray}
where $\tilde{\varepsilon}_{2}^{\alpha}=(e^{J})_{\beta}^{\alpha}\varepsilon_{1}^{\beta}$
is a new polarization state for $p_{2}$, that is, it satisfies $p_{2}\cdot\tilde{\varepsilon}_{2}=0$.
Thus we obtain the gravitational (Proca) amplitude as
\begin{equation}\label{12123ptcopy}
A_{3}^{\frac{1}{2}\otimes\frac{1}{2}}\propto\epsilon_{3}\cdot p_{1}\times\varepsilon_{2}\cdot e^{J}\cdot\varepsilon_{1}=\epsilon_{3}\cdot p_{1}\varepsilon_{2}\cdot\varepsilon_{1}\fb{-}k_{3\mu}\epsilon_{3\nu}\varepsilon_{2}^{\alpha}(J^{\mu\nu})_{\alpha}^{\beta}\varepsilon_{1\beta},
\end{equation}
where higher multipoles also vanish for $s=1$, in contrast with higher spins (see \eqref{eq:mass3pths}).This simple example shows that the exponential form is preserved under
double copy (this is particular of $n=3$), but more importantly it
shows the general fact that, as observed in \cite{Bautista:2019tdr}, the gravitational
amplitude is obtained in multipole form as well. The multipole operators
can be double copied via general rules, and in turn the resulting
multipole expansion can be used to decode the classical information
contained in the amplitude.

\subsection{General case with  \texorpdfstring{ $s,\tilde{s}\protect\leq1$}{sst1}}\label{sec:genspin}

Let us now give general considerations regarding the massive KLT construction
for $s,\tilde{s}\leq1$. Following the philosophy of \cite{Arkani-Hamed:2017jhn} we know massive
amplitudes in GR (for $s\leq2$) and QCD (for $s\leq1$) can be adjusted
so that they posses a smooth high energy limit, i.e. they are free
of $1/m$ terms. This criteria was used as a definition of minimal
coupling  \cite{Arkani-Hamed:2017jhn} in these cases and completely fixed the $n=3$
amplitudes. On the other hand, it is known that for higher
spins the situation changes drastically and such divergences cannot be avoided \cite{Cucchieri:1994tx,Deser:2001dt} (more recently, see \cite{Chung:2018kqs}), which reflects the
fact that interacting massless higher spins theories are inconsistent
\cite{Arkani-Hamed:2017jhn,Benincasa:2007xk}. Here we will exploit the fact that at low spins the
minimal coupling amplitudes are ``$1/m$-free'' to construct them
directly from their massless version via dimensional reduction. We will also see how this criteria interacts with the double copy and the natural value of $g$ (as defined
in the previous section), making contact with the results of e.g. \cite{PhysRevD.46.3529,Holstein:2006pq,Holstein:2006wi,Chung:2018kqs,Pfister_2002}.

From a purely group-theoretical perspective
it is direct to construct massive states in general dimensions for
spins $s=0,\frac{1}{2},1,\frac{3}{2},2$ out of products of two lower
spins. The cases $s=0=0+0$ and $s=\frac{1}{2}=0+\frac{1}{2}$ are
obvious while the cases for $s=\,1$ we have already introduced. The
remaining situations are $2=1+1$ and $\frac{3}{2}=\frac{1}{2}+1$,
i.e.\footnote{Even though these projections hold in arbitrary dimension, in general for
$d>4$ we will have an increased number of states labeled by additional
Casimirs of the Lorentz group and not just the spin quantum number
\cite{Bekaert:2006py}.}

\begin{eqnarray}\label{eq:replace}
\varepsilon^{\mu}\tilde{\varepsilon}^{\nu} & \to & \phi^{\mu\nu}=\varepsilon^{(\mu}\tilde{\varepsilon}^{\nu)}-\left(\eta^{\mu\nu}-\frac{p^{\mu}p^{\nu}}{m^{2}}\right)\varepsilon\cdot\tilde{\varepsilon},\label{eq:gdc}\\
\psi^{\alpha}\varepsilon^{\mu} & \to & \Psi^{\alpha\mu}=\psi^{\alpha}\varepsilon^{\mu}-\varepsilon^{\nu}\frac{(\gamma^{\mu}\gamma_{\nu}\psi)^{\alpha}}{d} .
\end{eqnarray}
Our goal is to construct an interacting theory containing only such
massive states (e.g. $\phi$ or $\Psi$) but no other massive particle,
i.e. as consistent truncation of the full double copy. For the case
$s=\frac{3}{2}$ we note that we will only consider the product $1+\frac{1}{2}$
and not $0+\frac{3}{2}$. This is because, at the massless level, theories
with an interacting gravitino field are well known to be inconsistent
unless coupled to GR, and hence the factor $\,\frac{3}{2}$
in this construction cannot correspond to a QCD theory.\footnote{Even though spin-$\frac{3}{2}$ QED can be made free of $1/m$ divergences \cite{PhysRevD.46.3529} as opposed to $s\geq 2$, unitarity and causality inconsistencies (related to the Velo-Zwanziger problem \cite{Velo:1969bt}) have been stressed in e.g. \cite{Deser:2000dz,Pfister_2002}.} A similar
situation holds for the case $0+2$, see e.g. \cite{Deser:2001dt,Cortese:2013lda,Porrati:2010hm}.

\subsubsection*{Detour: Arbitrary spin at $n=3$}

The massless origin of all these constructions should be by now clear. Let us take a brief detour to emphasize some remarkable properties at $n=3$.
In $D=4$, the massless three-point amplitude is fixed from helicity
weights as \cite{Benincasa:2007xk},

\begin{equation}
A_{3}^{h_{3},h}\sim\left(\frac{\langle13\rangle}{\langle23\rangle}\right)^{2h}\left(\frac{\langle13\rangle\langle32\rangle}{\langle12\rangle}\right)^{h_{3}},\label{eq:massless3pths}
\end{equation}
for an state of arbitrary $h$ emitting a gluon ($h_{3}=1$) or a
graviton $(h_{3}=2$). Consequently, it directly satisfies the double
copy relation
\begin{equation}
A_{3}^{{\rm gr},h+\bar{h}}=K_{3}A_{3}^{{\rm QCD},h}A_{3}^{{\rm QCD},\bar{h}}.\label{eq:3ptgen}
\end{equation}
On the other hand, by implementing the multipole expansion, in \cite{Bautista:2019tdr} 
we have found that the same relation can be imposed for massive
amplitudes of arbitrary spin, and fixes their full form as
\begin{equation}
A_{3}^{h_{3},s}\sim(\epsilon_{3}\cdot p_{1})^{h_{3}}\varepsilon_{2}{\cdot}\exp\left(-\frac{k_{3\mu}\epsilon_{3\nu}}{\epsilon_{3}\cdot p_{1}}J_{s}^{\mu\nu}\right){\cdot}\varepsilon_{1},\label{eq:mass3pths}
\end{equation}
where $J_{s}^{\mu\nu}$ is the generator in e.g. (\ref{eq:gens12})
naturally adapted to higher spin $s$.\footnote{A local form of this amplitude can be found in \cite{Lorce:2009br,Lorce:2009bs,Bautista:2019tdr}, which however features $1/m$ divergences.}
Observe that this form does not depend explicitly on the mass and,
as noted in \cite{Guevara:2017csg}, reduces to (\ref{eq:massless3pths}) when
written in terms of the $D=4$ spinor helicity variables.\footnote{For a quick derivation of this fact write the polarization tensors
as $\varepsilon_{1}\propto\left(\frac{|1\rangle[\mu_{1}|}{[1\mu_{1}]}\right)^{h}$
and $\varepsilon_{2}\propto\left(\frac{|2]\langle\mu_{2}|}{\langle2\mu_{2}\rangle}\right)^{h}$, together with $\frac{k_{3\mu}\epsilon_{3\nu}}{\epsilon_{3}\cdot p_{1}}J^{\mu\nu}=\frac{\langle12\rangle}{\langle32\rangle}\langle3\frac{\partial}{\partial\lambda_{1}}\rangle$
as in e.g. \cite{Cachazo:2014fwa}. Then,
\begin{align*}
\varepsilon_{2}\cdot e^{\fb{-}\frac{\langle12\rangle}{\langle32\rangle}\langle3\frac{\partial}{\partial\lambda_{1}}\rangle}\cdot\varepsilon_{1} & =\langle\mu_{2}|e^{\frac{\langle21\rangle}{\langle32\rangle}\langle3\frac{\partial}{\partial\lambda_{1}}\rangle}|1\rangle^{h}\left(\frac{[\mu_{1}2]}{[1\mu_{1}]\langle\mu_{2}2\rangle}\right)^{h}\\
 & =\left(\langle\mu_{2}1\rangle-\frac{\langle12\rangle\langle\mu_{2}3\rangle}{\langle32\rangle}\right)^{h}\left(\frac{[\mu_{1}2]}{[1\mu_{1}]\langle\mu_{2}2\rangle}\right)^{h}=\left(\frac{\langle31\rangle}{\langle32\rangle}\right)^{2h}\,,
\end{align*}
where we have used that $e^{-\frac{\langle12\rangle}{\langle32\rangle}\langle3\frac{\partial}{\partial\lambda_{1}}\rangle}$
acts as a Lorentz boost on $|1\rangle$, see Appendix B in \cite{Bautista:2019tdr}. Finally, the $h_3$ dependence is also the same in \eqref{eq:massless3pths} and \eqref{eq:mass3pths}.} Hence (\ref{eq:mass3pths}) is nothing but the natural
extension of (\ref{eq:massless3pths}) to generic dimension and helicities, whose
dimensional reduction in the sense of the previous section is trivial.
Curiously, when interpreted as a $D=4$ massless amplitude this object is
known to be inconsistent with locality for $|h|>1$ (or analogously
$s>1$) whereas in the massive case it has the physical interpretation
given in \cite{Guevara:2018wpp,Chung:2018kqs,Arkani-Hamed:2019ymq}. On the other hand, these
inconsistencies will only appear in the ``four-point test''  \cite{Arkani-Hamed:2017jhn,Benincasa:2007xk}, namely by computing $A_{4}^{{\rm QCD}}$ or $A_{4}^{\text{{\rm gr}}}$.
In the massive case they can be cured by including
contact interactions \cite{Chung:2018kqs}.

\subsubsection*{Arbitrary multiplicity at low spins}

From the above discussion we see that at least at low spins we can extend
the relation (\ref{eq:3ptgen}) and its compactification to arbitrary
multiplicity, since the massless theory is healthy. The starting QCD theories for scalars, Dirac fermions and gluons are standard and catalogued in the next section. Let us then write
\begin{equation}
A_{n}^{h+\bar{h}}(\varphi_{1}^{h+\bar{h}}H_{3}\cdots H_{n}\varphi_{2}^{-h-\bar{h}}):=\frac{1}{2}\sum_{\alpha\beta}K_{\alpha\beta}A_{n,\alpha}^{{\rm QCD}}(\varphi_{1}^{h}g_3{\cdots}g_n\varphi_{2}^{-h})A_{n,\beta}^{{\rm QCD}}(\varphi_{1}^{\bar{h}}g_3{\cdots}g_n\varphi_{2}^{-\bar{h}}),\label{eq:def}
\end{equation}
where we have denoted by $\varphi_{i}^{h}$ the state of helicity
$h$ and particle label $i$. This extends the relation (\ref{eq:dconeline})
for the cases $h,\bar{h}\leq1$. We can also uplift it to arbitrary
dimensions. Following the previous section we first rewrite the amplitudes
in terms of the corresponding polarization vectors/spinors and the
implement the tensor products $\otimes$ between representations (besides
the trivial cases, these are just the massless versions of (\ref{eq:gdc})).
For simplicity of the argument we regard (\ref{eq:def}) as a \textit{definition}
of the object $A_{n}^{{\rm gr}}$, and we claim that it corresponds
to a tree-level amplitude in a certain QFT coupled to gravity. We recall from the previous section that this
is because 1) diffeomorphism (gauge) invariance and crossing-symmetry
are manifest and 2) tree-level unitarity follows from general arguments \cite{Bern:2010yg,Chiodaroli:2017ngp}. This means that we just need to construct
a corresponding Lagrangian to identify the theory, which we will do
for most cases in Section \ref{constructing the lagrangians}.

We have already explained how under the dimensional reduction $D=d+1\to d$
we obtain massive momenta and the corresponding propagators. We have
also shown how the $D$-dimensional polarization vectors/spinors of
the compactified particles, $\varepsilon^{\mu}$ and $u^{\alpha}$,
can now be regarded as satisfying the corresponding massive wave equations.
The result of (\ref{eq:def}) after this procedure leads to the general
formula for one-massive line

\begin{equation}
\boxed{A_{n}^{s+\tilde{s}}(\varphi_{1}^{s+\tilde{s}}H_{3}\cdots H_{n}\varphi_{2}^{s+\tilde{s}}):=\frac{1}{2}\sum_{\alpha\beta}K_{\alpha\beta}A_{n,\alpha}^{{\rm QCD}}(\varphi_{1}^{s}g_3{\cdots}g_n\varphi_{2}^{s})\otimes A_{n,\beta}^{{\rm QCD}}(\varphi_{1}^{\tilde{s}}g_3{\cdots}g_n\varphi_{2}^{\tilde{s}}).}\label{eq:genklt}
\end{equation}
which holds for $s,\tilde{s}\leq1$ and has a smooth high-energy limit
by construction. Thus, this gives a double-copy formula for the minimally-coupled
partial amplitudes defined in the sense of \cite{Arkani-Hamed:2017jhn}. 

Even though
we have not yet specified the theory, let us momentarily restrict
the states $H_{i}$ to gravitons. We have explicitly checked, by inserting
massive spinor-helicity variables, that in $D=4$ we can obtain the
gravitational and QCD amplitudes given in \cite{Arkani-Hamed:2017jhn} for $n=3,4$, see \eqref{eq:s14d} below. This establishes a $D=4$ double-copy formula between these amplitudes, analogous to the one studied in Appendix \ref{4 d double copy}. In general
dimensions, we have also checked that this agrees with the amplitudes and
double copy for $s=0,\tilde{s}\neq0$ pointed out in \cite{Bjerrum-Bohr:2013bxa}. More generally we can use \eqref{eq:replace} to recover the $1\otimes 1$  construction studied by us in \cite{Bautista:2019tdr} or extend it with the case $1 \otimes \frac{1}{2}$. We remark that these
are precisely the gravitational amplitudes used to obtain perturbative
black hole observables in \cite{Vaidya:2014kza,Guevara:2017csg,Guevara:2018wpp,Maybee:2019jus,Guevara:2019fsj}, and that for the all-graviton
case the LHS of (\ref{eq:genklt}) is unique given the sum $s+\tilde{s}$.

We now provide simple examples to illustrate these points. In the rest of this section we shall indistinctly use $\varepsilon_2$ or $\varepsilon^{*}_2$ to refer to the (conjugated) polarization of the outgoing massive state.
\subsubsection{Non-universality of Dilaton Couplings}\label{sec221}
As opposed to gravitons, we have anticipated that the dilaton field
couples differently in the $0\otimes 1$ than in the \12x12
case. So let us compute the amplitude $A_{3}(W_{1}\phi W_{2}^{*})$
via double copy of $s=0$ and $s=1$. This is to say, we take the
trace of

\begin{equation}
A_{3}^{{\rm 0\otimes1}}(W_{1}H^{\mu\nu}W_{2}^{*})=A_{3}^{{\rm QCD},s=0}(\varphi_{1}g^{\mu}\varphi_{2})A_{3}^{{\rm QCD},s=1}(W_{1}g^{\nu}W_{2}^{*})\label{eq:01dc-2}
\end{equation}
i.e. the \0x1 double copy, and contrast it with (\ref{eq:dila3pt})
from the \12x12 double copy. The spin-1 QCD
factor arising from dimensional reduction is equivalent to a covariantized Proca action plus a correction
on the gyromagnetic ratio $g$, see next section. Explicitly,

\begin{equation}\label{eq:prevqcd}
A_{3}^{{\rm QCD},s=1}(W_{1}g^{\mu}W_{2}^{*})= p_{1}^{\mu}\varepsilon_{1}\cdot\varepsilon_{2}- \varepsilon_{1}^{\alpha}(J^{\mu\nu})_{\alpha}^{\beta}\varepsilon_{2\beta} k_{3\nu}
\end{equation}
where we used that $(J^{\mu\nu})_{\beta}^{\alpha}=\eta^{\alpha[\mu}\delta_{\beta}^{\nu]}$ according to our conventions in \eqref{12123ptcopy}. Recalling that for spin-0 $A_3^{\mu}\propto p_1^{\mu }$ , the trace of \eqref{eq:01dc-2} gives 
\begin{equation}\label{eq:01dila}
{\color{black}A_{3}^{{\rm 0\otimes1}}(W_{1}\phi W_{2}^{*})=\frac{\kappa}{\sqrt{d-2}}\left(m^{2}\varepsilon_{1}{\cdot}\varepsilon_{2}+k_{3}{\cdot}\varepsilon_{1}\,k_{3}{\cdot}\varepsilon_{2}\right)},
\end{equation}
where we restored the coupling $\kappa$ in order to be more precise.
We now observe that this differs from (\ref{eq:dila3pt}) in a term proportional
to $\varepsilon_{1}{\cdot}k_{3}\,\varepsilon_{2}{\cdot}k_{3}$, controlled
by a coupling $\phi F^{2}$ with the matter field that we derive in the next section. At first this may look like a contradiction given that we pinpointed the massless origin of this double copy, namely eq. \eqref{eq:3ptgen}. Here $A_{3}^{{\rm }}(W_{1}\phi W_{2}^{*})$ should be uniquely fixed by little-group as happened for the graviton case \eqref{eq:mass3pths}. The difference however lies in the coupling constant, which vanishes in the $d{\to} 4\,, m{\to}0$ limit for  $A_{3}^{{\rm \frac{1}{2}\otimes\frac{1}{2}}}(W_{1}\phi W_{2}^{*})$ but not for $A_{3}^{{ 0\otimes 1}}(W_{1}\phi W_{2}^{*})$. Hence \textit{the reason why graviton amplitudes are the same in both \12x12 and \0x1  double-copies is not only because of its massless form \eqref{eq:massless3pths}, but also because the coupling $\kappa$ is fixed by the equivalence principle.}

A final and crucial remark is as follows. From general considerations it is known that the dilaton cannot couple linearly to the spin of a matter line \cite{Goldberger:2017ogt,Li:2018qap}. This is consistent, as we will see that \eqref{eq:01dila} contains only a quadrupole $\sim J^2$ term, but appears in contradiction with the fact that $A^{s=1}$ in \eqref{eq:prevqcd}, which carries the spin dependence, seems to have a dipole and no quadrupole. The resolution of this puzzle comes from distinguishing two types of multipoles. The first type are the covariant multipoles carrying the action of the full Lorentz group $\text{SO}(d-1,1)$, as generated by $J^{\mu\nu}$. The second type are the rotation multipoles defined by the condition $p_\mu S^{\mu\nu}=0$ with respect to e.g. the average momentum $p=\frac{p_1 + p_2}{2}$. They generate the $\text{SO}(d-1)$ rotation subgroup and in the classical limit represent the classical spin-tensor of compact objects. The relation between the two multipoles is the decomposition $\text{SO}(d-1,1)\to \text{SO}(d-1)$ explained in Appendix A of \cite{Bautista:2019tdr} (see also \cite{Maybee:2019jus}), such that one can write $J_{\mu \nu}=S_{\mu\nu}+\textrm{boost terms}.$ Using this, \eqref{eq:prevqcd} can be written as

\begin{equation}
A_{3}^{{\rm QCD},s=1,\mu}= p^{\mu}\left(1+\frac{k_{3\mu}S^{\mu \alpha}S_{\alpha}^{\,\,\nu}k_{3\nu} }{m^2 (d-3)}\right)- S^{\mu\nu} k_{3\nu}\,,
\end{equation}
where the quadrupole term $S^{\mu \alpha}S_{\alpha}^{\,\,\nu}$ is obtained precisely from the boost piece and we have stripped polarization states.\footnote{Here the massive polarization vectors have been removed and the quantum amplitude is  understood to be  an operator acting on them. On the other hand, in the classical context, $S^{\mu\nu}$ is interpreted as a spin tensor (c-number) describing the intrinsic rotation of the classical object. } The double copy now gives
\begin{align}
A_{3}^{{\rm 0\otimes1}}(W_{1}\phi W_{2}^{*}) =& \frac{\kappa}{\sqrt{d-2}}p_\mu \left[p^{\mu}\left(1+\frac{k_{3\mu}S^{\mu \alpha}S_{\alpha}^{\,\,\nu}k_{3\nu} }{m^2 (d-3)}\right)- S^{\mu\nu} k_{3\nu}\right] \nonumber \\
 =& \frac{\kappa\,m^2}{\sqrt{d-2}}\left(1+\frac{k_{3\mu}S^{\mu \alpha}S_{\alpha}^{\,\,\nu}k_{3\nu} }{m^2 (d-3)}\right)\,.
\end{align}
Comparing this to our previous result, it is clear that the term $k_3{\cdot} \varepsilon_1 k_3 {\cdot} \varepsilon_2$ in \eqref{eq:01dila} is in direct correspondence with the quadrupole operator. A similar argument holds for the \12x12 theory: In this case there is genuinely no quadrupole contribution in the QCD factor,

\begin{equation}
A_{3}^{{\rm QCD},s=\frac{1}{2},\mu}= p^{\mu}- S^{\mu\nu} k_{3\nu}\,,
\end{equation}
whereas in the double copy $A^{\frac{1}{2}\otimes \frac{1}{2}}$ the linear-in-spin terms again cancel due to $p_\mu S^{\mu \nu}=0$. We are left again with a quadrupole term $\sim S^2$, as can be also seen from \eqref{eq:dila3pt}. We conclude that the \12x12 and \0x1 theories differ in the dilaton coupling only at the level of the matter quadrupole. We come back to this point during section \ref{GGT} in the context of classical double copy.

\subsubsection{Compton Amplitude and the $g$ factor}

Moving on to $n=4$, we can explore the interplay between the double copy and the multipole expansion. Let us first quote here the spin-1 QCD result for general
gyromagnetic factor $g$ computed  by Holstein in \cite{Holstein:2006pq}

\begin{equation}
\begin{aligned}A_{4}^{{\rm QCD},s=1}(1324)= & \frac{1}{4}\bigg\{-2\varepsilon_{1}\cdot\varepsilon_{2}\left[\frac{\epsilon_{3}\cdot p_{1}\epsilon_{4}\cdot p_{2}}{p_{1}\cdot k_{3}}+\frac{\epsilon_{3}\cdot p_{2}\epsilon_{4}\cdot p_{1}}{p_{1}\cdot k_{4}}+\epsilon_{3}\cdot\epsilon_{4}\right]\\
 & -g\left[\varepsilon_{1}\cdot F_{4}\cdot\varepsilon_{2}\left(\frac{\epsilon_{3}\cdot p_{1}}{p_{1}\cdot k_{3}}-\frac{\epsilon_{3}\cdot p_{2}}{p_{1}\cdot k_{4}}\right)+\varepsilon_{1}\cdot F_{3}\cdot\varepsilon_{2}\left(\frac{\epsilon_{4}\cdot p_{2}}{p_{1}\cdot k_{3}}-\frac{\epsilon_{4}\cdot p_{1}}{p_{1}\cdot k_{4}}\right)\right]\\
 & +g^{2}\left[\frac{1}{2p_{1}\cdot k_{3}}\varepsilon_{1}\cdot F_{3}\cdot F_{4}\cdot\varepsilon_{2}-\frac{1}{2p_{1}\cdot k_{4}}\varepsilon_{1}\cdot F_{4}\cdot F_{3}\cdot\varepsilon_{2}\right]\\
 & -\frac{(g-2)^{2}}{m^{2}}\bigg[\frac{1}{2p_{1}\cdot k_{3}}\varepsilon_{1}\cdot F_{3}\cdot p_{1}\varepsilon_{2}\cdot F_{4}\cdot p_{2}\\
 & -\frac{1}{2p_{1}\cdot k_{4}}\varepsilon_{1}\cdot F_{4}\cdot p_{1}\varepsilon_{2}\cdot F_{3}\cdot p_{1}\bigg]\,\bigg\} ,
\end{aligned}
\label{eq:hlqcd}
\end{equation}
where $F_{i}^{\mu\nu}=2k_{i}^{[\mu}\epsilon_{i}^{\nu]}$. Here all momenta are outgoing and satisfy the on-shell conditions $p_{1}^{2}=p_{2}^{2}=m^{2}$
and $k_{3}^{2}=k_{4}^{2}=0$. The covariantized Proca theory is obtained
by setting $g=1$ and hence contains a $1/m$ divergence. On the other
hand, if the Proca field is identified with a $W^{\pm}$ boson of
the electroweak model we obtain $g=2$ and completely cancel the $1/m$ term. This is a general feature of the $g=2$ theory at any multiplicity \cite{PhysRevD.46.3529}.  Moreover, in this case we observe not only a well behaved high
energy limit, but also not apparent dependence on $m$ at all! This
means that the amplitude is essentially equal to its massless limit,
which corresponds to a $n=4$ color-ordered gluon amplitude, see Sec. \ref{sec:universality_scalars}  below. 

From the above we find that for this amplitude setting $g=2$ will automatically yield to the double copy relation \eqref{eq:genklt}. This is the underlying reason for the result found in \cite{Holstein:2006pq,Holstein:2006wi} \fb{for the natural value of $g$}. The converse is also true as gravitational amplitudes always have $g=2$, thus imposing the same value on its QCD factors. The universality of $g$ is a feature of the gravitational Lagrangians, independently of the covariantization or the couplings considered. It was checked explicitly in  \cite{Chung:2018kqs} and is a direct
consequence of the universal subleading soft theorem in gravity \cite{Bautista:2019tdr}.
This contrasts to QCD in that only the leading soft factor is universal
there and hence $g$ becomes a parameter. Finally, it can also be understood
from the fact that both rotating black hole or neutron stars also
yield $g=2$ indistinctly in classical GR \cite{Pfister_2002}.

Let us elaborate on the relation between \eqref{eq:hlqcd} and the 4-gluon amplitude. Pretend
that (\ref{eq:hlqcd}) (with $g=2$) is indeed the massless amplitude.
As we compactify we must send $p_{i}\to P_{i}=(p_{i},\pm m)$ and
$k_{i}\to(k_{i},0)$, while setting the polarizations $\varepsilon_{i},\epsilon_{i}$
to lie also in $D-1$ dimensions. As the amplitude itself only depends
on $p_{i}$ through $P_{i}\cdot k_{j}$ and $P_{i}\cdot\epsilon_{j}$
the extra dimensional component of $P_{i}$ drops and the mass $m$
simply does not appear. More generally, the reader can convince themselves that the only appearances of $m$ are through 1) $P_{1}\cdot P_{2}=p_{1}\cdot p_{2}+m^{2}$
or 2) $P_{1}\cdot\epsilon_{i}P_{2}\cdot\epsilon_{j}$, which we have
seen lead to $p_{1}\cdot p_{2}$ after dilaton projection. In the
first case we can use momentum conservation to write $P_{1}\cdot P_{2}=\sum_{i<j}k_{i}\cdot k_{j}$
and effectively cancel the mass dependence. Hence, if we choose a basis
of kinematic invariants that excludes $P_{1}\cdot P_{2}$ the compactification
will be trivial: The amplitudes $A_{n}$ will essentially be identical
to their massless limit \textit{except} in the cases of dilaton amplitudes,
since they contain terms like $p_{1}\cdot p_{2}=-m^{2}+\sum_{i<j}k_{i}\cdot k_{j}$.
The same observation applies to the KLT construction (\ref{eq:genklt})
and the KLT kernel introduced in the previous section. We will extend
these observations to more matter lines in Section \ref{two matter lines}.

Note also that the explicit mass dependence can as well be hidden by means of $d=4$ massive spinor-helicity variables.\footnote{See  \cite{Arkani-Hamed:2017jhn}
for the details on this formalism and \cite{Guevara:2018wpp,Bautista:2019tdr}  for a construction
of these amplitudes via soft factors.} For instance, using these variables eq. (\ref{eq:hlqcd})
with $g=2$ reads

\begin{equation}\label{eq:s14dqcd}
A_{4}^{{\rm QCD},s=1}(1324)\propto \frac{\langle 3|1|4]^2}{p_1 \cdot k_3 \, p_1 \cdot k_4 } \left([1^a 3]\langle 42^b\rangle + \langle 1^a4 \rangle [2^b3]\right)^2
\end{equation}
In this form the double copy can be performed as in Appendix \ref{4 d double copy}. For instance, from two copies of the previous spin-1 amplitude we obtain the following spin-2 amplitude:

\begin{equation}\label{eq:s14d}
A_{4}^{{\rm gr},s=2}\propto  \frac{\langle 3|1|4]^4}{p_1 \cdot k_3 \, p_1 \cdot k_4 \, k_3 \cdot k_4} \left([1^a 3]\langle 42^b\rangle + \langle 1^a 4 \rangle [2^b 3]\right)^4
\end{equation}
This result has been used to construct observables associated to the Kerr Black-Hole in \cite{Guevara:2018wpp,Chung:2018kqs}. Here we can conclude that such amplitude is nothing but the 4-graviton amplitude in higher dimensions.  Again, since there are no massless higher spin particles in flat space, this framework provides a natural explanation for the fact that $A_{4}^{{\rm gr},s>2}$ and $A_{4}^{{\rm QCD},s>1}$ must contain $\frac{1}{m}$ divergences.

\subsubsection{Universality of Scalar Multipole}\label{sec:universality_scalars}

We close this section with a final observation on the multipole expansion.
In \cite{Bautista:2019tdr}  we observed that the multipoles of section \ref{sec:multipoles}  were
universal with respect to spin for gluon or graviton emission. This
means for instance that we can consider $A_{4}^{{\rm QCD,}s}$ and
the multipole decomposition should be the same for $s=0$ and $s=1$.
Now we can prove this explicitly for the scalar multipole, which is
by definition the term proportional to the zeroth power of the angular
momentum $J^{\mu\nu}$. In (\ref{eq:hlqcd}) that would correspond
to $\varepsilon_{1}\cdot(J^{\mu\nu}){}^{0}\cdot\varepsilon_{2}=\varepsilon_{1}\cdot\varepsilon_{2}$, e.g.
\begin{equation}
\left.A_{4}^{{\rm QCD},s=1}\right|_{J^{0}}=\frac{1}{2}\left(\frac{\epsilon_{3}\cdot p_{1}\epsilon_{4}\cdot p_{2}}{p_{1}\cdot k_{3}}-\frac{\epsilon_{3}\cdot p_{2}\epsilon_{4}\cdot p_{1}}{p_{1}\cdot k_{4}}-\epsilon_{3}\cdot\epsilon_{4}\right).
\end{equation}
On the other hand for the spin-0 representation the multipole expansion
is trivial $J^{\mu\nu}\to1$. Thus the claim of universality becomes,
at any multiplicity, 
\begin{equation}
\boxed{\left.A_{n}^{{\rm QCD},s=1}\right|_{J^{0}}=A_{n}^{{\rm QCD},s=0}.}\label{eq:scalarmult}
\end{equation}
But this relation is now obvious from the massless perspective. In
fact the amplitude $A_{4}^{{\rm QCD},s=0}$ in the massless limit
is simply the special Yang-Mills Scalar (YMS) theory considered in
e.g. \cite{Brink:1976bc,Cachazo:2014xea} and reviewed in next section. It is known that amplitudes
with two scalars in such theory can be obtained from the pure Yang-Mills
amplitude by allowing two polarization vectors to explore a one-dimensional
internal space (see \cite{Cheung:2017ems} for a review). More specifically we obtain scalars by setting $\varepsilon_{i}=(0,\ldots,0,1)$
while the remaining polarizations and momenta are written as $\epsilon_{i}\to(\epsilon_{i},0)$
and $p_{i}\to(p_{i},0)$. Thus the only surviving contraction involving
$\varepsilon_{1}$, $\varepsilon_{2}$ is precisely $\varepsilon_{1}\cdot\varepsilon_{2}$.
Hence at any multiplicity the YMS amplitude is the coefficient of
$\varepsilon_{1}\cdot\varepsilon_{2}$ in the pure gluon amplitude
and (\ref{eq:scalarmult}) follows. In the case of the gravitational
theory the same situation arises for the case the fat states $H_{i}$
are only gravitons $h_{i}$. As explained in section 
\ref{sec221},
these amplitudes should depend on $s+\tilde{s}$ and not on $s,\tilde{s}$ individually. In particular
for $s+\tilde{s}\leq1$ they read

\begin{equation}
A_{n}^{0\otimes s+\tilde{s}}(\varphi_{1}^{s+\tilde{s}}h_{3}\cdots h_{n}\varphi_{2}^{s+\tilde{s}}):=\frac{1}{2}\sum_{\alpha\beta}K_{\alpha\beta}A_{n,\alpha}^{{\rm QCD}}(\varphi_{1}^{0}{\cdots}\varphi_{2}^{0})A_{n,\beta}^{{\rm QCD}}(\varphi_{1}^{s+\tilde{s}}{\cdots}\varphi_{2}^{s+\tilde{s}}),\label{eq:ssagre}
\end{equation}
i.e. the whole spin dependence can be put in a single QCD factor. Applying the construction
of the previous paragraph to such factor we find
\begin{equation}
\boxed{\left.A_{n}^{{\rm gr},s}\right|_{J^{0}}=A_{n}^{{\rm {\rm gr}},0},}
\end{equation}
for graviton emission. Observe that in the case of dilaton fields
this relation breaks down: First, the LHS of (\ref{eq:ssagre}) depends
not only on $s+\tilde{s}$ but on $s,\tilde{s}$ individually. Second,
for e.g. spin-1 terms of the type $\varepsilon_{1}\cdot\epsilon_{i}\,\varepsilon_{2}\cdot\epsilon_{j}\,\tilde{\epsilon}_{i}\cdot\tilde{\epsilon}_{j}$
in $A_{n,\beta}^{{\rm QCD}}$ will lead to extra pieces proportional
to $\varepsilon_{1}\cdot\varepsilon_{2}$ in $A_{n,\beta}^{{\rm gr}}$, thereby
altering its scalar piece. It would be interesting to generalize these
arguments for higher multipoles $(J^{\mu\nu})^{n}$.

\section{Constructing the Lagrangians} \label{constructing the lagrangians}

In this section we will provide the Lagrangians associated to the previous constructions, covering all the QCD theories and mainly focusing on the
\12x12 and \0x1 gravitational cases. This will allow us to gain further insight in the
corresponding amplitudes. On the QCD side we will employ the compactification method to obtain the actions. On the gravity side we will construct them from simple considerations in the string frame, including classical regime. 
We will check our proposal using CHY-like formulas in Appendix \ref{ap:chy}.
For two matter lines some of these Lagrangians acquire contact terms
which we further study in  Section \ref{two matter lines}.

\subsection{QCD Theories}

We start by considering the QCD factors associated to the double copy.
The cases of spin-0 and spin-$\frac{1}{2}$ are standard and we can
provide the Lagrangian for more than one matter line straight away.
The case of the QCD theory of spin-1 \cite{Holstein:2006wi,Holstein:2006pq} is more interesting
and will be treated in a separate subsection.

\subsubsection{Spins \texorpdfstring{ $s=0,\frac{1}{2}$}{s012}}\label{sec:qcdspin0}

We have explained in the previous section how the scalar theory coupled
to QCD arises from a particular compactification both in momenta and
polarization vectors. The compactification in polarization vectors
is obtained by considering a pure gluon amplitude and setting $\varepsilon_{i}=(0,\ldots0|1)$
where the non-zero component explores an ``internal space''. We
can immediately ask what happens if the internal space is enlarged
to $N$ slots, namely the scalars are obtained by setting

\begin{equation}
\varepsilon_{i}=(\underbrace{0,\ldots,0}_{D}|\underbrace{0,\ldots1,\ldots,0}_{N})\,.\label{eq:compeps}
\end{equation}
This construction is well known from string theory and the resulting
amplitudes correspond to $N$ scalars in QCD.
In other words, letting $I,J=1,\ldots,N$ the resulting amplitudes
for any number of scalar lines are given by the aforementioned ``special''
Yang-Mills scalar theory:

\begin{equation}
\mathcal{L}_{D}^{s=0}=-\frac{1}{4}{\rm tr}(F_{\mu\nu}F^{\mu\nu})+\frac{1}{2}{\rm tr}(D_{\mu}\varphi_{I}D^{\mu}\varphi^{I})\textcolor{black}{-}\frac{1}{4}{\rm tr}([\varphi^{I},\varphi^{J}][\varphi_{I},\varphi_{J}]).\label{eq:syms}
\end{equation}
The proof of this compactification is very simple and illustrative so we briefly outline it here. It follows from
decomposing the gluon polarization in $D+N$ dimensions as

\begin{equation}\label{eq:dim_red_YM}
A_{\mu}\to(A_{\mu}|\varphi_{1},\ldots,\varphi_{N}),
\end{equation}
which implies 
\begin{equation}
F_{\mu I}=D_{\mu}\varphi_{I}\,,\quad F_{IJ}=[\varphi_{I},\varphi_{J}],
\end{equation}
together with the $D$ dimensional $F_{\mu\nu}$ components. Then,
the resulting Lagrangian just follows from expanding ${\rm tr}(F^2)$. 
Note that the fields only depend on $D$ coordinates (see e.g. \cite{Cheung:2017yef}). Two key remarks which will be useful later are as follows: First, the extra dimensional (scalar) modes are always pair-produced and hence will assemble into matter lines in the Feynman diagrams. In particular this means that even after dimensional reduction the pure gluon amplitudes coincide with the ones of YM theory.
Second, as already pointed out in the original construction \cite{Brink:1976bc} of the compactified Yang-Mills action, the action (\ref{eq:syms}) indeed corresponds to
the bosonic sector of $\mathcal{N}=4$ Super Yang-Mills theory (in that case $D=4$ and $N=6$). 

Let us now provide masses to the scalars in the Lagrangian (\ref{eq:syms}). This requires to consider complex fields as is standard in KK reductions. There are a number of ways to achieve this. For instance, still following \cite{Brink:1976bc}, we can consider an even number of compact dimensions $N$ after which the scalars can be grouped as $\psi=\varphi_I+i\varphi_{I+1}$. 

Here we will instead take an alternative route that connects more directly to our previous amplitudes discussion, and therefore extends to particles with spin. Recall that so far we have constructed the double-copy formula for a single matter line \eqref{eq:genklt}. We can also consider scattering amplitudes for more matter lines as long as they have different flavors, a restriction that we impose throughout this paper. Now, for a given flavor $I$, the Lagrangian \eqref{eq:syms} takes the form $\mathcal{L_D}\supset  \textcolor{black}{\frac{1}{2}} \varphi_I \mathcal{D} \varphi^I$ (without summation) where $\mathcal{D}$ is a Hermitian operator that can depend on other fields. This Lagrangian generates the same Feynman rules than $\varphi_I^* \mathcal{D} \varphi^I$, which is the previous statement that the scalar fields are pair-produced. Repeating the argument for $I,J=1,\ldots,N$, we conclude that we can replace

\begin{equation}
\mathcal{L}_{D}^{s=0} \to -\frac{1}{4}{\rm tr}(F_{\mu\nu}F^{\mu\nu})+{\rm tr}(D_{\mu}\varphi^*_{I}D^{\mu}\varphi^{I})\textcolor{black}{-}{\rm tr}([\varphi^{*I},\varphi^{*J}][\varphi_{I},\varphi_{J}]).\label{eq:symsc}
\end{equation}
carrying a $U(1)^{N}$ flavour. After providing  masses to the complex fields, they can be turned into real fields again via the same argument. We will use this procedure in the remaining compactifications presented in this paper.  

We now proceed then  via KK reduction on a torus, $M_{D}=\mathbb{R}^{d}\times T^{N}$,
and we let each of $N$  scalars to have a non-zero momentum in one
of the circles $S^{1}$,
\begin{equation}
\varphi_{I}(x,\theta)=e^{im_{I}\theta_{I}}\varphi_{I}(x),
\end{equation}
where $0<\theta_{I}\leq\frac{2\pi}{m_{I}}$. The gluon field has no
momenta on $T^{N}$, i.e.  is $\theta$-independent, and its only non-zero
components are $A_{\mu}(x)$, where now $\mu=0,\ldots,d-1$. By acting
with the derivative $\partial_{\bar{\mu}}$, where $\bar{\mu}=0,\ldots,D-1=d+N-1$ ,
we can read off the momentum of the flavour $\varphi_{I}$:

\begin{equation}
p_{i\bar{\mu}}^{(I)}=(\underbrace{p_{i\mu}}_{d}|\underbrace{0,\ldots,m_{I},\ldots,0}_{N}).\label{eq:masscom}
\end{equation}
Thus the on-shell condition becomes $(p_{i}^{I})^{2}=p_{i}^{2}-m_{I}^{2}=0$
and, for $N=1$, this procedure is equivalent to the one described
in the previous section. It generalizes it to more massive lines by
imposing that the momenta of scalars of different flavour are orthogonal
in the KK directions, i.e. $p_{i}^{(I)}\cdot p_{j}^{(J)}=p_{i}\cdot p_{j}$
for $I\neq J$. By integration on $T^{N}$ we find the corresponding
massive action:

\begin{equation}\label{scalarqcd}
 \int d^{d}xd^{N}\theta\mathcal{L}_{D}^{s=0}\propto\int d^{d}x{\rm tr}(-\frac{1}{4}F_{\mu\nu}F^{\mu\nu}+\fb{\frac{1}{2}}D_{\mu}\varphi_{I}D^{\mu}\varphi^{I}+\fb{\frac{1}{2}}m_{I}^{2}\varphi_{I}\varphi^{I}\textcolor{black}{-}\fb{\frac{1}{4}}[\varphi^{I},\varphi^{J}][\varphi_{I},\varphi_{J}]),   
\end{equation}
which corresponds to a scalar QCD theory, with a sum over flavours $I$ implicit. Here the scalars inherit the adjoint representation from the higher-dimensional gluons. For one matter line we can nevertheless take them in the fundamental representation (see sec. \ref{sec:s1W} below) and also drop the quartic term from the Lagrangian: The double copy of the resulting theory has been studied in \cite{Luna:2017dtq} and we will come back to it in Section \ref{two matter lines}. On the other hand, by keeping the last term we have a non-trivial contact interaction between flavours. In the massless case the double copy of this theory with itself corresponds to Einstein-YM as first observed in \cite{Chiodaroli:2014xia}. In our case we will be interested in the double copy of \eqref{scalarqcd} with the spin-1 theory constructed in the next subsection, leading to the $0\otimes 1$ gravitational theory. For two matter lines we shall see how the procedure applies both with and without the quartic interaction in \eqref{scalarqcd}, see sections \ref{321alt} and \ref{sec:casee} respectively. In the classical regime we also anticipate that this distinction is irrelevant and both cases can be regarded as equivalent.
\\

Finally, we note that we can also apply the reduction procedure to massless QCD in order to get the massive theory, as discussed previously from the amplitudes perspective. Using the splitting \eqref{sigmad} we obtain, after dropping some irrelevant KK modes,

\begin{equation}\label{qcdfermions}
  \int d^{d}xd^{N}\theta\mathcal{L}_{D}^{s=\frac{1}{2}}\propto\int d^{d}x{\rm tr}\left(-\frac{1}{4}F_{\mu\nu}F^{\mu\nu}+i\bar{\psi}_{I}\Gamma_{\mu}D^{\mu}\psi^{I}-m \bar{\psi}_{I}\psi^{I}\right).
\end{equation}
In $d=4$ and for a single fermion line, we note that this reproduces
the fermion amplitudes of $\mathcal{N}=4$ SYM in the Coulomb branch. This will be useful for performing the double copy via the CHY-like formalism introduced in \cite{Cachazo:2018hqa,Geyer:2018xgb}, which we do in Appendix \ref{ap:chy}.

\subsubsection{Spin  \texorpdfstring{$s=1$}{s1}}\label{sec:s1W}

We now consider in detail the case of spin-1, that is, a complex Proca field
coupled to QCD. In order to motivate this theory we will reproduce
here the argument given by Holstein in \cite{Holstein:2006pq} regarding the natural value of $g$,  which we  used  in
\cite{Bautista:2019tdr} to derive  the three-point amplitude for spinning partilces in QED, but here we consider a slightly more general setup by promoting QED to QCD
amplitudes.

Consider first the (complex) Proca theory minimally coupled to $SU(N)$ Yang-Mills
theory,
\begin{equation}
\mathcal{L}=-\frac{1}{4}F_{\mu\nu}^{a}F_{a}^{\mu\nu}-\frac{1}{4}W_{\mu\nu}^{\bar{I}}W_{I}^{\mu\nu}+\frac{m^{2}}{2}W_{\bar{I}}^{\mu}W_{\mu}^{I},\label{eq:wym}
\end{equation}
where we have distinguished color indices $I,\bar{I}$ to emphasize
that ($W^{\bar{I}}$) $W^{I}$ transforms in the (anti)fundamental
representation. This is just a formal feature since for now we will
only consider one matter line (note also that the mass does not depend
on $I$). Here
\begin{equation}
\begin{split}
W_{\mu\nu}^{I} & = D_{\mu}W_{\nu}^{I}-D_{\nu}W_{\mu}^{I}\label{eq:defwuv},\\
D_{\mu}W_{\nu}^{I} & =  \partial_{\mu}W_{\nu}^{I}+A_{\mu}^{a}T_{a}^{I\bar{J}}W_{\nu\bar{J}}.
\end{split}
\end{equation}
Now consider the three point amplitude obtained from (\ref{eq:wym}),
\begin{equation}
A_{3}^{{\rm QCD,1}}(W_{1}^{I}A_{3}^{a}W_{2}^{\bar{J}}) =  2T^{aI\bar{J}}\times(\epsilon_{3}\cdot p_{1}\varepsilon_{1}\cdot\varepsilon_{2}^{*}-\,\epsilon_{3\mu}k_{3\nu}\varepsilon_{1}^{[\mu}\varepsilon_{2}^{*\nu]}).\label{eq:3ptqcd1}
\end{equation}
By recalling the example of (\ref{eq:3ptqcd}) we can easily identify
the scalar and dipole pieces in these two terms. Note that $\varepsilon_{1}\cdot J^{\mu\nu}\cdot\varepsilon_{2}^{*}=2 \varepsilon_{1}^{[\mu}\varepsilon_{2}^{*\nu]}$
and hence we obtain $g=1$. This is consistent with the value of $g=\frac{1}{s}$ obtained for minimally covariantized Lagrangians as conjectured by Belinfante \cite{Belinfante1953}. We
then proceed to modify the value of $g$ by adding the interaction
\begin{equation}
\mathcal{L}_{int}=\beta\,F_{\mu\nu}^{a}T_{a}^{I\bar{J}}W_{I}^{\mu}W_{\bar{J}}^{\nu}.\label{eq:lint}
\end{equation}
This interaction was studied in e.g. \cite{Holstein:2006pq} restricted to the context of QED. In such case we can take $T_{a}^{I\bar{J}}\to\delta^{+-}$ and $\mathcal{L}_{int}$ arises
from the spontaneous symmetry breaking in the $W^{\pm}$-boson model (with
$\beta=1$). In our case we need to promote this to QCD so that we
can perform the double copy at higher multiplicity. In any case, this
term precisely deforms the value of the dipole interaction to $g=1+\beta$,
because
\begin{equation}
\mathcal{L}_{int}\to -2\beta\,T_{a}^{I\bar{J}}\times\epsilon_{3\mu}k_{3\nu}\varepsilon_{1}^{[\mu}\varepsilon_{2}^{*\nu]}\,.\label{eq:bint}
\end{equation}
Now, we claim that in order for $A_{3}^{{\rm QCD}}$ to be consistent
with the double copy for the graviton states we will need to set $g=2$,
i.e. $\beta=1$ as in the electroweak model. This is because only
in such case we find\footnote{This is a slight simplification of the argument, which is what we
used in \cite{Bautista:2019tdr} at $n=3$, arbitrary spin. Actually, Holstein \cite{Holstein:2006wi} studied the double copy of $A_{4}^{{\rm QED}}$ with the purpose
of showing the $1/m$ cancellations which are equivalent to $g=2$
as we saw in \eqref{eq:hlqcd}. Of course, the amplitude $A_{4}^{{\rm gr}}$
did not feature any such divergences.}
\begin{eqnarray}
A_{3}^{{\rm QCD},0}\times A_{3}^{{\rm QCD},1} & = & A_{3}^{{\rm gr},1}(W_{1}h_{3}W_{2}), \nonumber \\
 & = & \epsilon_{3}\cdot p_{1}\times(\epsilon_{3}\cdot p_{1}\varepsilon_{1}\cdot\varepsilon_{2}^{*}-2\,\epsilon_{3\mu}k_{3\nu}\varepsilon_{1}^{[\mu}\varepsilon_{2}^{*\nu]})\label{eq:013pts}
\end{eqnarray}
Here we have stripped the coupling constants to make the comparison
direct and written the graviton polarization as $\epsilon_{3}^{\mu\nu}=\epsilon_{3}^{\mu}\epsilon_{3}^{\nu}$
for simplicity, which can then be promoted to a general polarization
$\epsilon_{3}^{\mu\nu}$. The fixing of $g=2$ follows then from the
fact that gravitational amplitudes for any spin will always lead to
$g=2$ as we outlined in the Compton example of Section \ref{sec:genspin}.

The fact that the double copy is satisfied for the $W$-boson model
but not for the ``minimally coupled'' Proca action is not a coincidence.
As we have explained, the concept of minimal coupling that we attain
here does not necessarily agree with the covariantization of derivatives
in (\ref{eq:wym}). Our condition for minimal coupling, and that of
\cite{Arkani-Hamed:2017jhn}, is that the $m\to0$ limit of $A_{n}^{{\rm QCD}}$ is
well defined at any multiplicity $n$. The $W$-boson model arises
from spontaneous symmetry breaking in $SU(2)_{L}\times U(1)_{Y}$
gauge theory, and as such, will be deformed back to Yang-Mills as
we take $m\to0$. This will precisely fix $\beta=1$ in (\ref{eq:bint})
and we now show how.

From a Feynman diagram perspective, we have already explained how
the QCD amplitudes we are after can be obtained from massive compactification
of YM amplitudes. In the case of spin-1 and a single matter line,
we interpret the cubic Feynman diagrams of $A_{n}^{{\rm YM}}$ as
associated to a color factor made of fundamental and adjoint structure
constants, following  \cite{Johansson:2014zca}. As an example, for partial amplitudes
in the half ladder (DDM) basis, we will consider the color factor
associated to the ordering $\alpha=(1\beta_{1}\ldots\beta_{n-2}2)$
as
\begin{equation}
f^{a_{1}a_{\beta_{1}}b_{1}}f^{b_{1}a_{\beta_{2}}b_{2}}\ldots f^{b_{n-3}a_{\beta_{n-2}}a_{2}}\to T_{a_{\beta_{1}}}^{I_{1}\bar{J}_{1}}T_{a_{\beta_{2}}}^{J_{1}\bar{J}_{2}}\ldots T_{a_{\beta_{n-2}}}^{J_{n-3}\bar{I}_{2}},
\end{equation}
where particles in $\{\beta_{1},\ldots,\beta_{n}\}$ are gluons and
particles 1 and 2 are bosons $W^{I_{1}},W^{\bar{I}_{2}}$ respectively.
The same operation can be repeated in any cubic color numerator of
YM, which in general means to replace $f^{abc}\to T_{a}^{I\bar{J}}$
for matter vertices or just leave them as $f^{abc}$ for the 3-gluon
vertices. This means we identify three types of color indices: $A=(a,I,\bar{I})$.\footnote{Formally we take $T_{a}^{I\bar{J}}=-T_{a}^{\bar{J}I}$ as in \cite{Johansson:2014zca}.
One must also be careful in that the structure constants $\{T_{a}^{I\bar{J}},f^{abc}\,\}$
do not form a Lie algebra (except in the $SU(2)$ case) and hence cannot be used as an input to construct a pure YM action. However, the inconsistency
appears in the Jacobi relation $T_{a}^{I\bar{J}}T_{a}^{K\bar{L}}+\ldots$
which is associated to two matter lines, which we are not interested
here: We drop such interactions in our resulting
Lagrangian.} After relabelling the structure constants and the fields accordingly,
the field strength $\mathcal{F}_{\mu\nu}^{A}$ can be split into the
components

\begin{equation}
\mathcal{F}_{\mu\nu}^{a}=F_{\mu\nu}^{a}+2\,T_{I\bar{J}}^{a}W_{[\mu}^{I}W_{\nu]}^{\bar{J}}\,,\quad\mathcal{F}_{\mu\nu}^{I}=W_{\mu\nu}^{I}\,,\quad\mathcal{F}_{\mu\nu}^{\bar{I}}=W_{\mu\nu}^{\bar{I}},
\end{equation}
where $W_{\mu\nu}$ is defined in (\ref{eq:defwuv}). Now consider
the YM action after relabelling
\begin{equation}
\frac{1}{4}\mathcal{F}_{\mu\nu}^{A}\mathcal{F}_{A}^{\mu\nu}=\frac{1}{4}F_{\mu\nu}^{a}F_{a}^{\mu\nu}+\frac{1}{4}W_{\mu\nu}^{\bar{I}}W_{I}^{\mu\nu}+F_{a}^{\mu\nu}T_{I\bar{J}}^{a}W_{\mu}^{I}W_{\nu}^{\bar{J}}+\ldots,
\end{equation}
where we have dropped the term with four $W$-bosons. Repeating the
compactification procedure, this time on a single circle $S^{1}$,
gives
\begin{equation}\label{qcdW-boson}
\mathcal{L}^{s=1}=-\frac{1}{4}F_{\mu\nu}^{a}F_{a}^{\mu\nu}-\frac{1}{4}W_{\mu\nu}^{\bar{I}}W_{I}^{\mu\nu}+\frac{m^{2}}{2}W_{\bar{I}}^{\mu}W_{\mu}^{I}-F_{a}^{\mu\nu}T_{I\bar{J}}^{a}W_{\mu}^{I}W_{\nu}^{\bar{J}},
\end{equation}
which is indeed the deformation of (\ref{eq:wym}) by the ``spin-dipole''
coupling (\ref{eq:lint}). Thus, we have shown that the massive spin-1
theory yielding the $g=2$ interaction when coupled to QCD is precisely
the compactification of Yang-Mills theory for a single matter line,
as described in Section \ref{sec:sec2}.

\subsection{Proposal for Gravitational Theories}

Let us now introduce the gravitational Lagrangians. We begin by a construction of both $0\otimes 1$ and \12x12 theories in the string
frame, following some simple guidelines. First, let us assume momentarily
that the base massless theory, leading to the amplitudes $A_{n}^{{\rm gr}}(\gamma^{-}h_{3}{\cdots}h_{n}\gamma^{+})$
is indeed Einstein-Maxwell in both \12x12 and
\0x1 cases,
\begin{equation}
\mathcal{L}_{{\rm base}}=-\sqrt{g}\left[\frac{2}{\kappa^{2}}R+\frac{1}{2}F_{\mu\nu}^{*}F^{\mu\nu}\right].
\end{equation}
This allow us to signal the crucial difference between the \12x12
and \0x1 theories in the dilaton coupling. Following \cite{9780511816079},
in the string frame this can be generated by adding the kinetic term
and promoting $\sqrt{g}\to\sqrt{g}e^{-\frac{\kappa}{2}\phi}$. Thus
we propose
\begin{eqnarray}
\mathcal{L}_{{\rm base}}^{0\otimes1} & = & \sqrt{g}e^{-\frac{\kappa}{2}\phi}\left[-\frac{2}{\kappa^{2}}R+\frac{1}{2}(\partial\phi)^{2}-\frac{1}{2}F_{\mu\nu}^{*}F^{\mu\nu}\right],\\
\mathcal{L}_{{\rm base}}^{\frac{1}{2}\otimes\frac{1}{2}} & = & \sqrt{g}e^{-\frac{\kappa}{2}\phi}\left[-\frac{2}{\kappa^{2}}R+\frac{1}{2}(\partial\phi)^{2}\right]-\sqrt{g}\times\frac{1}{2}F_{\mu\nu}^{*}F^{\mu\nu}.
\end{eqnarray}
We now see the difference lies in the fact that the Maxwell term has
been added before and after incorporating the dilaton, respectively.
The coupling of the dilaton is simpler and in a sense trivial in the
\12x12 theory, which is characteristic of the
Brans-Dicke-Maxwell action \cite{Frau:1997mq}. In fact, we can take such
theory into the so-called Jordan frame by setting
\begin{equation}
\phi=-\frac{2}{\kappa}\ln\Phi,
\end{equation}
which leads to the standard Brans-Dicke theory \cite{Sheykhi:2008tt}
\begin{equation}
\mathcal{L}_{{\rm base}}^{\frac{1}{2}\otimes\frac{1}{2}}=\frac{2}{\kappa^{2}}\sqrt{g}\left[-\Phi R+\frac{(\partial\Phi)^{2}}{\Phi}-\frac{\kappa^{2}}{2}\times\frac{1}{2}F_{\mu\nu}^{*}F^{\mu\nu}\right]\,.
\end{equation}
On the other hand, our proposal that the \0x1 action involves
a non-trivial coupling to the dilaton arises from a careful consideration
of the classical results of \cite{Goldberger:2017ogt}, which construction we
further realize in Section \ref{two matter lines} as a double copy of a spinning source (e.g.
$s=1$) in QCD with a scalar theory ($s=0$).

At this point we can generate a mass term by performing the compactification
on a circle, $M_{D}=\mathbb{R}^{d}\times S^{1}$, letting the Proca
field to have a non-zero (quantized) momentum on $S^{1}$
\begin{equation}
A_{\mu}(x,\theta)=e^{im\theta}A_{\mu}(x)\,,
\end{equation}
whereas the remaining fields have not, i.e. $h_{\mu\nu}(x)$ and $\phi(x)$
are $\theta$-independent. Notice we have also implicitly restricted
the polarizations to lie in $d=D-1$ dimensions. For instance, the
full metric reads
\begin{equation}
g_{\bar{\mu}\bar{\nu}}=\eta_{\bar{\mu}\bar{\nu}}+\frac{\kappa}{2}h_{\bar{\mu}\bar{\nu}}\,,
\end{equation}
but $h_{\bar{\mu}\bar{\nu}}$ only has non-zero components $h_{\mu\nu}$. This relies on the assumption, exemplified in section \ref{sec:qcdspin0}, that additional KK components will assemble into matter lines and hence can be decoupled. The only exception is the dilaton field, which would in principle obtain a contribution from the extra component $h_{DD}$ in $h_{\bar{\mu}\bar{\nu}}$. The reason we set this component to zero beforehand is precisely to reproduce our prescription \eqref{eq:pr2} as opposed to \eqref{eq:pr1} (which would lead to the standard dimensional reduction of the dilaton amplitudes).

After this clarification we can now readily perform the integration of the action
over the compact direction, leading to
\begin{equation}
\frac{1}{2\pi}\int d^{d}xd\theta\,\mathcal{L}_{{\rm base}}{=}\int d^{d}x\,\sqrt{g}\begin{cases}
e^{{-}\frac{\kappa}{2}\phi}\left[-\frac{2}{\kappa^{2}}R{-}\frac{1}{2}(\partial\phi)^{2}{-}\frac{1}{2}F_{\mu\nu}^{*}F^{\mu\nu}{+}m^{2}A_{\mu}^{*}A^{\mu}\right] & ,\,\text{for }0\otimes1\\
e^{{-}\frac{\kappa}{2}\phi}\left[-\frac{2}{\kappa^{2}}R{-}\frac{1}{2}(\partial\phi)^{2}\right]{-}\frac{1}{2}F_{\mu\nu}^{*}F^{\mu\nu}{+}m^{2}A_{\mu}^{*}A^{\mu} & ,\,\text{for }\frac{1}{2}\otimes\frac{1}{2}
\end{cases}
\end{equation}
The key point here is that we have performed the compactification
in the string frame, where the dilaton coupling is trivial. We can
move to the Einstein frame by setting $g_{\mu\nu}\to e^{-\frac{\kappa\phi}{d-2}}g_{\mu\nu}$.
Perturbatively, this is equivalent to a change of basis in the asymptotic
states, given by 
\begin{equation}
h_{\mu\nu}\to h_{\mu\nu}-\frac{\phi}{d-2}\eta_{\mu\nu}+\mathcal{O}(\kappa),
\end{equation}
which means the amplitudes in this frame can be computed as linear
combinations of the string frame ones. Returning to the Lagrangian,
we use
\begin{eqnarray}
R & \to & e^{-\frac{\kappa\phi}{d-2}}(R-\kappa\frac{d-1}{d-2}D^{2}\phi-\frac{d-1}{d-2}\frac{\kappa^{2}}{4}\partial_{\mu}\phi\partial^{\mu}\phi)
\end{eqnarray}
after which we perform a trivial rescaling ($\phi\to(d-2)\phi$) to
get
\begin{equation}
\mathcal{L}^{\frac{1}{2}\otimes\frac{1}{2}}=\sqrt{g}\left[-\frac{2}{\kappa^{2}}R+\frac{(d-2)}{2}(\partial\phi)^{2}-\frac{1}{2}e^{\frac{\kappa}{2}(d-4)\phi}F_{\mu\nu}^{*}F^{\mu\nu}+m^{2}e^{\frac{\kappa}{2}(d-2)\phi}A_{\mu}^{*}A^{\mu}\right],\label{eq:1212dc-1}
\end{equation}
and
\begin{equation}
\mathcal{L}^{0\otimes1}=\sqrt{g}\left[-\frac{2}{\kappa^{2}}R+\frac{(d-2)}{2}(\partial\phi)^{2}-\frac{1}{2}e^{-\kappa\phi}F_{\mu\nu}^{*}F^{\mu\nu}+m^{2}A_{\mu}^{*}A^{\mu}\right].\label{eq:01dc-1}
\end{equation}
Note that only in $d=4$ the dilaton is not sourced by matter in the
\12x12 theory. Indeed, consider momentarily
the massless limit $m=0$. A general Einstein-Maxwell-Dilaton theory
in four dimensions can be classified in the Einstein frame from the
coupling $e^{-\kappa\alpha\phi}F^{2}$, with $0\leq\alpha\leq\sqrt{3}$ \cite{PhysRevD.46.1340,Pacilio:2018gom}. The Brans-Dicke
theory corresponds to $\alpha=0$ whereas the low-energy limit of
string theory yields $\alpha=1$. This is not surprising as we will
soon identify the \0x1 with a dimensional extension of $\mathcal{N}=4$
Supergravity. We should mention that the $\alpha=\sqrt{3}$ case is
characteristic of the well-known five dimensional KK theory, whose
double copy structure was considered in \cite{Chiodaroli:2014xia}.

These actions would be enough for amplitudes involving only gravitons,
dilatons and two Proca fields as external states. However, in the
case of the \0x1 theory we have seen that axions can be sourced
by matter. Keeping the classical application in mind, this means that
for two matter lines we will need to compute such contributions, as
they will appear as virtual states. We begin by constructing the interaction
that reproduces single matter-line amplitudes with external axions.

In order to introduce the axion coupling in the \0x1 theory we
again resort to the classical results of \cite{Goldberger:2017ogt}, which found
that in the string-frame the axion couples to the matter through 
\begin{equation}
\kappa\int d\tau\,H_{\mu\nu\rho}v^{\mu}S^{\nu\rho}.\label{eq:3ptwl}
\end{equation}
Here $S^{\mu \nu}$ is the spin operator as introduced in section \ref{sec221}. This coupling can be reproduced in QFT by computing a ``three-point''
amplitude between the dipole and the axion,
\begin{equation}\label{eq:target}
A_{3}^{\mu\nu}\propto\kappa p^{[\mu}\times S^{\nu]\rho}q_{\rho},
\end{equation}
where $q^{\mu}$ and $p^{\mu}$ are the momentum of the axion and
the matter line respectively. As predicted, we identify the first
factor as the scalar 3pt. amplitude $A_{3}^{\mu,s=0}\propto p^{\mu}$
and the second factor as the dipole of the spin-1 amplitude $\left.A_{3}^{\mu,s=1}\right|_{J}\propto S^{\mu\rho}q_{\rho}$ \cite{Bautista:2019tdr}, which signals this corresponds to the \0x1 theory. The
overall proportionality factor can be adjusted accordingly. The QFT
3-pt. vertex leading to \eqref{eq:target} is then the direct analog of (\ref{eq:3ptwl}), That is, after identifying $S^{\mu \nu} \to J^{\mu \nu}$ up to longitudinal terms, (\ref{eq:3ptwl}) becomes
\begin{equation}
- B_{ \mu\nu}(q)\times \kappa p_2^{\mu}A^{*[\nu}(p_2)A^{\rho]}(p_1)q_{\rho}\rightarrow \frac{\kappa}{2}H_{\mu\nu\rho}\partial^{\mu}A^{*[\nu}A^{\rho]}=\frac{\kappa}{4}H_{\mu\nu\rho}A^{*\mu}F^{\nu\rho}.
\end{equation}
Attaching then the canonically normalized kinetic term $\frac{1}{6}H_{\mu\nu\rho}H^{\mu\nu\rho}$
we can readily take this vertex into the Einstein frame (also applying
the aforementioned rescaling to $\phi$), 
\begin{equation}
\sqrt{g}e^{-\frac{\kappa}{2}\phi}\times\frac{1}{6}H_{\mu\nu\rho}(H^{\mu\nu\rho}+\frac{3\kappa}{2}\left(A^{*\mu}F^{\nu\rho}{+}\rm{c.c.}\right))\to\sqrt{g}e^{-2\kappa\phi}\times\frac{1}{6}H_{\mu\nu\rho}(H^{\mu\nu\rho}+\frac{3\kappa}{2}\left(A^{*\mu}F^{\nu\rho}{+}\rm{c.c.}\right)).
\end{equation}
Note that this term is not deformed by the massive compactification
since the derivatives in $F^{\mu \nu}$ are contracted with $H_{\mu\nu\rho}$ living
in $d=D-1$ dimensions. We note that the complex character of the fields is important for the following compactification. However, once the compactification is done we are left with a quadratic action in the Proca field, which can then be turned into real invoking the argument above \eqref{eq:symsc}. 
% .\footnote{\label{transition_to_real}
% More explicitly, note that our Lagrangian for spin-1 fields can be written as $\mathcal{L}\supset A^*_\mu \mathcal{D}^{\mu\nu}A_{\nu} +\textrm{c.c.}$, where the operator $\mathcal{D}^{\mu\nu}$ is real and depends only on the other fields. This entails that there are no interaction terms between $Re(A_\mu)$ and $Im(A_\mu)$.
% Alternatively, note that the Lagrangian leads to Feynman rules for a single matter line of the form $\varepsilon^*_{2,\mu} (\mathcal{D}^{\mu \nu}+\mathcal{D}^{\nu \mu} )\varepsilon_{1,\nu}  $ (with a slight abuse of notation). On the other hand, for real fields the Lagrangian $\mathcal{L}\supset A_\mu \mathcal{D}^{\mu \nu }A_\nu$ leads to the Feynman rules $\varepsilon_{2,\mu} (\mathcal{D}^{\mu \nu}+\mathcal{D}^{\nu \mu} )\varepsilon_{1,\nu}  $ due to symmetrization of spin-1 states. We conclude that amplitudes in these theories then take the same form so we can omit the real/complex distinction hereafter. Note that the argument still applies to quartic matter interactions in our double copy theories such as \eqref{eq:final01} or \eqref{eq:10LAGRANGIAN} since the fields have different flavor contractions or simply correspond to different species. \label{fn15}} 
Thus we finally arrive at the action principle presented
in the introduction for one matter line:
\begin{equation}
\mathcal{L}^{0\otimes1}{=}\sqrt{g}\left[{-}\frac{2R}{\kappa^{2}}{+}\frac{(d{-}2)}{2}(\partial\phi)^{2}{-}\frac{e^{{-}2\kappa\phi}}{6}H_{\mu\nu\rho}(H^{\mu\nu\rho}{+}\frac{3\kappa}{2}A^{\mu}F^{\nu\rho}){-}\frac{1}{4}e^{{-}\kappa\phi}F_{\mu\nu}F^{\mu\nu}{+}\frac{m^{2}}{2}A_{\mu}A^{\mu}\right]\label{eq:final011line}
\end{equation}
Note that the massless sector corresponds to $\mathcal{N}=0$ Supergravity \cite{9780511816079}
as seen also in \cite{Goldberger:2017ogt}. We will rederive
this result from a pure on-shell point of view in the following subsubsection,
and extend it to two-matter lines. We will also perform various checks
in our proposals for both \12x12 and \0x1
actions. We can also already draw some conclusion regarding the interactions:
Even though the axion is sourced by the Proca field, it is pair produced
in the massless sector. This means that the axion is projected
out in amplitudes involving external gravitons and dilatons with a
single matter line, just as in the \12x12 theory. More importantly, an analogous
reasoning can be applied to dilatons to show that in both \0x1
and \12x12 theories the graviton emission
amplitudes are precisely the same, as we observed first in \cite{Bautista:2019tdr}.  Now, as we have mentioned, when the dilaton is included as an external state its coupling
differs in both theories: In particular, it follows from \eqref{eq:1212dc-1} that in the massless four dimensional
case the dilaton is not sourced by the photon in the \12x12 theory,
see e.g. the 4-pt. example in \cite{Johansson:2014zca}. 

\subsubsection{Alternative Construction of the  \texorpdfstring{\0x1}{01} Action}\label{321alt}

From the identifications we have performed we can provide an additional
argument to obtain $\mathcal{L}^{0\otimes1}$. First consider the massless
case $m=0$. Then the \0x1 theory in any dimension is obtained
from the massless version of (\ref{eq:genklt}); the double copy between
scalar QCD ($s=0$) and pure Yang-Mills ($s=1$). In section \ref{sec:qcdspin0}  we
have identified the scalar QCD theory as the bosonic sector of $\mathcal{N}=4$
SYM in four dimensions. It should be clear however that if these amplitudes
are computed without imposing any kind of Gram identities (or plugging
spinor-helicity variables) the resulting object is dimension independent
and trivially extends the bosonic sector of $\mathcal{N}=4$ SYM.
This is can be achieved for instance by computing the compactification
of YM explicitly via the CHY formulation \cite{Cachazo:2014xea} as in Appendix \ref{ap:chy}.

Now, as the double copy in (\ref{eq:genklt}) is precisely obtained
via the standard massless KLT, we know that in four dimensions it yields \cite{Bern:2010ue}

\begin{equation}
(\mathcal{N}=4\,{\rm SUGRA})=(\mathcal{N}=4\,{\rm SYM})\otimes({\rm pure}\,{\rm YM})\,,\label{eq:dcknown}
\end{equation}
Observe that the fermionic content in this theory only comes from
the SYM factor. This means we can consistently truncate to the bosonic
sector at both sides of the equality. Thus we learn that the \0x1
theory in four dimensions corresponds to the bosonic sector of $\mathcal{N}=4$
SUGRA specialized to a single matter line. Recall also that relation
(\ref{eq:dcknown}) follows from truncating the spectrum of $\mathcal{N=}8$
SUGRA, and the corresponding bosonic on-shell states are obtained
as
\begin{equation}
\{h^{\mu\nu},B^{\mu\nu},\phi\}\cup\{\gamma^{I\nu}\}=\{g^{\mu},\phi^{I}\}\otimes\{\tilde{g}^{\nu}\}\,,\label{eq:spec}
\end{equation}
where $I=1,\ldots,6$, with $\phi^I$ denoting six different adjoint scalars. Note that in this section we will consider a real photon or Proca field
interacting gravitationally, this distinction is only relevant in
QCD. Even though we have written the action for a single matter line,
the fact the spectrum has flavoured fields suggests that we
can promote the construction to more matter lines and yet employ the
standard KLT kernel as we will explain shortly. In the next section
we will contrast this with the \12x12 theory
and a different approach to the \0x1 construction obtained via
the BCJ prescription. However, this example will already illustrate
an important feature of these constructions: In general we will find
matter contact interactions when more massive lines are included;
these contact interactions however will not affect the classical limit.

Our strategy is analogous to the one in Section \ref{sec:sec2}. We first rewrite $\mathcal{N}=4$
SUGRA in a way in which the action is not sensitive to the dimension $d$. We then uplift it to general $d$ and compactify it on a torus. Starting
from the standard Lagrangian given in \cite{CREMMER197861,PhysRevD.15.2805}, i.e.
\begin{equation}
\mathcal{L}^{\mathcal{N=}4}=\sqrt{g}\left[R-2(\partial\phi)^{2}-2e^{4\phi}(\partial\chi)^{2}-e^{-2\phi}F_{\mu\nu}^{I}F_{I}^{\mu\nu}-2\chi F_{\mu\nu}^{I}\star F_{I}^{\mu\nu}\right] \,
\end{equation}
where $\chi$ is the dual axion field, we review in Appendix \ref{ap:niktow} the
construction of Nicolai and Townsend and obtain
\begin{equation}
\sqrt{g}\left[R-2(\partial\phi)^{2}+3e^{-4\phi}(A_{I}^{\nu}F^{I\rho\sigma}+\frac{1}{6}H^{\nu\rho\sigma})(A_{J\nu}F_{\rho\sigma}^{J}+\frac{1}{6}H_{\nu\rho\sigma})-e^{-2\phi}F_{\mu\nu}^{I}F_{I}^{\mu\nu}\right]\,,
\end{equation}
from dualizing the axion off-shell. This generates the term $\sim A^{2}F^{2}$, representing a new contact
interaction between flavours which will appear for two matter lines. Even though this action makes sense
in any dimension, we know it is in principle still four dimensional.
To obtain a faithful dimensional continuation of $\mathcal{N=}4$
SUGRA we need to show that the amplitudes are indeed dimension independent.
By looking at $2\to2$ scattering of photons (i.e. Proca fields) we
can conjecture that this is achieved by promoting the kinetic
term of the dilaton as $2(\partial\phi)^{2}\to(d-2)(\partial\phi)^{2}$
(such that it cancels the factor $\frac{1}{d-2}$ in the graviton
propagator). This is exactly what we obtained from adopting the Einstein
frame in the previous subsection. We can now swap to the string frame and
perform the compactification on a torus $T^{6}$ as in the scalar
case. After checking that the term $A_{I}^{\nu*}F^{I\rho\sigma}A_{J\nu}^{*}F_{\rho\sigma}^{J}$
will not generate mass dependence we get to the
following form

\begin{equation}
\sqrt{g}\left[R{-}(d{-}2)(\partial\phi)^{2}{+}3e^{{-}4\phi}(A_{I}^{\nu}F^{I\rho\sigma}{+}\frac{1}{6}H^{\nu\rho\sigma})(A_{J\nu}F_{\rho\sigma}^{J}{+}\frac{H_{\nu\rho\sigma}}{6}){-}e^{{-}2\phi}F_{\mu\nu}^{I}F_{I}^{\mu\nu}{+}2m_{I}^{2}A_{\mu}^{I}A_{I}^{\mu}\right]\, 
\end{equation}
%%%
(recall that we will only consider amplitudes for Proca pairs of different flavours). In order to construct perturbation theory in the gravitational constant
$\kappa$ we scale $\phi\to\frac{\kappa}{2}\phi,\,B_{\mu\nu}\to\kappa B_{\mu\nu},A_{\mu}\to\frac{\kappa}{2\sqrt{2}}A_{\mu}$
and attach and overall factor of $-\frac{2}{\kappa^{2}}$ leading
to

\begin{equation}\label{eq:final01}
\begin{split}
& \sqrt{g}\bigg[{-}\frac{2}{\kappa^{2}}R{+}\frac{(d{-}2)}{2}(\partial\phi)^{2}{-}\frac{1}{6}e^{{-}2\kappa\phi}(\frac{3\kappa}{4}A_{I}^{\nu}F^{I\rho\sigma}{+}H^{\nu\rho\sigma})(\frac{3\kappa}{4}A_{J\nu}F_{\rho\sigma}^{J}{+}H_{\nu\rho\sigma})\\
&\qquad \qquad {-}\frac{1}{4}e^{{-}\kappa\phi}F_{\mu\nu}^{I}F_{I}^{\mu\nu}{+}\frac{1}{2}m_{I}^{2}A_{\mu}^{I}A_{I}^{\mu}\bigg]\,.
\end{split}
\end{equation}
%%%
For one matter line this precisely agrees with (\ref{eq:final011line}).
For more Proca fields it features the aforementioned contact interaction.
Note that amplitudes for \eqref{eq:final01} can be computed by again compactifying the
KLT relation, where the massive photons in (\ref{eq:spec}) appear
as products of massive scalars with massive $W$-bosons, see Appendix \ref{ap:01chy} for an example with two matter lines.\footnote{This example will also illustrate how the prescription \eqref{eq:pr2} for dilaton dimensional reduction is automatically implemented also for internal dilatons. Namely, the KK components that could have contributed to internal dilatons decouple because they are orthogonal for different flavor lines.} Both the massive scalar and the $W$-boson theory will as well contain interactions between flavours: In the massive scalar case, recall the special YMS action \eqref{scalarqcd} which included a quartic term for scalars. In the W-boson case, recall that in Section \ref{sec:s1W} we established
the compactification of YM into the spin-1 theory only for a single matter
line. This is precisely because for two matter lines the following diagram appears from compactification (the KK momenta are given by the masses $m_a$ and $m_b$):
\begin{figure}[h!]
\begin{equation}\label{eq:ndiag}
  \includegraphics[width=56mm]{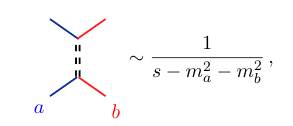}
\end{equation}
\end{figure}
% \begin{equation}\label{eq:ndiag}
%     \begin{fmffile}{m4new}
%      \parbox{40pt}{
%     \begin{fmfgraph*}(30,42)
%     \fmfstraight
%     \fmfleft{i1,i2}
%     \fmfright{o1,o2}
%     \fmf{plain,foreground=(0.035,,0.168,,0.623)}{i1,v1}
%     \fmf{dbl_dashes}{v1,v2}
%     \fmf{plain,foreground=(0.035,,0.168,,0.623)}{v2,i2}
%     \fmf{plain,foreground=(1,,0.1,,0.1)}{o1,v1}
%     \fmf{plain,foreground=(1,,0.1,,0.1)}{v2,o2}
%   %  \marrow{a}{right}{top}{}{v1,i1}
%     %\marrow{b}{right}{top}{}{v2,v1}
%     %\marrow{c}{right}{top}{}{o1,v2}
%     \fmflabel{\textcolor{blue}{$a$}}{i1}
%     \fmflabel{\textcolor{red}{$b$}}{o1}
%     %\fmflabel{$3,c$}{v3}
%     %\fmflabel{$4,d$}{v4}
%     \end{fmfgraph*}}
%     \end{fmffile}
%     \sim\frac{1}{s-m_{a}^{2}-m_{b}^{2}}\, ,
% \end{equation}\\
%
%\vspace{0.1cm}
where two interacting flavours generate an extra massive particle with $M^{2}=m_{a}^{2}+m_{b}^{2}$. This pole is however cancelled by the KLT kernel yielding only contact interactions between flavours in \eqref{eq:final01}.

At this point it may be desirable to remove flavor interactions such as the ones present in diagram \eqref{eq:ndiag}, hence also removing the extra massive pole. In the massless theory this diagram is required by gauge invariance, but given that for massive particles this is not a constraint the truncation would in principle define valid tree-level amplitudes. However, one must be careful in that without gauge invariance the longitudinal modes $\epsilon^\mu = \frac{p^\mu}{m}$ may lead to divergent amplitudes as we approach back to the massless case $m \to 0$.\footnote{We thank A. Ochirov and H. Johansson for this observation, see also the introduction of \cite{Johansson:2019dnu}.} On the other hand, it is known that flavor interactions such as \eqref{eq:ndiag} will not contribute to the classical regime (e.g. to a long-range potential) and hence the truncated amplitudes should lead to the same classical observables as the original theory, which will indeed be finite as $m\to 0$. Motivated by this, in the next section we will construct an alternative double copy for $0\otimes 1$ at two matter lines, leading to the same classical observables as the extension of $\mathcal{N=}4$ SUGRA.

Despite the evidence we have presented so far it is important to perform
explicit checks on our proposed Lagrangians for both \0x1 and
\12x12 theories. We have checked explicitly
the amplitudes, in arbitrary dimension, with all possible combinations
of external states for $n=3,4,5$ points, including the two matter-line
cases presented in Section \ref{two matter lines}. We have also performed six point checks
for dilaton amplitudes, allowing us to test the exponential in (\ref{eq:1212dc-1})
up to fourth order. The analytical results for some of  these checks are too long to be printed in this paper, we therefore provide a Mathematica notebook with those amplitudes. These checks were also  done efficiently via the CHY-like
implementations we detail in Appendix \ref{ap:chy}.

\section{Two matter lines from the  BCJ construction}\label{two matter lines}

So far we have used the KLT double copy mostly to compute the
amplitudes $A_{n}$, i.e. those involving one matter line. To test the extent of the double copy it is important to include more matter lines transforming in the fundamental representation. In our case it will be enough to consider two matter lines of different flavours in order to make contact with the classical results mentioned in the introduction. These amplitudes lose many nice features of the $A_n$ amplitudes: For instance we cannot trivially remove the dilaton-axion propagation nor write the multipole expansion directly. We shall anyhow conclude that the relevant classical information is already contained in the $A_n$ amplitudes, as pointed out in e.g. \cite{Neill:2013wsa}, which we have used to remove the dilaton/axion from the classical perspective in \cite{Bautista:2019tdr}.

For more than one matter line a basis of amplitudes based on Dyck words was introduced by Melia \cite{Melia:2013bta,Melia:2013epa} and later refined by Johansson and Ochirov \cite{Johansson:2015oia,Melia:2015ika}.\footnote{The amplitudes in Melia basis satisfy a restricted set of BCJ relations \cite{Johansson:2015oia,delaCruz:2015dpa}, and consequently a \textit{generalized} KLT construction has been recently introduced in \cite{Brown:2018wss,Johansson:2019dnu}, see also \cite{delaCruz:2016wbr}. For loop level extensions of colour-kinematics duality in this context see \cite{Ochirov:2019mtf,Kalin:2018thp,Johansson:2017bfl,Kalin:2017oqr}.} Since we only consider here two matter lines we choose to resort instead to the BCJ representation, thereby extending the approach of \cite{Luna:2017dtq}. The equivalence between the approaches has been detailed, including spin-$\frac{1}{2}$ applications, in e.g. \cite{delaCruz:2016wbr}.

Consider the two matter  lines to have mass $m_a$  and $m_b$, and spin $s_a$ and $s_b$. 
For QCD scattering, the two  massive particles have different flavours, and  we restrict their spins to  lie 
in $\left\{0,\frac{1}{2},1\right\}$. These amplitudes are defined by the Lagrangians provided in Section \ref{constructing the lagrangians}: For the spin-0 case we use the scalar QCD Lagrangian $(\ref{scalarqcd})$ with the removed quartic term as per our previous discussion; for spin-1 we use the $W-$boson model $(\ref{qcdW-boson})$
and for spin-$1/2$ we use the standard QCD Lagrangian for massive Dirac
fermions $(\ref{qcdfermions})$. \\
Following the BCJ prescription we arrange the QCD
amplitudes into a sum of the form
\begin{equation}
M_{n}^{\rm{QCD}}=\sum_{i\in\Gamma}\frac{c_{i}n_{i}^{(s_{a},s_{b})}}{d_{i}},\label{eq:qcd bcj}
\end{equation}
running over the set $\Gamma$ of all cubic diagrams, with denominators $d_i$. The superscript $(s_{a},s_{b})$ here denotes the spin of the lines and may be omitted. For a given triplet $(i,j,k)$, if the color factors satisfy the Jacobi identity
\begin{equation}\label{jacobi}
c_{i}\pm c_{j}=\pm c_{k},
\end{equation}
then colour kinematics duality requires there is a choice of numerators $n_i$ such that
\begin{equation}\label{color-kinematic}
  n_{i}\pm n_{j}=\pm n_{k}.  
\end{equation}
The gravitational amplitudes can be computed starting  from $(\ref{eq:qcd bcj})$ by replacing the color factors with further kinematic factors, which can be associated to a different QCD theory. In this section
we will explore some of the choices for QCD theories, and write the explicit form of the resulting gravitational Lagrangians. With this in mind, the $n-$point
gravitational amplitude, where now the massive lines have spins $s_a + \tilde{s}_a$ and $s_b+\tilde{s}_b$ respectively, reads
\begin{equation}
M_{n}^{(s_a \otimes \tilde{s}_a, s_b \otimes \tilde{s}_b)}=\sum_{i\in\Gamma}\frac{n^{(s_a,s_b)}_{i}\otimes \tilde{n}^{(\tilde{s}_a,\tilde{s}_b)}_{i}}{d_{i}},\label{eq:BCJ gravity}
\end{equation}
 where the product $\otimes$  depends on the spin of the massive particles in the QCD theory.
For instance, 
for $s_a=\tilde{s}_a=s_b=\tilde{s}_b=1/2$ we define it in an analogous way to the case of only  one matter line  $(\ref{tensor product })$; that is: consider the spin $\frac{1}{2}$ operators $\mathcal{X}_{i}$ and $\mathcal{Y}_{i}$, entering in a
QCD numerator $n^{\rm{QCD}}$ with four external fermions whose momenta we choose to be all
 outgoing as follows
\begin{equation}
n^{(\frac{1}{2},\frac{1}{2})}=\bar{u}_{2}\mathcal{X}_{i}v_{1}\bar{u}_{4}\mathcal{Y}_{i}v_{3},
\end{equation}
analogously, the charge conjugated numerator reads 
\begin{equation}
\bar{n}^{(\frac{1}{2},\frac{1}{2})}=\bar{u}_{1}\bar{\mathcal{X}}_{i}v_{2}\bar{u}_{3}\bar{\mathcal{Y}}_{i}v_{4}.
\end{equation}
 We define the spin-1 gravitational numerator as the tensor product of the two QCD numerators as follows:
\begin{equation}
n^{(\frac{1}{2},\frac{1}{2})}\otimes\bar{n}^{(\frac{1}{2},\frac{1}{2})}=\frac{1}{2^{2\left\lfloor d/2\right\rfloor -2}}\text{tr}\left[\mathcal{X}_{i}\slashed{\varepsilon}_{1}(\slashed{p}_{1}{+}m_{a})\bar{\mathcal{X}}_{i}\slashed{\varepsilon}_{i}(\slashed{p}_{2}{+}m_{a})\right]\text{tr}\left[\mathcal{Y}_{i}\slashed{\varepsilon}_{3}(\slashed{p}_{3}{+}m_{b})\bar{\mathcal{Y}}_{i}\slashed{\varepsilon}_{4}(\slashed{p}_{4}{+}m_{b})\right],\label{eq:doublestar-1-1}
\end{equation}
 Notice that the generalization
of $(\ref{eq:doublestar-1-1})$ to an arbitrary number of massive lines could be done analogously by introducing one Dirac trace for each matter line.

 In this section we focus on elastic scattering, given by $M_4$, and inelastic scattering, given by $M_5$, firstly from a QFT perspective and then from a classical perspective. Nevertheless, we propose Lagrangians for arbitrary multiplicity as long as we keep two matter lines.
 
 Setting conventions, the momenta of the particles are taken as follows: For the $2\rightarrow2$ elastic scattering, the two incoming momenta are $p_1$ and $p_3$, and the outgoing momenta are $p_2=p_1-q$ and $p_4=p_3+q$, for $q$ the momentum transfer. For the $2\rightarrow3$ inelastic scattering, again the two incoming momenta are  $p_1$ and $p_3$, whereas the momenta for the two outgoing massive particles are $p_2=p_1-q_1$ and $p_4=p_3-q_3$, and the outgoing gluon or graviton has momentum $k$.

\subsection{Elastic scattering}

The simplest example of the scattering of two massive particles of
mass $m_{a}$ and $m_{b}$ , and spin $s_{a}$ and $s_{b}$, is the
elastic scattering, which we call $M_{4}^{(s_{a},s_{b})}$ amplitudes.
Let us illustrate how the double copy works for some choices of $s_{a}$
and $s_{b}$.

\subsubsection{Spinless case}

As a warm up consider the case of two scalar particles of incoming momenta $p_{1}$
and $p_{3}$, and momentum transfer $q$. The gravitational amplitude \eqref{eq:BCJ gravity}
can be computed from the double copy of the gluon exchange amplitude between two scalar particles. The kinematic numerator
and denominator for the gauge theory have the explicit form
\begin{equation}\label{numM4scalar}
 n^{(0,0)}=-e^{2}\left(4p_{1}{\cdot}p_{3}+q^{2}\right),\qquad d_{4}=q^{2}.
\end{equation}
In this case there is no kinematic Jacobi identities \eqref{color-kinematic}, hence color-kinematics duality is trivially satisfied. It is therefore straightforward to use the double copy $(\ref{eq:BCJ gravity})$ to write the gravitational amplitude
\begin{equation}\label{m4scalars}
  M_{4}^{(0\otimes 0,0\otimes 0)}=\frac{\kappa^2}{16}\frac{\left(4p_{1}{\cdot}p_{3}{+}q^{2}\right)^{2}}{q^{2}}.
\end{equation}
As the double copy is symmetric in the two numerators, by looking at the cut $q^2\to 0$ we can see that the axion field does not propagate and instead there is  only the propagation of the  graviton
and the  dilaton. Also, the fact that the amplitude is a perfect square is non-trivial from a Feynman diagram perspective. This factorization can be understood by decomposing $(\ref{m4scalars})$ into three pieces:

\begin{equation}
\label{decomposition m5 scalar}
    \frac{\left(4p_{1}{\cdot}p_{3}{+}q^{2}\right)^{2}}{q^{2}} =  \underbrace{\frac{\left(4p_{1}{\cdot}p_{3}{+}q^{2}\right)^{2}- 8 m_a^2 m_b^2-q^4}{q^{2}}}_{\textrm{Pure Gravity}} + \underbrace{\frac{8 m_a^2 m_b^2}{q^{2}}}_{\phi\,\, \rm{exchange}} + \underbrace{q^2}_{\rm{contact\,\, term}}
\end{equation}
Thus, we find that in order to achieve the factorization a quartic term must be included in the action. This is a general feature even with spin.\\
Likewise  the one matter line case, we can write a gravitational Lagrangian from which the amplitude \eqref{m4scalars} can be computed. In the Einstein frame it takes the form  
\begin{equation}
\mathcal{L}^{(0\otimes 0,0\otimes 0)}=\sqrt{g}\bigg[{-}\frac{2}{\kappa^{2}}R{+}\frac{2(d{-}2)}{\kappa^{2}}(\partial\phi)^{2}{+}\frac{1}{2}(\partial\varphi_{I})^{2}{+}\frac{1}{2}e^{-2\phi}m_{I}^{2}\varphi_{I}^{2}{+}\frac{\kappa^{2}}{16}\varphi_{1}\varphi_{2}(\partial\varphi_{1}){\cdot}(\partial\varphi_{2})\bigg],\label{eq:00LAGRANGIAN}
\end{equation}
 which we call the scalar-gravitational theory. Here the
$\varphi_{I}$ fields correspond to  the two massive scalars fields, and it is understood that $I\in \{1,2\}$, with  $m_1=m_a$ and $m_2=m_b$.

Guided from the decomposition \eqref{decomposition m5 scalar}, we have introduce a contact interaction  term in the Lagrangian to match precisely the double copy result. We have implicitly extended this contact term to arbitrary multiplicity by providing a covariant action. We will find that this covariantization is respected when computing the  double copy for the inelastic amplitude, $M_{5}^{(0\otimes 0,0\otimes 0)}$.

As a final short remark, let us point out that the contact interaction does not contribute to the classical limit (see Sec. \ref{GGT} for its definition). Indeed, by removing the dilaton exchange, the first and last terms of \eqref{decomposition m5 scalar} were used in \cite{Luna:2017dtq} to recover classical gravitational radiation.

\subsubsection{Case  \texorpdfstring{$s_{a}=0+1$}{s10} and  \texorpdfstring{$s_{b}=0+0$}{s00}}
Now we can add spin to one of the massive lines. In the gravitational theory, particle $a$ has spin $1$, whereas particle $b$ remains spinless. The gravitational amplitude $(\ref{eq:BCJ gravity})$ for this case  can be computed from the double copy of the scalar numerators $(\ref{numM4scalar})$ and the numerator $n^{(1,0)}$. The later corresponds to the gluon exchange between a scalar and a spin-$1$ particle, using the interactions from \eqref{scalarqcd} and \eqref{qcdW-boson}. It takes the simple form
\begin{equation}
n^{(1,0)}=-e^{2}\left[\left(4p_{1}{\cdot}p_{3}+q^{2}\right)\varepsilon_{1}{\cdot}\varepsilon_{2}-4\left(p_{3}{\cdot}\varepsilon_{2}\,q{\cdot}\varepsilon_{1}{-}p_{1}{\cdot}\varepsilon_{2}\,p_{3}{\cdot}\varepsilon_{1}\right)\right],
\end{equation}
 where $\varepsilon_{1(2)}$ is  the polarization vector for the incoming
 (outgoing) spining particle. This is precisely the full numerator used in e.g. \cite{Holstein:2008sw}. However, we warn the reader that it arises from a single matter line truncation of Yang-Mills theory \eqref{qcdW-boson} and hence should be regarded only as a thoroughfare to the classical limit, as discussed below eq. \eqref{eq:ndiag}.\\
 The simplicity of the construction enables us to readily write down a Lagrangian for the double copy. More precisely, an analogous decomposition to \eqref{decomposition m5 scalar} of the gravitational amplitude allows us to identify the contact interaction.  The gravitational Lagrangian in the Einstein frame is simply given by 
\begin{align}
\mathcal{L}^{(0\otimes1,0\otimes 0)} & =\sqrt{g}\left[{-}\frac{2}{\kappa^{2}}R{+}\frac{2(d{-}2)}{\kappa^{2}}(\partial\phi)^{2}{-}\frac{1}{4}e^{-2\phi}F_{\mu\nu}F^{\mu\nu}{+}\frac{m_{a}^{2}}{2}A_{\mu}A^{\mu}{+}\frac{1}{2}(\partial\varphi)^{2}\right.\nonumber \\
 & \left.\qquad\quad+\frac{1}{2}m_{b}^{2}e^{-2\phi}\varphi^{2}{+}\frac{\kappa^{2}}{16}\left(2A^{\mu}\partial_{\mu}\varphi\,A^{\nu}\partial_{\nu}\varphi{+}\varphi A^{\nu}\partial_{\mu}\varphi\partial^{\mu}A_{\nu}\right)\right],\label{eq:10LAGRANGIAN}
\end{align}
where $A_\mu $ corresponds to the massive spin-1 field and $\varphi$ is the massive scalar field.  Again, we will find that the covariantization of the contact term will be confirmed by the respective $M_5$ and we conjecture the same for higher multiplicity. Indeed, we have already found the term $A^{\mu}\partial_{\mu}\varphi\,A^{\nu}\partial_{\nu}\varphi$ previously! We obtained it in \cite{Bautista:2019tdr} and identified it as certain quantum contributions to the quadrupole arising in our classical double copy.
\subsubsection{Case  \texorpdfstring{$s_a=s_b=0+1$}{sasb01}}\label{sec:casee}
The natural generalization of the previous cases is to consider that both matter lines have  spin-$1$  in the gravitational theory. For this case there are two possible theories: The first one is dictated  by the factorization   $s_a=s_b=0+1$, whereas for the second theory, the factorization is  $s_a=s_b=\frac{1}{2}+\frac{1}{2}$. In this subsection we consider the former, and leave the  latter to be explored in the  next subsection.

The construction we consider here is an alternative to the two matter lines amplitudes obtained from \eqref{eq:final01}, i.e. the extension of $\mathcal{N}=4$ SUGRA. The difference lies in that here we will chop the flavour-interaction terms in the QCD Lagrangians, also leading to no such interaction on the gravity side. As explained below \eqref{eq:ndiag} this prevents the appearance of the additional massive particles arising in the dimensional reduction of $\mathcal{N}=4$ SUGRA. It is also interesting to explore because for only one line with spin we have found that this truncation leads to a simplified double copy construction.

The gravitational scattering amplitude $\eqref{eq:BCJ gravity}$ at four points  can be obtained from the double copy of the scalar numerators  $\eqref{numM4scalar}$ and the spin-$1$ numerator $n^{(1,1)}$. This numerator  can be computed from the gluon exchange between two massive spin-1 fields, each described by the matter part of \eqref{qcdW-boson}, and results into  
\begin{equation}\begin{split}
     n^{(1,1)}&=-4e^2\bigg[\frac{1}{4}\left(4p_1{\cdot}p_3{+}q^2\right)\varepsilon_1{\cdot}\varepsilon_2\,\varepsilon_3{\cdot}\varepsilon_4-\left(p_1{\cdot}\varepsilon_3\,p_3{\cdot}\varepsilon_4{+}p_1{\cdot}\varepsilon_3\,q{\cdot}\varepsilon_4\right)\varepsilon_1{\cdot}\varepsilon_2 \\
     &\qquad\qquad-\left(p_1{\cdot}\varepsilon_2\,p_3{\cdot}\varepsilon_1{-}p_3{\cdot}\varepsilon_2\,q{\cdot}\varepsilon_1\right)\varepsilon_3{\cdot}\varepsilon_4
     +p_1{\cdot}\varepsilon_2\,p_3{\cdot}\varepsilon_4\,\varepsilon_1{\cdot}\varepsilon_3\\
     &\qquad\qquad
    -q{\cdot}\varepsilon_1\,q{\cdot}\varepsilon_3\,\varepsilon_2{\cdot}\varepsilon_4 -p_3{\cdot}\varepsilon_4\,q{\cdot}\varepsilon_1\,\varepsilon_2{\cdot}\varepsilon_3+p_1{\cdot}\varepsilon_2\,q{\cdot}\varepsilon_3\,\varepsilon_1{\cdot}\varepsilon_4\bigg].
\end{split}
\end{equation}
The gravitational Lagrangian for this theory has a more intricate structure than the ones we have considered so far for the case of  two matter lines, which is natural due to additional propagation of the axion coupling to the spin of the matter lines. It can be shown that the Lagrangian is given by  
\begin{equation}\label{lagrangian01x01}
\begin{split}
    \mathcal{L}^{(0\otimes1,0\otimes1)} &=\mathcal{L}_{ct}+\sqrt{g}\bigg[{-}\frac{2}{\kappa^{2}}R{+}\frac{2(d{-}2)}{\kappa^{2}}(\partial\phi)^{2}-\frac{e^{-4\phi}}{6\kappa^{2}}H_{\mu\nu\rho}H^{\mu\nu\rho}\\&\qquad -\frac{e^{-4\phi}}{6\kappa^{2}}H_{\mu\nu\rho}A_I^{\mu}F^{I\nu\rho}-\frac{1}{4}e^{-2\phi}F_{I,\mu\nu}F^{I\mu\nu}{+}\frac{m_{I}^{2}}{2}A_{I,\mu}A^{I\mu}\bigg] ,
\end{split}
\end{equation}
where the flavour index  $I\in\{1,2\}$, and once again the masses $m_1=m_a$ and $m_2=m_b$. The contact interaction Lagrangian for this case  has the form
\begin{equation}\label{contact term 01}
    \begin{split}
        \mathcal{L}_{ct}&\sim \sqrt{g}\big[ 2A_{1}{\cdot} A_{2}\,(\partial_{\mu}A_{1,\nu}{-}3\partial_{\nu}A_{1,\mu})\partial^{\mu}A_{2}^{\nu}-2A_{2}{\cdot} F_{1}{\cdot}F_{2}{\cdot} A_{1}  \\
        &-2A_{2}^{\mu}\partial_{\mu}A_{1}^{\alpha}A_{2}^{\nu}\partial_{\nu}A_{1,\alpha} {-}A_{1}^{\mu}\partial_{\mu}A_{2}^{\alpha}A_{1}^{\nu}\partial_{\nu}A_{2\alpha}{-}A_{1}^{\mu}\partial_{\alpha}A_{1,\mu}A_{2}^{\nu}\partial^{\alpha}A_{2,\nu}\big],
    \end{split}
\end{equation}
where the  product of field strength tensors reads explicitly
\begin{equation}\begin{split}
     A_{2}{\cdot} F_{1}{\cdot} F_{2}{\cdot} A_{1}&=A_{2}^{\mu}\partial_{\mu}A_{1}^{\alpha}\partial_{\alpha}A_{2,\nu}A_{1}^{\nu}{-}A_{2}^{\mu}\partial^{\alpha}A_{1,\mu}\partial_{\alpha}A_{2,\nu}A_{1}^{\nu}\\
     &{-}A_{2}^{\mu}\partial_{\mu}A_{1}^{\alpha}\partial_{\nu}A_{2,\alpha}A_{1}^{\nu}{-}A_{2}^{\mu}\partial^{\alpha}A_{1,\mu}\partial_{\nu}A_{2,\alpha}A_{1}^{\nu} .
\end{split}
\end{equation}
Thus in this case, for two particles including spin, we have found an elevated level of complexity even for the four-point terms in the Lagrangian. Because of this in the following we simply resort to the extension of $\mathcal{N}=4$ SUGRA, eq. \eqref{eq:final01}, in order to write the interaction Lagrangian for two or more matter lines. Once again, the same classical limit of the amplitude can be obtained from the above construction (i.e. via \eqref{eq:BCJ gravity}) or from the extension of $\mathcal{N}=4$ SUGRA (i.e. via KLT), the equivalence of which will prove useful for inelastic scattering in the next section.

\subsubsection{Case  \texorpdfstring{$s_{a}=s_{b}=\frac{1}{2}+\frac{1}{2}$}{sasb12p12}}

We  finish the discussion  for the  elastic scattering considering the simplest gravitational theory for both of the massive lines with spin-1. As we mentioned previously, this theory is  dictated  by the factorization $s_{a}=s_{b}=\frac{1}{2}{+}\frac{1}{2}$. The gravity amplitude \eqref{eq:BCJ gravity} at $4$ pt. is computed 
from the double copy of the QCD spin $\frac{1}{2}$ numerator $n^{(\frac{1}{2},\frac{1}{2})}$, and its charge conjugated pair. They have a simple form
\begin{equation}\begin{split}
      n^{(\frac{1}{2},\frac{1}{2})} & =e^{2}\bar{u}_{2}\gamma^{\mu}u_{1}\bar{u}_{4}\gamma_{\mu}u_{3},\\
\bar{n}^{(\frac{1}{2},\frac{1}{2})} & =e^{2}\bar{v}_{1}\gamma^{\mu}v_{2}\bar{v}_{3}\gamma_{\mu}v_{4},
\end{split}
\end{equation}
where we use the condition for momentum conservation  $p_{2}=p_{1}{-}q$ and $p_{4}=p_{3}{+}q$.
Now, using the double copy operation for two matter lines $(\ref{eq:doublestar-1-1})$,
the gravitational amplitude takes the compact form 
\begin{equation}
\begin{split}
M_{4}^{(\frac{1}{2}\otimes\frac{1}{2},\frac{1}{2}\otimes\frac{1}{2})}&=\frac{4}{2^{2\left\lfloor D/2\right\rfloor }}\frac{\kappa^{2}}{q^{2}}\text{tr}\big[\gamma^{\mu}\slashed{\varepsilon}_{1}(\slashed{p}_{1}{-}m_{a})\gamma^{\nu}\slashed{\varepsilon}_{2}(\slashed{p}_{2}{+}m_{a})\big]\text{tr}\big[\gamma_{\mu}\slashed{\varepsilon}_{3}(\slashed{p}_{3}{-}m_{b})\gamma_{\nu}\slashed{\varepsilon}_{4}(\slashed{p}_{4}{+}m_{b})\big],
\end{split}
\end{equation}
 Notice the momenta $p_1$ and $p_3$  are incoming, therefore the sign in the projector  changes. After taking  the traces the amplitude reads
\begin{equation}
\begin{split}
M_{4}^{(\frac{1}{2}\otimes\frac{1}{2},\frac{1}{2}\otimes\frac{1}{2})} & =\frac{4\kappa^{2}}{q^{2}}\bigg\{\big[\varepsilon_{1}{\cdot}\varepsilon_{2}\left((d{-}6)p_{1}^{\nu}p_{2}^{\mu}{+}(d{-}2)p_{1}^{\mu}p_{2}^{\nu}\right){-}p_{1}{\cdot}\varepsilon_{2}\left((d{-}6)\varepsilon_{1}^{\nu}p_{2}^{\mu}{+}(d{-}2)\varepsilon_{1}^{\mu}p_{2}^{\nu}\right){-}\\
 & p_{2}{\cdot}\varepsilon_{1}\left((d{-}6)p_{1}^{\nu}\varepsilon_{2}^{\mu}{+}(d{-}2)p_{1}^{\mu}\varepsilon_{2}^{\nu}\right){+}\left((d{-}6)p_{1}{\cdot}p_{2}{+}(d{-}4)m_{a}^{2}\right)\left(\varepsilon_{1}^{\mu}\varepsilon_{2}^{\nu}{-}\varepsilon_{1}{\cdot}\varepsilon_{2}\eta^{\mu\nu}\right)\\
 & +\left((d{-}2)p_{1}{\cdot}p_{2}{+}d\,m_{a}^{2}\right)\varepsilon_{1}^{\mu}\varepsilon_{2}^{\nu}{+}(d{-}6)p_{1}{\cdot}\varepsilon_{2}\,p_{2}{\cdot}\varepsilon_{1}\eta^{\mu\nu}\big]\times\big[{\rm {line}}\,a\rightarrow{\rm {line}}\,b\big]_{\mu \nu}\bigg\},\\
\end{split}
\end{equation}
where the change $\big[{\rm {line}}\,a\rightarrow{\rm {line}}\,b\big]$
means to do $\big[1\rightarrow3,\,2\rightarrow4,\,a\rightarrow b\big].$
Likewise for the two previous cases, we can write the gravitational
Lagrangian for this theory, surprisingly it has a very simple form 
\begin{equation}
\mathcal{L}^{(\frac{1}{2}\otimes\frac{1}{2},\frac{1}{2}\otimes\frac{1}{2})}=\sqrt{g}\bigg[{-}\frac{2}{\kappa^{2}}R{+}\frac{2(d{-}2)}{\kappa^{2}}(\partial\phi)^{2}{-}\frac{1}{4}e^{(d{-}4)\phi}F_{I,\mu\nu}F_{I,}^{\mu\nu}{+}\frac{1}{2}e^{(d{-}2)\phi}m_{I}^{2}A_{I\mu}A_{I}^{\mu}\bigg]\, ,\label{eq:1212LAGRANGIAN}
\end{equation}
We say  that  this is the simplest theory for spinning particles coupled to gravity in two senses:
First, even thought the two massive lines have spin, there is no propagation of the axion. This  confirms that in the \12x12 double copy setup the spin-$1$ field does not source the axion. Second and more importantly, there is no need for adding a contact interaction between matter lines, a feature we will confirm also in $M_5$. This is the only gravitational theory we have found for which this happens and reflects its underlying fermionic origin.

\subsection{Inelastic Scattering}
Moving on to the inelastic scattering, we consider the emission of a gluon or a (fat) graviton in the final state. The relevance of this amplitude is that it allows us to make contact with classical double copy for radiation at the end of this section.

The QCD amplitude obtained from Feynman diagrams can be arranged into the color decomposition \eqref{eq:qcd bcj} with only five terms. The color factors and numerators will satisfy 
\begin{equation}
c_{1}-c_{2}=-c_{3},\,\,\,c_{4}-c_{5}=c_{3},\label{eq:color identities-1}
\end{equation}
\begin{equation}
n_{1}-n_{2}=-n_{3},\,\,\:n_{4}-n_{5}=n_{3}.\label{eq:color-kinematics duality}
\end{equation}
The gravitational amplitude will be given again by $(\ref{eq:BCJ gravity})$, with the sum running from $1$ to $5$.
The product of polarization vectors of the external gluon $\epsilon_{\mu}\tilde{\epsilon}_{\nu}$
corresponds to a fat graviton state $H_5$. To extract the graviton amplitude we replace $\epsilon_{\mu}\tilde{\epsilon}_{\nu}\rightarrow\epsilon_{\mu\nu}^{\rm{TT}}$
i.e. the symmetric, transverse and traceless polarization tensor for the graviton.
If on the other hand we want to compute the dilaton amplitude,  we replace
$\epsilon_{\mu}\tilde{\epsilon}_{\nu}\rightarrow\frac{\eta_{\mu\nu}}{\sqrt{D{-}2}}.$
In the same way that  for the case of elastic scattering,  let us consider
different cases for the spin of the massive lines. In the case with spin we will also consider an external axion state with the help of a suitable BCJ gauge.

\subsubsection{Spinless case }
We first review the spinless case as considered in e.g. \cite{Luna:2017dtq}. The tree-level QCD amplitude \eqref{eq:qcd bcj} in BCJ form can be taken from there. We use the momentum conservation condition $k=q_{1}+q_{3},$ to write the 
the numerators, color factors and denominators in \eqref{eq:qcd bcj} as follows
\begin{equation}
\begin{split}
  n_{1}^{(0,0)} & =e^{3}\left[2\left(2p_{3}{-}q_{3}\right){\cdot}\left(2p_{1}{+}q_{3}\right)\left(p_{1}+q_{3}\right){\cdot}\epsilon{-}\left(2p_{1}{\cdot}q_{3}{+}q_{3}^{2}\right)\left(2p_{3}{-}q_{3}\right){\cdot}\epsilon\right],\\
n_{2}^{(0,0)} & =e^{3}\left[2p_{1}{\cdot}\epsilon\left(2p_{1}{-}k{-}q_{1}\right){\cdot}\left(2p_{3}{-}q_{3}\right){+}2p_{1}{\cdot}k\left(2p_{3}{-}q_{3}\right){\cdot}\epsilon\right],\\
n_{3}^{(0,0)} & =e^{3}\left(2p_{1}{-}q_{1}\right)^{\mu}\left(2p_{3}{-}q_{3}\right)^{\rho}\left[\left(k+q_{3}\right)_{\mu}\eta_{\nu\rho}+\left(q_{1}-q_{3}\right)_{\nu}\eta_{\mu\rho}-\left(k+q_{1}\right)_{\rho}\eta_{\mu\nu}\right]\epsilon^{\nu},\\
n_{4}^{(0,0)} & =e^{3}\left[2\left(2p_{1}{-}q_{1}\right){\cdot}\left(2p_{3}{+}q_{1}\right)\left(p_{3}{+}q_{1}\right){\cdot}\epsilon{-}\left(2p_{3}{\cdot}q_{1}{+}q_{1}^{2}\right)\left(2p_{1}{-}q_{1}\right){\cdot}\epsilon\right],\\
n_{5}^{(0,0)} & =e^{3}\left[2p_{3}{\cdot}\epsilon\left(2p_{3}{-}k{-}q_{3}\right){\cdot}\left(2p_{1}{-}q_{1}\right){+}2p_{3}{\cdot}k\left(2p_{1}{-}q_{1}\right){\cdot}\epsilon\right].  
\end{split}
\end{equation}
\begin{equation}\label{color and denominators}
\begin{split}
c_{1} & =(T_{1}^{a}.T_{1}^{b})T_{3}^{b},\qquad d_{1}=q_{3}^{2}\left(2p_{1}{\cal \cdot}k-q_{1}^{2}+q_{3}^{2}\right),\\
c_{2} & =(T_{1}^{b}.T_{1}^{a})T_{3}^{b},\qquad d_{2}=-2\left(p_{1}{\cdot}k\right)q_{3}^{2},\\
c_{3} & =f^{abc}T_{1}^{b}T_{3}^{c},\qquad\,\,\, d_{3}=q_{1}^{2}q_{3}^{2},\\
c_{4} & =(T_{3}^{a}.T_{3}^{b})T_{1}^{b},\qquad d_{4}=q_{1}^{2}\left(2p_{3}{\cal \cdot}k+q_{1}^{2}-q_{3}^{2}\right),\\
c_{5} & =(T_{3}^{b}.T_{3}^{a})T_{1}^{b}.\qquad\, d_{5}=-2\left(p_{3}{\cdot}k\right)q_{1}^{2}.
\end{split}
\end{equation}
It is now straightforward to compute the gravitational amplitude $M_5^{\rm gr}$ using  $(\ref{eq:BCJ gravity})$.
The comparison of the double copy result with the Feynman diagrammatic
computation from the Lagrangian $(\ref{eq:00LAGRANGIAN})$ shows complete
agreement for both the graviton ($\epsilon_{\mu}\tilde{\epsilon}_{\nu}\rightarrow\epsilon_{\mu\nu}^{\rm{TT}}$)
and dilaton ($\epsilon_{\mu}\tilde{\epsilon}_{\nu}\rightarrow\frac{\eta_{\mu\nu}}{\sqrt{d-2}}$)
amplitude. This agreement provides a non trivial check of the contact
interaction we included to match the double copy result for the elastic scattering amplitude.

\subsubsection{Case  \texorpdfstring{$s_{a}=s_{b}=\frac{1}{2}+\frac{1}{2}$}{sasb1212}}
For the case of the inelastic  scattering of two fermions with different flavour, and
the emission of one gluon, the QCD amplitude in the BCJ form \eqref{eq:qcd bcj} can be computed directly from Feynman diagrams. As in the scalar case the numerators also take a compact form  
\begin{equation}\begin{split}
n_{1}^{(\frac{1}{2},\frac{1}{2})} & =e^{3}\bar{u}_{2}\slashed{\epsilon}(\slashed{p_{1}}{+}\slashed{q_{3}}{+}m_{a})\gamma^{\mu}u_{1}\bar{u}_{4}\gamma_{\mu}u_{3},\\
n_{2}^{(\frac{1}{2},\frac{1}{2})} & =e^{3}\bar{u}_{2}\gamma^{\mu}(\slashed{p}_{1}{-}\slashed{k}{+}m_{a})\slashed{\epsilon}u_{1}\bar{u}_{4}\gamma_{\mu}u_{3},\\
n_{4}^{(\frac{1}{2},\frac{1}{2})} & =e^{3}\bar{u}_{2}\gamma^{\mu}u_{1}\bar{u}_{4}\slashed{\epsilon}(\slashed{p_{3}}{+}\slashed{q_{1}}{+}m_{b})\gamma_{\mu}u_{3},\\
n_{5}^{(\frac{1}{2},\frac{1}{2})} & =e^{3}\bar{u}_{2}\gamma^{\mu}u_{1}\bar{u}_{4}\gamma^{\mu}(\slashed{p}_{3}{-}\slashed{k}{+}m_{b})\slashed{\epsilon}u_{3},\\
n_{3}^{(\frac{1}{2},\frac{1}{2})} & =-2e^{3}\left(\bar{u}_{2}\slashed{\epsilon}u_{1}\bar{u}_{4}\slashed{k}u_{3}-\bar{u}_{2}\slashed{k}u_{1}\bar{u}_{4}\slashed{\epsilon}u_{3}-\bar{u}_{2}\gamma^{\mu}u_{1}\bar{u}_{4}\gamma_{\mu}u_{3}q_{1}{\cdot}\epsilon\right).
\end{split}
\end{equation}
Analogously, their charge conjugated pairs read
\begin{equation}
\begin{split}
\bar{n}_{1}^{(\frac{1}{2},\frac{1}{2})} & =e^{3}\bar{v}_{1}\gamma^{\mu}(\slashed{p_{1}}{+}\slashed{q_{3}}{-}m_{a})\tilde{\slashed{\epsilon}}v_{2}\bar{v}_{3}\gamma_{\mu}v_{4},\\
\bar{n}_{2}^{(\frac{1}{2},\frac{1}{2})} & =e^{3}\bar{v}_{1}\tilde{\slashed{\epsilon}}(\slashed{p}_{1}{-}\slashed{k}{-}m_{a})\gamma^{\mu}v_{2}\bar{v}_{3}\gamma_{\mu}v_{4},\\
\bar{n}_{4}^{(\frac{1}{2},\frac{1}{2})} & =e^{3}\bar{v}_{1}\gamma^{\mu}v_{2}\bar{v}_{3}\gamma_{\mu}(\slashed{p_{3}}{+}\slashed{q_{1}}{-}m_{b})\tilde{\slashed{\epsilon}}v_{4},\\
\bar{n}_{5}^{(\frac{1}{2},\frac{1}{2})} & =e^{3}\bar{v}_{1}\gamma^{\mu}v_{2}\bar{v}_{3}\tilde{\slashed{\epsilon}}(\slashed{p}_{3}{-}\slashed{k}{-}m_{b})\gamma^{\mu}v_{4},\\
\bar{n}_{3}^{(\frac{1}{2},\frac{1}{2})} & =-2e^{3}\left(\bar{v}_{1}\tilde{\slashed{\epsilon}}v_{2}\bar{v}_{3}\slashed{k}v_{4}-\bar{v}_{1}\slashed{k}v_{2}\bar{v}_{3}\tilde{\slashed{\epsilon}}v_{4}-\bar{v}_{1}\gamma^{\mu}v_{2}\bar{v}_{3}\gamma_{\mu}v_{4}q_{1}{\cdot}\tilde{\epsilon}\right).
\end{split}
\end{equation}
The color factors and denominators are the same as the scalar case \eqref{color and denominators}. The spin-$\frac{1}{2}$ numerators are readily seen to satisfy the kinematic Jacobi relation \eqref{eq:color-kinematics duality}. Now, to compute the gravitational amplitude \eqref{eq:BCJ gravity},
we need to compute the double copy of these numerators using $(\ref{eq:doublestar-1-1})$. For instance we have 
\begin{equation}
\begin{split}
n_{1}^{(\frac{1}{2},\frac{1}{2})}\otimes\bar{n}_{1}^{(\frac{1}{2},\frac{1}{2})} & =\frac{\kappa^3}{2^{2\left\lfloor d/2\right\rfloor +4}}\text{tr}\big[\slashed{\epsilon}(\slashed{p_{1}}{+}\slashed{q_{3}}{+}m_{a})\gamma^{\mu}\slashed{\varepsilon}_{1}(\slashed{p}_{1}{-}m_{a})\gamma^{\nu}(\slashed{p_{1}}{+}\slashed{q_{3}}{-}m_{a})\tilde{\slashed{\epsilon}}\slashed{\varepsilon}_{2}(\slashed{p}_{2}{+}m_{a})\big]\times\\
 & \qquad\qquad\quad \text{tr}\big[\gamma_{\mu}\slashed{\varepsilon}_{3}(\slashed{p}_{3}{-}m_{b})\gamma_{\nu}\slashed{\varepsilon}_{4}(\slashed{p}_{4}{+}m_{b})\big].
 \end{split}
\end{equation}
Analogous expressions follow for the double copy of the  remaining numerators. Although this is a very compact way to write
the gravitational numerator, the final result for the gravitational amplitude \eqref{eq:BCJ gravity} after doing all the
traces is too long to be printed in this paper, therefore we provide a Mathematica Notebook with the result. 

As we did for the case of elastic  scattering, we have checked that
the same amplitude can be computed with the Feynman rules derived
from the Lagrangian $(\ref{eq:1212LAGRANGIAN}),$ for both graviton
and dilaton radiation. This is in fact a non-trivial check of the Lagrangian, and discards the  need for a quartic   term, or  the  presence of the axion field. 

Finally, we leave the discussion for  the case  $s_{a}=0+1$ and $s_{b}=0+0$, and the case  $s_{a}=s_b=0+1$, to be illustrated in section \ref{GGT}. Using a more  convenient BCJ gauge, these amplitudes take compact expressions and the numerator can be put in a  manifestly gauge invariant form.

\subsection{Generalized Gauge Transformations and Classical Radiation}\label{GGT}

Now that we have computed the amplitude for different Bremsstrahlung processes we might ask what sort of classical information can be extracted from it. The answer is given by the radiated momentum which is carried by long range fields (photons, gravitons, dilatons and axions) to null infinity \cite{Goldberger:2016iau,Kosower:2018adc}. This momentum is determined by a phase space integral,

\begin{equation}
     K^{\mu} = \int \text{dLIPS}(k)\, k^{\mu} \,|J(k)|^2 \,,
\end{equation}
where $J(k)$ is the radiative piece of the stress energy tensor (or current) related to the amplitude via the LSZ formula. This also requires to implement a prescription for the classical limit, $J(k)=\lim_{\hbar \to 0 } M_5 $ \cite{Luna:2017dtq,Kosower:2018adc}. In light of the promising recent developments of \cite{Luna:2016due,PV:2019uuv,Chester:2017vcz,Shen:2018ebu} it is desirable to understand how a double copy structure turns out to be realized in classical radiation, and more specifically, how it follows from the BCJ construction in QFT. 

We would like to extract the classical piece of the amplitude in such a way that the double copy structure is preserved untouched in the final result.  Taking the classical limit of  \eqref{eq:BCJ gravity} however  does not show explicitly the double copy form of the classical amplitude as we will see in a moment. This was first observed for scalar sources in \cite{Luna:2017dtq}, but is also true for the spinning case.  We find that the problem can be fixed if we  write the double copy for inelastic scattering in a more convenient \textit{generalized} gauge. This further provides an alternative derivation to the classical double copy formula presented in \cite{Bautista:2019tdr}.

\subsubsection{Classical radiation from the standard BCJ double copy }

For scalar radiation the comparison
of the classical construction derived by Goldberger and Ridgway \cite{Goldberger:2016iau} and the classical limit of the BCJ double copy result was presented in \cite{Luna:2017dtq}.
There the limit was taken by means of a large mass
expansion. Here we follow will \cite{Kosower:2018adc} instead, where it was shown that 
the classical piece of the amplitude can be obtained by re-scaling
the massless transfer momenta with $\hbar$ and take the leading
order piece in the $\hbar\rightarrow0$ limit. With this in mind it is convenient for us to introduce the average momentum transfer $q=\frac{q_1{-}q_3}{2}$.  The re-scaled momenta
can be interpreted as a classical wave vector $q\rightarrow\hbar\bar{q}$.
Notice that momentum conservation implies that the radiated on-shell momenta
needs to be re-scaled as well $k\rightarrow\hbar\bar{k}.$ For spinning radiation the classical limit was outlined in \cite{Bautista:2019tdr} and requires to introduce the angular momentum operator, performing the multipole expansion as we have described in the previous sections. We then scale such operator as $J\rightarrow\hbar^{-1}\bar{J}$  \cite{Guevara:2018wpp,Bautista:2019tdr} and strip the respective polarization states \cite{Guevara:2019fsj}, see \cite{Maybee:2019jus} for a formal derivation. Finally, for the
case of QCD amplitudes, one further scaling needs to be done in order
to correctly extract the classical piece. In reminiscence of the color-kinematics duality, we find that the generators of the color group
$T^a$ must also scale as those of angular momentum, i.e. $T^a\rightarrow\hbar^{-1}T^a$.

In order to motivate our procedure let us first consider the 5-pt. amplitudes for both QCD and gravity in the standard BCJ form we have provided. In other words, we want to see how the ingredients in \eqref{eq:qcd bcj} and \eqref{eq:BCJ gravity} behave in the
$\hbar$-expansion. By inspection, the leading order of the numerators $n_{i}$ goes
as $\hbar^{0},$ and the sub-leading correction is of order $\hbar$. 
Let us denote the expansion of the numerators as $n_{i}=\langle n_{i}\rangle +\delta n_{i}\hbar+{\cdots}$.
The denominators can also be expanded as $d_{i}= \langle d_{i} \rangle\hbar^{3}+\delta d_{i}\hbar^{4}+\cdots$.
At leading order, it is easy to check that $\langle n_{3}\rangle $=0, $\langle n_{1}\rangle=\langle n_{2}\rangle$
and $\langle n_{4}\rangle =\langle n_{5}\rangle$, whereas for  the denominators we have $\langle d_{1}\rangle=-\langle d_{2}\rangle$
and $\langle d_{4}\rangle=-\langle d_{5}\rangle .$ At sub-leading order $\delta d_{2}=\delta d_{5}=0.$
Finally, for the color factors we have $c_{i}\rightarrow\hbar^{-3}c_{i}$
for $i=1,2,4,5$ and $c_{3}\rightarrow\hbar^{-2}c_{3}$.\\

 With this in mind, the classical piece of the QCD amplitude for gluon radiation reads
\begin{equation}\label{M5qcd classical}
\langle M_{5}^{\rm{QCD}}\rangle
  ={-}c_{1}\left[\frac{\langle n_{1}\rangle \delta d_{1}}{\langle d_{1}\rangle^{2}}{-}\frac{\delta n_{1}{-}\delta n_{2}}{\langle d_{1}\rangle}\right]{-}c_{3}\left[\frac{\langle n_{1}\rangle}{\langle d_{1}\rangle}{-}\frac{\delta n_{3}}{\delta d_{3}}{-}\frac{\langle n_{4}\rangle}{\langle d_{4}\rangle}\right]{-}c_{4}\left[\frac{\langle n_{4}\rangle \delta d_{4}}{\langle d_{4}\rangle^{2}}{-}\frac{\delta n_{4}{-}\delta n_{5}}{\langle d_{4}\rangle}\right],
\end{equation}
where  $\langle M_{n}\rangle:{=}\lim_{\hbar\rightarrow0}M_{n}$.  A similar expansion can be done for the gravitational amplitude
given by the double copy \eqref{eq:BCJ gravity}
\begin{equation}\label{eq:m5gr classical}
\begin{split}
\langle M_{5}^{gr}\rangle & ={-}\frac{\langle n_{1}\rangle \otimes\langle \tilde{n}_{1}\rangle}{\langle d_{1,0}\rangle^{2}}\delta d_{1}{+}\frac{\langle n_{1}\rangle\otimes\left(\delta\tilde{n}_{1}{-}\delta\tilde{n}_{2}\right)+\left(\delta n_{1}{-}\delta n_{2}\right)\otimes\langle \tilde{n}_{1}\rangle}{\langle d_{1}\rangle}{+}\frac{\delta n_{3}\otimes\delta\tilde{n}_{3}}{\langle d_{3}\rangle}\\
 &\qquad-\frac{\langle n_{4}\rangle \otimes\langle \tilde{n}_{4}\rangle}{\langle d_{4,0}\rangle^{2}}\delta d_{4}{+}\frac{\langle n_{4}\rangle \otimes\left(\delta\tilde{n}_{4}{-}\delta\tilde{n}_{5}\right)+\left(\delta n_{4}{-}\delta n_{5}\right)\otimes \langle \tilde{n}_{4}\rangle}{\langle d_{4}\rangle}
 \end{split}
\end{equation}

Hence, we find that the classical piece of the gravitational amplitude \eqref{eq:m5gr classical} does not reflect
the BCJ double copy structure as expected. This can be traced back to the presence of $\frac{1}{\hbar}$ terms which will still contribute to the expansion even though the overall leading order (as $\hbar{\to}0$) cancels. We shall find a way to make such limit smooth and preserve the double copy structure.

In \cite{Bautista:2019tdr}  we provided a classical double copy formula to compute gravitational radiation at leading order in the coupling from photon Bremsstrahlung. The formula was obtained by looking at specific cuts carrying the classical information. Here we will see it follows naturally from a particular BCJ gauge. In fact, in such gauge we will further find no $\frac{1}{\hbar}$ terms and hence a smooth classical limit.

Before we start let us summarize the results. The classical expressions for gluon and "fat graviton"  radiation  \eqref{M5qcd classical} and \eqref{eq:m5gr classical}  agrees with the result of Goldberger and Ridgway for scalar sources 
\cite{Goldberger:2016iau}, as shown in \cite{Luna:2017dtq}. Here we will further recover the results of Goldberger, Li and Prabhu \cite{Goldberger:2017ogt,Li:2018qap} for spinning sources up to dipole order, including the full axion-dilaton-graviton classical radiation. We also make contact with our own results in \cite{Bautista:2019tdr} which already recovered graviton radiation in such case. This will be achieved by providing the classical numerators for each of these cases in the aforementioned gauge.

\subsubsection{Generalized gauge transformation}

In order to rewrite the quantum amplitudes \eqref{eq:qcd bcj}  and \eqref{eq:BCJ gravity} in a convenient
gauge we proceed as follows.  Observe that the non-abelian contribution to the QCD  amplitude
\eqref{eq:qcd bcj} comes from the diagram whose color factor  \eqref{color and denominators} is $c_{3}$,
which is proportional to the structure constants of the gauge group.
We can however gauge away this non-abelian piece of the amplitude
using a \textit{Generalized Gauge Transformation} (GGT) \cite{Bern:2008qj}. Recall that a  GGT is
a transformation on the kinematic numerators that leaves the amplitude invariant. This transformation allow us to move terms
between diagrams. For the case of the inelastic scattering, consider
the following shift on the numerators entering in \eqref{eq:qcd bcj}
\begin{equation}\label{eq:shifted numeratos}
\begin{split}
n_{1}^{\prime} & =n_{1}-\alpha d_{1},\\
n_{2}^{\prime} & =n_{2}+\alpha d_{2}, \\
n_{3}^{\prime} & =n_{3}-\alpha d_{3}+\gamma d_{3},\\
n_{4}^{\prime} & =n_{4}-\gamma d_{4}, \\
n_{5}^{\prime} & =n_{5}+\gamma d_{5}. 
\end{split}
\end{equation}
This shift leaves invariant the amplitude \eqref{eq:qcd bcj} since
\begin{equation}
\Delta M_{5}^{\rm{QCD}} =-\alpha(c_{1}-c_{2}+c_{3})-\gamma(c_{4}-c_{5}-c_{3})=0,
\end{equation}
where we have use the color identities $(\ref{eq:color identities-1})$
in the last equality. We can now solve for the values of $\alpha$ and
$\gamma$ that allow to impose $n_{3}^{\prime}=0$ , while satisfying
the color-kinematic duality for the shifted numerators
\begin{equation}
n_{1}^{\prime}-n_{2}^{\prime}=-n_{3}^{\prime}=0,\,\,\:n_{4}^{\prime}-n_{5}^{\prime}=n_{3}^{\prime}=0.
\end{equation}
The solution can be written as
\begin{align}
\alpha & =-\frac{n_{3}}{d_{1}+d_{2}},\,\,\,\,\,\gamma=-\frac{d_{1}+d_{2}+d_{3}}{d_{1}+d_{2}}\frac{n_{3}}{d_{3}}.\label{eq:alpha gamma parameters}
\end{align}
Explicitly these  parameters 
take the simple form
\begin{equation}
  \alpha=\frac{n_{3}}{2q{\cdot}k\left(q^{2}{-}q{\cdot}k\right)},\qquad\gamma=\frac{n_{3}}{2q{\cdot}k\left(q^{2}{+}q{\cdot}k\right)},  
\end{equation}
Importantly, this solution is general and independent of the spin of scattered particles.

The new numerators \eqref{eq:shifted numeratos} will be non-local since they have absorbed $n_3$. However, they exhibit nice features: They are independent, gauge invariant, and in the classical limit they will lead to a remarkably simple (and local!) form. Indeed, the QCD amplitude \eqref{eq:qcd bcj} for inelastic scattering takes already a more compact form 
\begin{equation}\label{m5qcd GGT}
M_{5}^{\rm{QCD}}=\left[\frac{c_{1}}{d_{1}}+\frac{c_{2}}{d_{2}}\right]n_{1}^{\prime}+\left[\frac{c_{4}}{d_{4}}+\frac{c_{5}}{d_{5}}\right]n_{4}^{\prime}.
\end{equation}
The gravitational amplitude \eqref{eq:BCJ gravity} then is given by the double copy of \eqref{m5qcd GGT}  as follows 
\begin{equation}
M_{5}^{\rm {gr}}=\frac{n_{1}^{\prime}\otimes\tilde{n}_{1}^{\prime}}{d_{1}^{\prime}}+\frac{n_{4}^{\prime}\otimes\tilde{n}_{4}^{\prime}}{d_{4}^{\prime}},\label{m5 gr CBCJ}
\end{equation}
 where 
\begin{equation}
d_{1}^{\prime}=\frac{d_{1}d_{2}}{d_{1}+d_{2}},\qquad d_{4}^{\prime}=\frac{d_{4}d_{5}}{d_{4}+d_{5}}.
\end{equation}
Explicitly, this gives
\begin{equation}\label{denominators explicit}
\frac{1}{d_{1}^{\prime}}{=}-\frac{q{\cdot}k}{p_{1}{\cdot}k\,q{\cdot}(q-k)\left(2q{\cdot}k{-}2p_{1}{\cdot}k\right)},\,\,\,\,\frac{1}{d_{4}^{\prime}}{=}-\frac{q{\cdot}k}{p_{3}{\cdot}k\,q{\cdot}(q+k)\left(2q{\cdot}k{+}2p_{3}{\cdot}k\right)},
\end{equation}
 When performing the double copy, there will in principle be a
 pole in $q{\cdot}k$  both in $(\ref{m5 gr CBCJ})$ and in the classical formula  \eqref{M5 classical CBCJ} below, which is nevertheless spurious and cancels out
in the final result.
 Notice we have reduced the problem of doing the double copy of five
numerators to do the double copy of  just two (the dimension of the BCJ basis). Indeed, now we can take $c_3\to 0$, setting $c_2 \to c_1$ and $c_5 \to c_4$. Further fixing $c_1=c_4=1$ we obtain the QED case (see \eqref{color and denominators}) with

\begin{equation}\label{m5qedd}
M_{5}^{\rm {QED}}=\frac{n_{1}^{\prime}}{d_{1}^{\prime}}+\frac{n_{4}^{\prime}}{d_{4}^{\prime}},
\end{equation}

The double copy formula \eqref{m5 gr CBCJ} agrees with \eqref{eq:BCJ gravity}. Remarkably, we can use \eqref{m5qedd} as a starting point for the (classical) double copy since the numerators $n_1'$ and $n_4'$ can be read off from $M_5^{\rm QED}$ from its pole structure. This has the advantage that the
classical limit of the amplitude will be smooth and will preserve the double copy form.

\subsubsection{Classical limit and Compton Residue}

In the gauge $(\ref{eq:shifted numeratos})$, extracting the classical
piece of the gravitational amplitude \eqref{m5 gr CBCJ} is straightforward. The shifted numerators
scale as $n_{i}^{\prime}=\langle n_{i}^{\prime}\rangle+\delta n_{i}^{\prime}\hbar $,
whereas the denominators scale as $d_{i}^{\prime}=\langle d_{i}^{\prime}\rangle\hbar^{2}+\delta d_{i}^{\prime}\hbar^3$. With this
in mind, the classical piece of the gravitational amplitude \eqref{m5 gr CBCJ} is simply 
\begin{equation}\label{M5 classical CBCJ}
\boxed{
\langle M_{5}^{(s_a\otimes\tilde{s}_a,s_b\otimes \tilde{s}_b)}\rangle =\frac{\langle n_{1}^{\prime\,(s_a,s_b)}\rangle \otimes\langle \tilde{n}_{1}^{\prime
\,(\tilde{s}_a,\tilde{s}_b)}\rangle}{\langle d_{1}^{\prime}\rangle}+\frac{\langle n_{4}^{\prime\,(s_a,s_b)}\rangle \otimes\langle\tilde{n}_{4}^{\prime\,(\tilde{s}_a,\tilde{s}_b)}\rangle}{\langle d_{4}^{\prime}\rangle},}
\end{equation}
 which shows explicitly the double copy structure. Indeed, the classical limit of the QED amplitude is naturally identified as the single copy in this gauge:
 
 \begin{equation}\label{qedclas}
 \boxed{
\langle M_{5}^{\text{QED},(s_a,s_b)}\rangle =\frac{\langle n_{1}^{\prime\, (s_a,s_b)}\rangle }{\langle d_{1}^{\prime}\rangle}+\frac{\langle n_{4}^{\prime \,(s_a,s_b)}\rangle }{\langle d_{4}^{\prime}\rangle}.}
\end{equation}
Taking the classical piece of the denominators \eqref{denominators explicit} leads to   
\begin{equation}\label{classical denominators}
\frac{1}{\langle d_{1}^{\prime}\rangle}{=}\frac{q{\cdot}k}{2\left(p_{1}{\cdot}k\right)^{2}(q^2-q{\cdot}k)},\,\,\,\,\frac{1}{\langle d_{4}^{\prime}\rangle}{=}-\frac{q{\cdot}k}{2\left(p_{3}{\cdot}k\right)^{2}(q^2+q{\cdot}k)}.
\end{equation}
 As a whole, the formulas \eqref{M5 classical CBCJ}, \eqref{qedclas} and \eqref{classical denominators} correspond to the construction given in \cite{Bautista:2019tdr}. The conversion can be done via $\langle n_i^{\prime} \rangle=\frac{2}{q\cdot k}n_i^{\rm{there}} $, where $n_i^{\rm{there}}$ is a local numerator in the classical limit. We have thus found here an alternative derivation which follows directly from the standard BCJ double copy of $M_5$, up to certain details we now describe.
 
Suppose first that the numerators $\langle n_i^{\prime} \rangle$ do not depend on $q^2$. Then we find they can be read off from the QED Compton residues at $q^2 \to \pm q\cdot k$. Indeed, using that \eqref{qedclas}-\eqref{classical denominators} should factor into the Compton amplitude $A_4$ together with a 3-pt. amplitude $A_3$, we get
 \begin{equation}
     \langle n_{i}^{\prime(s_{a},s_{b})}\rangle=\frac{2(p{\cdot}k)^2}{q{\cdot}k}\langle A_{4}^{{\rm QED},s_{a},\mu}\rangle\times\langle A_{3}^{{\rm QED},s_{b},\mu},\rangle
 \end{equation}
where the contraction in $\mu$ denotes propagation of photons. This guarantees the same is true for the gravitational numerators in \eqref{M5 classical CBCJ}, that is

\begin{align}
    \langle n_{i}^{\prime(s_{a},s_{b})}\rangle\otimes\langle n_{i}^{'(s_{a},s_{b})}\rangle&=\frac{4(p{\cdot}k)^4}{(q{\cdot}k)^2}\langle A_{4}^{{\rm QED},s_{a},\mu}\rangle\otimes\langle A_{4}^{{\rm QED},\tilde{s}_{a},\nu}\rangle\times\langle A_{3}^{{\rm QED},s_{b},\mu}\rangle\otimes\langle A_{3}^{{\rm QED},\tilde{s}_{b},\nu}\rangle, \nonumber \\&=\frac{4(p{\cdot}k)^4}{(q{\cdot}k)^2}\langle A_{4}^{s_{a}\otimes\tilde{s}_{a},\mu\nu}\rangle\times\langle A_{3}^{s_{b}\otimes\tilde{s}_{b},\mu\nu}\rangle,
\end{align}
where the contracted indices denote propagation of fat states. Thus we conclude that \textit{the classical limit is controlled by $A_4$ and $A_3$ via the Compton residues} provided the numerators do not depend on $q^2$. Considering the scaling of the multipoles $J\rightarrow\hbar^{-1} \bar{J}$ and that $q\rightarrow \hbar \bar{q}$, we see that this is true up to dipole ${\sim} J$ order. We will confirm this explicitly in the cases below. 

At quadrupole order ${\sim} J^2$, associated to spin-1 particles, we will find explicit dependence on $q^2$ in the numerators. Nevertheless, it is still true that the classical multipoles are given by the the Compton residues as we have exemplified already in \cite{Bautista:2019tdr}. Indeed, as a quick analysis shows, the $q^2$ dependence in $M_5$ that is not captured by them can only arise from 1) contact terms in $M_5$ or 2) contact terms in $M_4$ entering through the residues at $p\cdot k \to 0$. Both contributions can be canceled by adding appropriate (quantum) interactions between the matter particles. Note that canceling such contributions in the QCD side will automatically imply their cancellation on the gravity side.

Let us now see some specific examples of how to write the amplitudes \eqref{m5 gr CBCJ} and their classical pieces \eqref{M5 classical CBCJ}-\eqref{qedclas}, in the gauge \eqref{eq:shifted numeratos}.
\subsubsection{Spinless case}

 In the gauge $(\ref{eq:shifted numeratos})$, the scalar
numerators take an explicit gauge invariant form
\begin{align}
n_{1}^{\prime(0,0)} & =e^{3}\frac{8p_{1}{\cdot}k\left(p_{1}{\cdot}F{\cdot}p_{3}{-}q{\cdot}F{\cdot}p_{3}\right){+}2\left(4p_{3}{\cdot}k{-}4p_{1}{\cdot}p_{3}{-}q{\cdot}(q-k)\right)q{\cdot}F{\cdot}p_{1}}{q{\cdot}k},\label{eq:numeratorBCJ1}\\
n_{4}^{\prime(0,0)} & =e^{3}\frac{8p_{3}{\cdot}k\left(p_{1}{\cdot}F{\cdot}p_{3}{-}q{\cdot}F{\cdot}p_{1}\right){+}2\left(4p_{1}{\cdot}k{-}4p_{1}{\cdot}p_{3}{-}q{\cdot}(q+k)\right)q{\cdot}F{\cdot}p_{3}}{q{\cdot}k}.\label{eq:numerator BCJ2}
\end{align}
Observe these numerators contain $q^2$ dependence. Nevertheless it is completely quantum as the only classical piece is the leading order in $q$, given by
\begin{equation}
   \langle n_{1}^{\prime(0,0)}\rangle{=}\frac{8e^{3}}{q{\cdot}k} p_1{\cdot}R_3{\cdot}F{\cdot}p_1, \quad 
   \langle n_{4}^{\prime(0,0)}\rangle{=}\frac{8e^{3}}{q{\cdot}k} p_3{\cdot}R_1{\cdot}F{\cdot}p_3, \label{n1primeclassical}
\end{equation}
where $R_i^{\mu\nu}{=}p_{i}^{[\mu}(\eta_i 2q{-}k)^{\nu]}$, and  $\eta_1{=}{-}1,\eta_3=1$. It is very easy to check that these numerators reproduce the classical photon radiation of \cite{Goldberger:2016iau} and in fact can be read from there by looking at the pole structure. 
The classical "fat graviton" radiation \eqref{M5 classical CBCJ} for  scalar sources can be computed from the double copy of the classical scalar numerators (\ref{n1primeclassical}) with themselves. It can be shown that these results agree with  \cite{Goldberger:2016iau}. Here however we have taken advantage of the GGT to keep the
double copy structure of the classical gravitational amplitude.

\subsubsection{Case  \texorpdfstring{$s_{a}=0+1$}{s10} and  \texorpdfstring{$s_{b}=0+0$}{s00}}
Now that we have understood how to compute classical radiation for scalars, we can add spin to particle $a$, whereas particles $b$ remains spinless. The gravitational amplitude  $M_{5}^{(0\otimes1,0\otimes0)}$ computed from \eqref{m5 gr CBCJ} can be computed from the double copy of the spinless numerators  (\ref{eq:numeratorBCJ1}-\ref{eq:numerator BCJ2}) with the following numerators:
\begin{equation}
\label{n101}
\begin{split}
n_{1}^{\prime(0,1)} & =\frac{2e^{3}}{q{\cdot}k}\bigg\{ \left[\left(q^{2}{-}q{\cdot}k{+}4p_{1}{\cdot}p_{3}\right)q{\cdot}F{\cdot}p_{1}{+}4(q{-}p_{1}){\cdot}k\,p_{1}{\cdot}F{\cdot}p_{3}\right]\varepsilon_{1}{\cdot}\varepsilon_{2}-\\
 &\qquad\qquad \big[4\left(p_{1}{\cdot}k\,p_{3\mu}F_{\alpha\nu}q^{\alpha}{-}q{\cdot}k\,p_{3\mu}F_{\nu\alpha}p_{1}^{\alpha}\right){+}8\left(p_{1}{\cdot}k\,q_{\mu}F_{\nu\alpha}p_{3}^{\alpha}{-}q_{\mu}p_{3\nu}q{\cdot}F{\cdot}p_{1}\right)\\
 &\qquad\qquad -2\left(2p_{3}{\cdot}k\left(p_{1}{-}q\right){\cdot}k{+}q{\cdot}k\left(q^{2}{-}q{\cdot}k{+}4p_{1}{\cdot}p_{3}\right)\right)F_{\mu\nu}\big]\varepsilon_{1}^{[\mu}\varepsilon_{2}^{\nu]}\\
 &\qquad\qquad -2q{\cdot}k\left(4q^{\alpha}F_{\alpha\mu}p_{3\nu}{-}4p_{3}^{\alpha}F_{\alpha\mu}q_{\nu}{+}2k_{\mu}p_{3}^{\alpha}F_{\alpha\nu}\right)\varepsilon_{1}^{\mu}\varepsilon_{2}^{\nu}\bigg\},
 \end{split}
 \end{equation}
 \begin{equation}\label{n401}
 \begin{split}
n_{4}^{\prime(0,1)} & =\frac{2e^{3}}{q{\cdot}k}\bigg\{ \left[\left(q^{2}{+}q{\cdot}k{+}4p_{1}{\cdot}p_{3}\right)q{\cdot}F{\cdot}p_{3}{-}4(q{+}p_{3}){\cdot}k\,p_{1}{\cdot}F{\cdot}p_{3}\right]\varepsilon_{1}{\cdot}\varepsilon_{2}+\\
 & \qquad\qquad\left[2\left(q{+}p_3\right){\cdot}k\left(4p_{3}^{\alpha}q_{\mu}F_{\alpha\nu}{+}p_{3}{\cdot}k\,F_{\mu\nu}\right){+}4p_{3\mu}\left(2q{+}k\right)_{\nu}\right]\varepsilon_{1}^{[\mu}\varepsilon_{2}^{\nu]}\bigg\} .
 \end{split}
\end{equation}

These numerators can be obtained from the QCD action \eqref{qcdW-boson}, by considering a $W$-boson interacting with a scalar particle through gluons, and then applying the GGT \eqref{eq:shifted numeratos}. Alternatively, they correspond to the QED theory as already explained.

The amplitude  $M_{5}^{(0\otimes1,0\otimes0)}$ is in complete agreement with the Feynman diagrammatic
computation from the Lagrangian $(\ref{eq:10LAGRANGIAN})$ , which
as in the case of scalar sources, provides a strong check of the
contact interaction introduced to match the double copy for the elastic
scattering. 

Now we can take the classical limit of numerators \eqref{n101} and \eqref{n401}. To make contact with the multipole expansion we write the results in terms of the Lorentz generator $J_1^{\mu \nu}$ acting on a spin-1 representation. We strip the matter polarization vectors for simplicity and keep the operators. Up to dipole order, the operators are given by 
\begin{align}
\langle n_{1}^{\prime(0,1)}\rangle{=} & \langle n_{1}^{\prime(0,0)}\rangle{-}\frac{4e^{3}}{q{\cdot}k}\left[p_{1}{\cdot}R_{3}{\cdot}kF{\cdot}J_{1}{-}F_{1q}R_{3}{\cdot}J_{1}{+}p_{1}{\cdot}k\,[F,R_{3}]{\cdot}J_{1}\right], \label{eq:numerator 177/2} \\
\langle n_{4}^{\prime(0,1)}\rangle{=} & \langle n_{4}^{\prime(0,0)}\rangle{+}\frac{4e^{3}}{q{\cdot}k}p_{3}{\cdot}F{\cdot}\hat{R}_1{\cdot}p_{3},\label{eq:numerators 1/2}
\end{align}
where we have used $F_{iq}=\eta_{i}(p_{i}{\cdot}F{\cdot}q),$  and $\hat{R}_{i}=(\eta_{i}2q-k)^{[\mu}J_{i}^{\nu]\alpha}(\eta_{i}2q-k)_{\alpha}$. Also $[F,R]_{\mu \nu}= F_{\mu \alpha} R^{\alpha}_{\,\,\nu} - (\mu \leftrightarrow \nu)$, etc. These classical numerators agree with the ones given in \cite{Bautista:2019tdr}. Note that the $q^2$ dependence in \eqref{n101},\eqref{n401} has become has become suppressed by powers of $J$. In \cite{Bautista:2019tdr} we have nevertheless managed to cancel it by adding flavor interactions, hence obtaining also the form of the classical gravitational quadrupole ${\sim} J^2$.

Finally, the QED and gravitational radiation constructed from \eqref{eq:numerator 177/2},\eqref{eq:numerators 1/2} agrees with the results of Li and Prahbu \cite{Li:2018qap} when we set one of the massive objects to be spinless.

\subsubsection{Case  \texorpdfstring{$s_{a}=s_b=0+1$}{s10}}\label{sec:sasb01}
We can now move  on to the case in which both massive lines  have spin-$1$ in the gravitational theory. We want to compute  the gravitational amplitude for  inelastic scattering $M_5^{(0\otimes1,0\otimes1)}$ using \eqref{m5 gr CBCJ}. The scalar numerators are given in \eqref{eq:numeratorBCJ1} and \eqref{eq:numerator BCJ2}. The numerators for the spining case are constructed following the considerations of Sec. \ref{sec:casee} and give
\begin{equation}\label{n1p11}
\begin{split}
n_{1}^{\prime(1,1)} &{=}\frac{2e^{3}}{q{\cdot}k}\bigg\{\left[\left(q^{2}{-}q{\cdot}k{+}4p_{1}{\cdot}p_{3}\right)q{\cdot}F{\cdot}p_{1}{+}4(q{-}p_{1}){\cdot}k\,p_{1}{\cdot}F{\cdot}p_{3}\right]\varepsilon_{1}{\cdot}\varepsilon_{2}\,\varepsilon_{3}{\cdot}\varepsilon_{4}+\\
 & \bigg[8\left(q{-}p_{1}\right){\cdot}k\,q{\cdot}\epsilon_{2}\,q_{\mu}\epsilon_{1}^{\alpha}F_{\alpha\nu}+4\left[q{\cdot}k\left(2p_{1\mu}q_{\nu}{+}(q{-}p_{1})_{\mu}k_{\nu}\right){+}p_{1}{\cdot}k\,k_{\mu}q_{\nu}\right]\varepsilon_{1}{\cdot}F{\cdot}\varepsilon_{2}+\\
 & 4q_{\mu}\left(2q{\cdot}\varepsilon_{1}\,p_{1}{\cdot}k{-}k{\cdot}\varepsilon_{1}\,q{\cdot}k\right)\varepsilon_{2}^{\alpha}F_{\alpha\nu}{-}\left[4p_{1}{\cdot}k\,q_{\alpha}k_{\beta}\varepsilon_{1}^{[\alpha}\varepsilon_{2}^{\beta]}{+}q{\cdot}k\,k{\cdot}\varepsilon_{1}\left(2q{-}k\right){\cdot}\varepsilon_{2}\right]F_{\mu\nu}+\\
 & \left[2\left(q{-}p_{1}\right){\cdot}k\left(4q_{\mu}p_{1}^{\alpha}F_{\nu\alpha}{+}p_{1}{\cdot}k\,F_{\mu\nu}\right){+}4p_{1\mu}\left(2q-k\right)_{\nu}q{\cdot}F{\cdot}p_{1}\right]\varepsilon_{1}{\cdot}\varepsilon_{2}-\\
 & 4\left(p_{1}{\cdot}k\,q_{\rho}F^{\rho\sigma}{+}q{\cdot}k\,p_{1\rho}F^{\rho\sigma}{+}2q^{\sigma}\,q{\cdot}F{\cdot}p_{1}\right)\varepsilon_{1[\mu}\varepsilon_{2\sigma]}\left(2q{-}k\right)_{\nu}{-}4q{\cdot}k\,q{\cdot}F{\cdot}\varepsilon_{1}\varepsilon_{2\mu}\left(2q{-}k\right)_{\nu}\bigg]\\
 &\times \varepsilon_{3}^{[\mu}\varepsilon_{4}^{\nu]}{+} \bigg[4q{\cdot}\varepsilon_{2}\left(q{-}p_{1}\right){\cdot}k\,p_{3}{\cdot}F{\cdot}\varepsilon_{1}{+}2\left(2q{\cdot}\varepsilon_{1}\,p_{1}{\cdot}k{-}q{\cdot}k\,k{\cdot}\varepsilon_{1}\right)p_{3}{\cdot}F{\cdot}\varepsilon_{2}\\&{-}8p_{3\mu}q_{\nu}q{\cdot}F{\cdot}p_{1}\varepsilon_{1}^{[\mu}\varepsilon_{2}^{\nu]}-
  2p_{1}{\cdot}k\,p_{3}{\cdot}\varepsilon_{1}q{\cdot}F{\cdot}\varepsilon_{2}{-}4q{\cdot}k\,p_{3\mu}p_{1}^{\alpha}F_{\alpha\nu}\varepsilon_{1}^{[\mu}\varepsilon_{2}^{\nu]}{-}\\
 & 2\left(2q{-}p_{1}\right){\cdot}k\,p_{3}{\cdot}\varepsilon_{2}q{\cdot}F{\cdot}\varepsilon_{1}+ \left(q{\cdot}k\left(q^{2}{-}q{\cdot}k{+}4p_{1}{\cdot}p_{3}\right)-2\left(q{-}p_{1}\right){\cdot}k\,p_{3}{\cdot}k\right)\varepsilon_{1}{\cdot}F{\cdot}\varepsilon_{2}\bigg]\varepsilon_{3}{\cdot}\varepsilon_{4}\bigg\}.
 \end{split}
\end{equation}
The numerator $n_{4}^{\prime(1,1)}$ is given by exchanging particles $ a\leftrightarrow b$ in $n_{1}^{\prime(1,1)}$, with $q\rightarrow-q$. The result expressed in terms of these numerators is far more compact than the Feynman diagram expansion obtained from the covariantized Lagrangian \eqref{lagrangian01x01}.

Now, by taking the classical limit of the numerators \eqref{n1p11} we  can compute the amplitude $\langle M_5^{(0\otimes1,0\otimes1)}\rangle$ via \eqref{M5 classical CBCJ}, using also \eqref{n1primeclassical} and \eqref{classical denominators}. In the multipole form of the previous section, the numerators read, up to dipole order,

\begin{equation}\label{spin1classical numerators}
    \begin{split}
\langle n_{1}^{\prime(1,1)}\rangle & =\langle n_{1}^{\prime(0,0)}\rangle{-}4e^{3}\left[p_{1}{\cdot}R_{3}{\cdot}kF{\cdot}J_{1}{-}F_{1q}R_{3}{\cdot}J_{1}{+}p_{1}{\cdot}k\,[F,R_{3}]{\cdot}J_{1}-p_{1}{\cdot}F{\cdot}\hat{R}_{3}{\cdot}p_{1}\right],\\
\langle n_{4}^{\prime(1,1)}\rangle & =\langle n_{4}^{\prime(0,0)}\rangle{-}4e^{3}\left[p_{3}{\cdot}R_{1}{\cdot}kF{\cdot}J_{3}{-}F_{3q}R_{1}{\cdot}J_{3}{+}p_{3}{\cdot}k\,[F,R_{1}]{\cdot}J_{3}-p_{3}{\cdot}F{\cdot}\hat{R}_{1}{\cdot}p_{3}\right]. 
\end{split}
\end{equation}
Classical radiation computed in this way agrees with the classical double copy result of Goldberger, Li and Prabhu \cite{Li:2018qap,Goldberger:2017ogt} for the whole Fat Graviton radiative field given in  eq. (51) of \cite{Li:2018qap} up to dipole order.

Now, we have seen that an alternative Lagrangian construction for the $0\otimes 1$ double copy at two matter lines is given by the extension of $\mathcal{N}=4$ Supergravity. We know that the amplitude $M_5^{\mathcal{N}=4\, \rm{SUGRA}}$ of this theory differs from $ M_5^{(0\otimes1,0\otimes1)}$ given here only in terms arising from (quantum) flavour interactions, c.f. \eqref{contact term 01} vs \eqref{eq:final01}. Equivalently both amplitudes have the same residues as $q^2 \to \pm q\cdot k$ and hence the same classical limit. As explained, these cuts correspond to Compton amplitudes and in this case feature the propagation of the axion, dilaton and graviton. We have checked these cuts explicitly by comparing the numerators \eqref{n1p11} against the amplitude $M_5^{\mathcal{N}=4\, \rm{SUGRA}}$ obtained via CHY (see Appendix \ref{ap:chy}).
\subsubsection{Case  \texorpdfstring{$s_{a}=s_b=\frac{1}{2}+\frac{1}{2}$}{s1212}}
The final case for inelastic scattering in the gauge \eqref{eq:shifted numeratos} is given by the factorization of the  gravitational amplitude \eqref{m5 gr CBCJ} as $s_{a}=s_b=\frac{1}{2}+\frac{1}{2}$. For the QCD theory, the shifted numerators entering in \eqref{m5qcd GGT} are
\begin{align}
n_{1}^{\prime(\frac{1}{2},\frac{1}{2})} & =\frac{4e^{3}F_{\alpha\beta}}{q{\cdot}k}\big[q^{[\alpha}p_{1}^{\beta]}\bar{u}_{2}\gamma^{\mu}u_{1}\bar{u}_{4}\gamma_{\mu}u_{3}{+}(q{-}p_{1}){\cdot}k\,\bar{u}_{2}\gamma^{[\alpha}u_{1}\bar{u}_{4}\gamma^{\beta]}u_{3}{-}\frac{q{\cdot}k}{4}\,\bar{u}_{2}\gamma^{[\alpha}\gamma^{\beta]}\gamma^{\mu}u_{1}\bar{u}_{4}\gamma_{\mu}u_{3}\big],\\
n_{4}^{\prime(\frac{1}{2},\frac{1}{2})} & =\frac{4e^{3}F_{\alpha\beta}}{q{\cdot}k}\big[q^{[\alpha}p_{3}^{\beta]}\bar{u}_{4}\gamma^{\mu}u_{3}\bar{u}_{2}\gamma_{\mu}u_{1}{+}(q{+}p_{3}){\cdot}k\,\bar{u}_{4}\gamma^{[\alpha}u_{3}\bar{u}_{2}\gamma^{\beta]}u_{1}{-}\frac{q{\cdot}k}{4}\,\bar{u}_{4}\gamma^{[\alpha}\gamma^{\beta]}\gamma^{\mu}u_{3}\bar{u}_{2}\gamma_{\mu}u_{1}\big].
\end{align}
Analogously, their  charge conjugated pairs read
\begin{align}
\bar{n}_{1}^{\prime(\frac{1}{2},\frac{1}{2})} & =\frac{4e^{3}}{q{\cdot}k}F_{\alpha\beta}\big[q^{[\alpha}p_{1}^{\beta]}\bar{v}_{1}\gamma^{\mu}v_{2}\bar{v}_{3}\gamma_{\mu}v_{4}{+}(q{-}p_{1}){\cdot}k\,\bar{v}_{1}\gamma^{[\alpha}v_{2}\bar{v}_{3}\gamma^{\beta]}v_{4}{+}\frac{q{\cdot}k}{4}\,\bar{v}_{1}\gamma^{\mu}\gamma^{[\alpha}\gamma^{\beta]}v_{2}\bar{v}_{3}\gamma_{\mu}v_{4}\big],\\
\bar{n}_{4}^{\prime(\frac{1}{2},\frac{1}{2})} & =\frac{4e^{3}}{q{\cdot}k}F_{\alpha\beta}\big[q^{[\alpha}p_{3}^{\beta]}\bar{v}_{3}\gamma^{\mu}v_{4}\bar{v}_{1}\gamma_{\mu}v_{2}{+}(q{+}p_{3}){\cdot}k\,\bar{v}_{3}\gamma^{[\alpha}v_{4}\bar{v}_{1}\gamma^{\beta]}v_{2}{+}\frac{q{\cdot}k}{4}\,\bar{v}_{3}\gamma^{\mu}\gamma^{[\alpha}\gamma^{\beta]}v_{4}\bar{v}_{1}\gamma_{\mu}v_{2}\big].
\end{align}

The  gravitational amplitude $M_5^{(\frac{1}{2}\otimes\frac{1}{2},\frac{1}{2}\otimes\frac{1}{2})}$ can be computed from the double copy of the above  numerators with their charge conjugated pairs, using  the operation defined in  \eqref{eq:doublestar-1-1}. The result, which we provide in the Mathematica ancillary notebook,  is in complete agreement with the Feynman diagrammatic computation from the  Lagrangian  \eqref{eq:1212LAGRANGIAN}. 

On the classical side, although the classical limit of these QCD numerators agrees with \eqref{spin1classical numerators} (with appropiate conjugated numerators and up to dipole order), it is clear that the double copy  $\langle M_5^{(\frac{1}{2}\otimes\frac{1}{2},\frac{1}{2}\otimes\frac{1}{2})}\rangle$ differs  from  $\langle M_5^{(0\otimes1,0\otimes1)}\rangle$. For instance, as the double copy for the former is symmetric in the numerators the axion field has no radiative amplitude, whereas for the latter it is unavoidably present.

We do not provide the explicit result for  $\langle M_5^{(\frac{1}{2}\otimes\frac{1}{2},\frac{1}{2}\otimes\frac{1}{2})}\rangle$, but let us mention that it is naturally computed using the symmetric double copy product defined in \cite{Bautista:2019tdr}, which preserves the multipole structure of the amplitude.

%%%%%%%%%%%%%%%%%%%%%%%%%%%%%%%%%%%%%%%%%%%%%%%%%%%%%%%%%%%%%%%%%%%%%%%%%%%%%%%%%%%%%%%%%%%%%%%%%%%%%%%%%%%%%%%%%%%%%%%%%%%%%%%%%%%%%%%%%%%%%

\section{Discussion}

Based on the analysis performed along refs. \cite{Porrati:2010hm,osti_4073049,Deser:2000dz,Holstein:2006pq,Pfister_2002} and in the current work we can draw an equivalence for lower spins between the following three statements:
\begin{enumerate}
\item The cancellation of $\frac{1}{m}$ divergences in the tree-level high-energy
limit of single matter lines.
\item The ``natural value'' of the gyromagnetic ratio
$g=2$.
\item The double copy construction for the single matter line ($A_{n}$) amplitudes.
\end{enumerate}

Let us remark that this equivalence not only seems to show up in QFT amplitudes but also in classical solutions \cite{Pfister_2002}. One instance of this is the so-called $\sqrt{\text{Kerr}}$ solution in electrodynamics which has been the focus of recent studies \cite{Arkani-Hamed:2019ymq,Guevara:2019fsj}. This EM solution can be double-copied into the Kerr metric via the Kerr-Schild ansatz \cite{Monteiro:2014cda}, and also features $g=2$. Since these classical solutions contain the full tower of spin-multipoles, and so do higher spin particles in QFT, a natural question that arises is: \textit{How much of the above equivalence can be promoted to higher spins?} 

A hint of the answer may come from the 3-pt amplitudes first derived in \cite{Arkani-Hamed:2017jhn} which are directly related to the aforementioned classical solutions \cite{Guevara:2017csg,Guevara:2018wpp,Guevara:2019fsj,Chung:2018kqs,Chung:2019duq,Arkani-Hamed:2019ymq}, at least at leading order in the coupling. In \cite{Bautista:2019tdr} we have emphasized their double copy structure, which fixes not only $g=2$ but also the full tower of multipoles in both gravity and QCD side. Here we have pointed out that these objects are in correspondence with higher spins massless amplitudes, thereby providing an underlying reason for double copy. Quite paradoxically, the latter are known to be inconsistent \cite{Benincasa:2007xk} whereas the former have an striking physical realization. To elucidate this contradiction we recall that massless higher spin amplitudes only fail at the level of the "4-particle" test \cite{Benincasa:2007xk,Arkani-Hamed:2017jhn}. 

Indeed, the higher spin 4-point (Compton) $A_4$ amplitudes suffer from ambiguities in the form of contact terms and from $\frac{1}{m}$ divergences, although recent progress to understand these has been made in \cite{Guevara:2018wpp,Chung:2018kqs,Chung:2019duq,Bautista:2019tdr}. The importance of this object at higher spins was emphasized by one of us in \cite{Guevara:2017csg} and proposed to control the subleading order associated to gravitational and EM classical potentials. These potentials emerge in the two-body problem \cite{Chung:2018kqs,Chung:2019duq,Damgaard:2019lfh,Maybee:2019jus,Neill:2013wsa,Holstein:2008sw,Holstein:2008sx}  (particularly outside the test body limit) and hence their understanding could have not only theoretical but practical implications. In fact, the relevance of the full tower of $A_n$ amplitudes lies in that they have been proposed to control the classical piece of conservative potentials at deeper orders in the coupling \cite{Neill:2013wsa,Cheung:2018wkq,Bern:2019crd,Bern:2019nnu,Bjerrum-Bohr:2019nws}. 

In \cite{Bautista:2019tdr} we demonstrated the latter fact is true also for radiation: At least at order ${\sim} \kappa^3 $ and at spins $s\leq 2$ the non-conservative observables are controlled by $A_4$ and $A_3$ instead of the full $M_5$ amplitude. Here we have rederived this construction from a BCJ double-copy perspective and use it to make contact with the results of Goldberger et al. \cite{Goldberger:2016iau,Goldberger:2017frp,Goldberger:2017vcg,Goldberger:2017ogt,Li:2018qap} for the full massless spectrum including dilatons, axions and gravitons. As we have mentioned it is remarkable how via QFT double copy we have found the precise couplings of these fields to matter, besides the aforementioned $g=2$ condition. One such extensions has naturally led us to $\mathcal{N}=4$ Supergravity seen as a classical theory of long-range forces. On the practical side it is important to evaluate the relevance of these additional fields, as well as string theory corrections, from the perspective of effective classical potentials arising from amplitudes, see e.g. \cite{Emond:2019crr,Brandhuber:2019qpg} for recent related results.

As a last point, let us mention that even though the cancellation of $\frac{1}{m}$ divergences at higher spins has been ruled out  \cite{Deser:2000dz,Cucchieri:1994tx} it is still true that the choice of $g=2$ is preferred from both a causality perspective and the counting of degrees of freedom in certain cases \cite{Deser:2000dz,Deser:2001dt,Porrati:2010hm,Pfister_2002}. It would be interesting to see if such cases are to exhibit a double copy structure: In fact the higher-spin 3-pt. amplitudes \eqref{eq:mass3pths}, when written in a local form using polarization tensors, also feature such $1/m$ terms \cite{Lorce:2009br,Lorce:2009bs,Bautista:2019tdr}. On the other hand, it is true that string theory provides a consistent tower of higher spin states exhibiting double copy \cite{Kawai:1985xq}. In fact, on the open string side such states have $g=2$ \cite{Porrati:2010hm} as implied by the double copy relations \cite{Chung:2018kqs}. It remains to understand whereas a truncation of the full string spectrum to only certain higher spin states (for instance by isolating a single Regge trajectory \cite{Porrati:2010hm}) is possible. 

\acknowledgments
We thank Freddy Cachazo, Henrik Johansson and Alex Ochirov for useful discussions. We thank Emilio Ojeda and Shan-Ming Ruan for their help with a Mathematica implementation. A.G. acknowledges support via Conicyt grant 21151647. Y.F.B.  acknowledges the Natural Sciences and Engineering Research Council of Canada (NSERC) the financial support.  Research at Perimeter Institute is supported by the Government of Canada through the Department of Innovation, Science and Economic Development Canada and by the Province of Ontario through the
Ministry of Research, Innovation and Science.

%%%%%%%%%%%%%%%%%%%%%%%%%%%%%%%%%%%%%%%%%%%%%%%%%%%%%%%%%%%%%%%%%%%%%%%%%%%%%%%%%%%%%%%%%%%%%%%%%%%%%%%%%%%%%%%%%%%%%%%%%%%%%%%%%%%%%%%%%%%%%
\appendix 
%%%%%%%%%%%%%%%%%%%%%%%%%%%%%%%%%%%%%%%%%%%%%%%%%%%%%%%%%%%%%%%%%%%%%%%%%%%%%%%%%%%%%%%%%%%%%%%%%%%%%%%%%%%%%%%%%%%%%%%%%%%%%%%%%%%%

\section{Double Copy in  \texorpdfstring{$d=4$}{d4}}\label{4 d double copy}

In this appendix we outline the \12x12 construction in $d=4$. 
It is interesting to make connection with the spinor formalism
for massive particles \cite{Arkani-Hamed:2017jhn}, recently implemented for obtaining a massive double copy in \cite{Johansson:2019dnu}. Let us briefly sketch how our operation will read in such variables. For this, observe that we can write

\begin{equation}
E_{\mu}^{ab}\sigma^{\mu}={\color{black}\frac{\sqrt{2}}{m}}|p^{(a}]\langle p^{b)}|\,\quad E_{\mu}^{ab}\tilde{\sigma}^{\mu}={\color{black}\frac{\sqrt{2}}{m}}|p^{(a}\rangle [ p^{b)}|.\label{eq:pols1}
\end{equation}
where $E_{\mu}^{ab}$ is a spin-1 polarization vector, $\tilde{E}^{ab}{\cdot}P=0$,
with the little group indices $\{a,b\}=\{1,1\},\{2,2\},\{1,2\}$. Note its spinors satisfy
the Dirac equation

\begin{equation}
P|p^{a}\rangle=m|p^{a}],\,\quad\tilde{P}|p^{a}]=m|p^{a}\rangle,\label{eq:Diraceq}
\end{equation}
where $P=P_{\mu}\sigma^{\mu}$ and $\tilde{P}=P_{\mu}\tilde{\sigma}^{\mu}$. Then it is true that $[1^a,1^b]=-m\epsilon^{ab}$, and $\langle 1^a,1^b\rangle=m\epsilon^{ab}$
Now, in terms of the Dirac matrices note that

\begin{eqnarray}
(\slashed P+m\mathbb{I}_{4\times4})\slashed{E}^{ab} & = & {\color{black}\frac{\sqrt{2}}{m}}\left(\begin{array}{cc}
m\mathbb{I}_{2\times2} & P\\
\tilde{P} & m\mathbb{I}_{2\times2}
\end{array}\right)\left(\begin{array}{cc}
0 & |1^{(a}]\langle 1^{b)}|\\
|1^{(a}\rangle [1^{b)}| & 0
\end{array}\right),\label{projector}\\
 & = & {\color{black}\sqrt{2}}\left(\begin{array}{cc}
 |1^{(a}][1^{b)}|&|1^{(a}]\langle1^{b)}|\\
|1^{(a}\rangle [1^{b)}| &|1^{(a}\rangle \langle1^{b)}| 
\end{array}\right),\\
 & = & {\color{black}\sqrt{2}}\left(\begin{array}{c}
|1^{(a}]\\
|1^{(a}\rangle
\end{array}\right)([1^{b)}|\,\,\langle1^{b)}|),\\
 & = & {\color{black}\sqrt{2}}\,u^{(a}\bar{v}^{b)}\label{product projector},
\end{eqnarray}
where $u$ and $v$ are Dirac spinors satisfying $\slashed{P}u=m u$, $\slashed{P}v=-m v$,
as follows from (\ref{eq:Diraceq}). Note that the spin-1 polarization
can be recovered from \eqref{product projector} via
\begin{equation}
E_{\mu}^{ab}=\frac{1}{{\color{black}\sqrt{2}}m}\bar{v}^{(a}\gamma_{\mu}u^{b)}.
\end{equation}
In this sense the spin-1 polarization vector is constructed out of spin-$\frac{1}{2}$ polarizations, analogously to the rules \eqref{eq:replace} for higher spins.

We see that in $d=4$ the choice of polarizations given by
(\ref{eq:pols1}) turns the product \eqref{tensor product } into
\begin{equation}
X\otimes Y=\bar{v}_{2}^{(b_{2}}Xu_{1}^{(a_{1}}\times\bar{v}_{1}^{b_{1})}\bar{Y}u_{2}^{a_{2})},
\end{equation}
which is simple multiplication together with symmetrization over the
spin-$\frac{1}{2}$ states. Since this operation coincides with the one given in \cite{Johansson:2019dnu} we conclude that the amplitudes for a spin-1 field will agree in $d=4$.

For instance, for one matter line we will write
\begin{equation}\label{eq:a9}
A_{n}^{{\rm gr}}(E_{1}^{a_{1}b_{1}},E_{2}^{a_{2}b_{2}})=\sum_{\alpha,\beta}K_{\alpha\beta}\left(A_{n,\alpha}^{\text{QCD}}\right){}^{(a_{1}(b_{2}}\left(A_{n,\beta}^{\text{QCD}}\right){}^{a_{2})b_{1})}.
\end{equation}
which exhibits the symmetry properties of the indices explicitly. In particular it can be used to streamline the argument given in Section \ref{sec:sec2} for axion pair-production.

In an analogous way to \eqref{projector}, in the representation where $\gamma^{5}=\left(\begin{array}{cc}
-\mathbb{I}_{2\times2} & 0\\
0 & \mathbb{I}_{2\times2}
\end{array}\right)$,  we have 
\begin{align}
(\slashed P+m\mathbb{I}_{4\times4})\gamma^{5} & =\left(\begin{array}{cc}
-m\mathbb{I}_{2\times2} & |1]^{a}\langle 1|^{b}\epsilon_{ab}\\
|1\rangle^{a}[1|^{b}\epsilon_{ab} & m\mathbb{I}_{2\times2}
\end{array}\right)\\
 & =u^{[a}\bar{{v}}^{b]}\epsilon_{ab}\,.
\end{align}
By inserting the projector on the LHS instead of \eqref{projector} into our double copy, we find that antisymmetrizing little group indices from the Dirac spinors leads to a pseudoscalar. This antisymmetrization will necessarily require an odd number of axion fields in \eqref{eq:a9}. Hence the axion can be sourced by matter if the Proca field decays to a pseudoscalar, which is again consistent with the Lagrangian of \cite{Johansson:2019dnu}. Further analysis in general dimensions is done in the next Appendix.\\

\subsection*{Lagrangian comparison with \cite{Johansson:2019dnu}}

The results  of \cite{Johansson:2019dnu} consider the full spectrum of the $1/2 \otimes 1/2$ double copy restricted to four-dimensions. In contrast, our work shows that there exists a truncated spectrum in general dimensions. It is interesting  to analyze the overlap by comparing the interactions in our Lagrangian \eqref{eq:1212dc} with a truncated version of the one in \cite{Johansson:2019dnu}. Note that in principle the matching at the level of amplitudes does not guarantee such an off-shell agreement due to diverse field redefinitions. However, in our case it is possible since 1) both actions are written on the the Einstein frame for the graviton-dilaton couplings and 2) It can be shown that the axion and massive pseudoscalar fields decouple in the amplitudes of \cite{Johansson:2019dnu}, hence the corresponding interaction terms can be ignored in their Lagrangian.

With the previous considerations the Lagrangian of \cite{Johansson:2019dnu} leads to the following explicit couplings of the dilaton to the Proca field at $\mathcal{O}(\kappa^2)$
\begin{equation}\label{eq:lagrangian_qcds}
    \mathcal{L}_{\rm{QCD^2}} = -\frac{2}{\kappa^2}R + \frac{\partial_\mu \bar{Z}\partial^\mu Z}{\left(1-\frac{\kappa^2}{4}\bar{Z}Z\right)}-\frac{1}{2}F_{\mu\nu }^* F^{\mu\nu}+m^2A_\mu ^*A^\mu \left(1-\frac{\kappa}{2}(\bar{Z}+Z)+\frac{\kappa^2}{2}\bar{Z}Z+\mathcal{O}(\kappa^3)\right)
\end{equation}
The kinetic term for $Z$ can be cast into the standard form when we identify the dilaton component. Indeed, recall the field $Z$ was defined by

\begin{equation}
    Z=\frac{2a+i(e^{-2\phi}-1)}{2a+i(e^{-2\phi}+1)}.
\end{equation}
Where the axion corresponds to the parity-odd piece, i.e. the field $a$. Setting $a\to0$ implies $\bar{Z}=Z=-\tanh{\phi}$. Doing the further field redefinition $Z\to \frac{\kappa}{2}Z $, the Lagrangian \eqref{eq:lagrangian_qcds} becomes

\begin{equation}
     \mathcal{L}_{\rm{QCD^2}} =  -\frac{2}{\kappa^2}R+ \frac{4}{\kappa^2}(\partial\phi)^2-\frac{1}{2}F_{\mu\nu }^* F^{\mu\nu} +m^2A_\mu ^*A^\mu \left(1+2\tanh{\phi} +2(\tanh{\phi})^2+\mathcal{O}(\kappa^3) \right).
\end{equation}
Finally, we do the field re-definition  $\phi\to \frac{\kappa}{2}\phi $,  expanding up to second order in $\phi$, which is the order of the validity of the Lagrangian \eqref{eq:lagrangian_qcds}; in addition, we  turn $A^\mu$ into a real  field   using the argument made above \eqref{eq:symsc},
to arrive at
\begin{equation}
 \mathcal{L}_{\rm{QCD^2}} =  -\frac{2}{\kappa^2}R+ (\partial\phi)^2-\frac{1}{4}F_{\mu\nu } F^{\mu\nu} +\frac{m^2}{2}A_\mu A^\mu \left(1+ \kappa\phi +\frac{\kappa^2}{2}\phi^2 +\mathcal{O}(\kappa^3) \right),
\end{equation}
which precisely agrees with the Lagrangian \eqref{eq:1212dc} for $d=4$ if we truncate at  $\mathcal{O}(\kappa^2)$.

%%%%%%%%%%%%%%%%%%%%%%%%%%%%%%%%%%%%%%%%%%%%%%%%%%%%%%%%%%%%%%%%%%%%%%%%%%%%%%%%%%%%%%%%%%%%%%%%%%%%%%%%%%%%%%%%%%%%%%%%%%%%%%%%%%%
%%%%%%%%%%%%%%%%%%%%%%%%%%%%%%%%%%%%%%%%%%%%%%%%%%%%%%%%%%%%%%%%%%%%%%%%%%%%%%%%%%%%%%%%%%%%%%%%%%%%%%%%%%%%%%%%%%%%%%%%%%%%%%%%%%%%
 
\section{Tree-level Unitarity at $n=4$}\label{residues at 4 points}

In this appendix we compute the residues of the gravitational
amplitude $A_4^{\frac{1}{2}\otimes\frac{1}{2}}$.
The aim of this is twofold. On the one hand this checks explicitly that the operation $(\ref{eq:massiveklt})$ defines a QFT amplitude for $n=4$ and outlines the argument for general $n$. On the other hand, we want to find the matter fields that propagate in a given factorization channel.
For two dilaton emissions we find only the propagation of the Proca field, which is consistent with our Lagrangian \eqref{eq:1212dc-1}. For two axion emissions we find the propagation of tensor structures of rank four and five. The former can be interpreted as a pseudoscalar in $d=4$. In general dimension, the propagation of these structures makes it more
involved to write the Lagrangian of the full \12x12 theory including axions. \\
Consider then the Compton amplitude
from the \12x12 theory $(\ref{eq:massiveklt})$
\begin{equation}
A_{4}^{\frac{1}{2}\otimes\frac{1}{2}}(W_{1}H_{3}^{\mu_{3}\nu_{3}}H_{4}^{\mu_{4}\nu_{4}}W_{2}^{*})=\frac{K_{1324,1324}}{2^{\left\lfloor d/2\right\rfloor -1}}\text{tr}\left[A_{4,1324}^{\rm{QCD},\mu_{3}\mu_{4}}\slashed{\varepsilon}_{1}\left(\slashed{p_{1}}{+}m\right)\bar{A}_{4,1324}^{\rm{QCD},\nu_{3}\nu_{4}}\slashed{\varepsilon}_{2}\left(\slashed{p_{2}}{+}m\right)\right],\label{A4 grklt}
\end{equation}
 where the $4$ pt. QCD partial amplitudes are given in $(\ref{eq:A41324})$,
and the massive  KLT kernel at four points was given in $(\ref{eq:klt 1324})$. \\
We have claimed that $(\ref{A4 grklt})$ defines a tree-level amplitude. First, from the standard argument it is clear that the RHS is local. Let us then argue that unitarity of the gravitational amplitude follows from
the unitarity of the QCD amplitudes.
Consider for instance the factorization channel $2p_{1}{\cdot}k_{3}\rightarrow0$. We know that in such case
the QCD amplitude factorizes as
\begin{equation}
A_{4,1324}^{\rm{QCD},\mu_{3}\mu_{4}}\rightarrow\frac{1}{2p_{1}{\cdot}k_{3}}A_{3,\text{R}}^{\rm{QCD},\mu_{3}}\left(\slashed{p}_{13}-m\right)A_{3,\text{L}}^{\rm{QCD},\mu_{4}}{+}\cdots,
\end{equation}
 Analogously, the  charge conjugated amplitude  factorizes as 
\begin{equation}
\bar{A}_{4,1324}^{\rm{QCD},\nu_{3}\nu_{4}}\rightarrow\frac{1}{2p_{1}{\cdot}k_{3}}\bar{A}_{3,\text{L}}^{\rm{QCD},\nu_{4}}\left(\slashed{p}_{13}+m\right)\bar{A}_{3,\text{R}}^{\rm{QCD},\nu_{3}}{+}\cdots,
\end{equation}
This implies that $(\ref{A4 grklt})$ behaves as
\begin{equation}\label{res1}
\begin{split}
A_{4}^{\frac{1}{2}\otimes\frac{1}{2}}(W_{1}H_{3}^{\mu_{3}\nu_{3}}H_{4}^{\mu_{4}\nu_{4}}W_{2}^{*})\rightarrow & -\frac{1}{2p_{1}{\cdot}k_{3}2^{\left\lfloor d/2\right\rfloor -1}}\text{tr}\bigg[A_{3,\text{L}}^{\rm{QCD},\mu_{4}}\slashed{\varepsilon}_{1}\left(\slashed{p_{1}}{+}m\right)\bar{A}_{3,\text{L}}^{\rm{QCD},\nu_{4}}\left(\slashed{p}_{13}+m\right)\\
 & \qquad\qquad\bar{A}_{3,\text{R}{}}^{\rm{QCD},\nu_{3}}\slashed{\varepsilon}_{2}\left(\slashed{p_{2}}{+}m\right)A_{3,\text{R}}^{\rm{QCD},\mu_{3}}\left(\slashed{p}_{13}-m\right)\bigg]{+}\cdots,
\end{split}
\end{equation}
We can examine the inner spectrum in the factorization channel by using the Fierz relations for the product of two
matrices $M$ and $N$ \cite{9781139026833},
\begin{equation}
\text{tr}[M\times N]=\frac{1}{2^{\left\lfloor d/2\right\rfloor }}\sum_{J}^{[d]}\frac{(-1)^{|J|}}{|J|!}\text{tr}\left[M\Gamma_{J}\right]\text{tr}\left[N\Gamma^{J}\right],\qquad[d]=\begin{cases}
d & {\rm {for\,\,even}\,}\,d\\
\frac{d-1}{2} & {\rm {for\,\,odd}\,\,}d
\end{cases}\label{eq:fierrz relations-1}
\end{equation}
where $\{\Gamma^{J}=\mathbb{I},\gamma^{\alpha},\gamma^{\alpha_{1}\alpha_{2}},\cdots,\gamma^{\alpha_{1}\cdots\alpha_{d}}\}$
is the Clifford algebra basis, with $\alpha_{1}<\alpha_{2}<\cdots<\alpha_{r}$. The gravitational amplitude \eqref{res1} then takes the form 
\begin{equation}\label{sumJ}
\begin{split}
&-\frac{1}{4p_{1}{\cdot}k_{3}2^{2\left\lfloor d/2\right\rfloor -2}}\sum_{J}^{[d]}\frac{(-1)^{|J|}}{|J|!}\text{tr}\left[A_{3,\text{L}}^{\rm{QCD},\mu_{4}}\slashed{\varepsilon}_{1}\left(\slashed{p_{1}}{+}m\right)\bar{A}_{3,\text{L}}^{\rm{QCD},\nu_{4}}\left(\slashed{p}_{13}+m\right)\Gamma_{J}\right]\times\\
 & \qquad\qquad \text{tr}\left[\bar{A}_{3\text{R}}^{\rm{QCD},\nu_{3}}\slashed{\varepsilon}_{2}\left(\slashed{p_{2}}{+}m\right)A_{3,\text{R}}^{\rm{QCD},\mu_{3}}\left(\slashed{p}_{13}-m\right)\Gamma^{J}\right]{+}\cdots,
\end{split}
\end{equation}
Now it is clear that each trace corresponds to the double copy for
the $3$pt amplitudes, therefore we have
\begin{equation}
A_{4}^{\frac{1}{2}\otimes\frac{1}{2}}(W_{1}H_{3}^{\mu_{3}\nu_{3}}H_{4}^{\mu_{4}\nu_{4}}W_{2}^{*})\rightarrow-\frac{1}{4p_{1}{\cdot}k_{3}}\sum_{J}^{[d]}\frac{(-1)^{|J|}}{|J|!}A_{3,\text{L}}^{\frac{1}{2}\otimes\frac{1}{2}}(W_{1}H_{3}^{\mu_{3}\nu_{3}}\varPhi_{J})\times A_{3,\text{R}}^{\frac{1}{2}\otimes\frac{1}{2}}(\varPhi^{J}H_{4}^{\mu_{4}\nu_{4}}W_{2}^{*}).\label{eq:resp1k3 general}
\end{equation}
Hence, we have shown that the gravitational 4-pt. amplitude factorizes
into the product of two 3-pt. amplitudes. Moreover, $\varPhi_{J}$
indicates all possible Lorentz structure propagating in the given
factorization channel. We can expand the sum to see the explicit form
of some of these structures propagating in this channel. To do so, first  notice
that since $\left(\slashed{p}_{13}+m\right)\mathbb{I}=\frac{p_{13,\alpha}}{m}\left(\slashed{p}_{13}+m\right)\gamma^{\alpha},$
we can identify the contribution from the terms $|J|=0$ and $|J|=1$ with the transverse and longitudinal modes of the spin-1 field. With this consideration \eqref{eq:resp1k3 general} takes the form
\begin{equation}
\begin{split}
 & -\frac{1}{p_{1}{\cdot}k_{3}2^{2\left\lfloor d/2\right\rfloor }}\bigg\{ \text{tr}\left[A_{3,\text{L}}^{\rm{QCD},\mu_{4}}\slashed{\varepsilon}_{1}\left(\slashed{p_{1}}{+}m\right)\bar{A}_{3,\text{L}}^{\rm{QCD},\nu_{4}}\left(\slashed{p}_{13}+m\right)\gamma^{\alpha}\right]D_{W,\alpha\beta}\\
 & \qquad\quad \text{tr}\left[\bar{A}_{3,\text{R}}^{\rm{QCD},\nu_{3}}\slashed{\varepsilon}_{2}\left(\slashed{p_{2}}{+}m\right)A_{3,\text{R}}^{\rm{QCD},\mu_{3}}\left(\slashed{p}_{13}-m\right)\gamma^{\beta}\right]\\
&\qquad +\frac{1}{2}\text{tr}\left[A_{3,\text{L}}^{\rm{QCD},\mu_{4}}\slashed{\varepsilon}_{1}\left(\slashed{p_{1}}{+}m\right)\bar{A}_{3,\text{L}}^{\rm{QCD},\nu_{4}}\left(\slashed{p}_{13}+m\right)\gamma^{\mu\nu}\right]\\
 & \qquad \eta_{[\mu\alpha}\eta_{\nu]\beta}\text{tr}\left[\bar{A,\text{R}}_{3}^{\rm{QCD},\nu_{3}}\slashed{\varepsilon}_{2}\left(\slashed{p_{2}}{+}m\right)A_{3,\text{R}}^{\rm{QCD},\mu_{3}}\left(\slashed{p}_{13}-m\right)\gamma^{\alpha\beta}\right]+\cdots\bigg\},
\end{split}
\end{equation}
 where 
\begin{equation}
D_{W,\alpha\beta}=\eta_{\alpha\beta}-\frac{p_{13,\alpha}p_{13,\beta}}{m^{2}},
\end{equation}
 and the $\cdots$ indicate the terms with higher value of $|J|$.
 
A similar analysis can be made at higher multiplicity starting from \eqref{eq:massiveklt}. The additional complication is that we have to deal with the factorization of the KLT kernel $K_{\alpha \beta}$, which is however standard. Once the dust settles we obtain

\begin{equation}
\begin{split}
&-\frac{1}{2(p_I^2 -m^2)2^{2\left\lfloor d/2\right\rfloor -2}}\sum_{J}^{[d]}\frac{(-1)^{|J|}}{|J|!}K_{\alpha_L \beta_L}\text{tr}\left[A_{n_L,\alpha_L}^{\rm{QCD}}\slashed{\varepsilon}_{1}\left(\slashed{p_{1}}{+}m\right)\bar{A}_{n_L,\beta_L}^{\rm{QCD}}\left(\slashed{p}_{I}+m\right)\Gamma_{J}\right]\times\\
 & \qquad\qquad K_{\alpha_R \beta_R}\text{tr}\left[\bar{A}_{n_R,\beta_R}^{\rm{QCD}}\slashed{\varepsilon}_{2}\left(\slashed{p_{2}}{+}m\right)A_{n_R,\beta_R}^{\rm{QCD}}\left(\slashed{p}_{I}-m\right)\Gamma^{J}\right]{+}\cdots,
\end{split}
\end{equation}
as $p_I^2\to m^2$, for $p_I$ any internal massive momenta. This means that unitarity of $A_n^{\frac{1}{2}\otimes \frac{1}{2}}$ should follow from that of $A_n^{\rm{QCD}}$ provided we correcltly account for the tensor structures $\Gamma^J$ as particles propagating in this channel.

Let us leave the analysis for general multiplicity for future work, and here instead focus in the internal spectrum at $n=4$. Next we consider two such cases and determine the fields propagating
in this channel. The first is the gravitational amplitude
for a massive line emitting two dilatons, whereas the second one corresponds
to the amplitude for the emission of two axions. 

\subsection*{Dilaton emission}

For this explicit
example the sum truncates at $|J|=3$.
Moreover, it can be checked that the terms $|J|=2$ and $|J|=3$
add up exactly to the contributions given by the $|J|=0$ and $|J|=1$
terms, namely, they account for a propagating spin-1 field. With this in mind,   $(\ref{eq:resp1k3 general})$ gives 
 
\begin{equation}
A_{4}^{\frac{1}{2}\otimes\frac{1}{2}}(W_{1}\phi_{3}\phi_{4}W_{2}^{*})\rightarrow\frac{\kappa^{2}}{32p_{1}{\cdot}k_{3}(2{-}d)}\left[\left(d{-}4\right)p_{1}^{\alpha}\,p_{3}{\cdot}\varepsilon_{1}{+}2m^{2}\varepsilon_{1}^{\alpha}\right]D_{W,\alpha\beta}\left[\left(d{-}4\right)p_{2}^{\beta}\,p_{4}{\cdot}\varepsilon_{2}{+}2m^{2}\varepsilon_{2}^{\beta}\right].\label{eq:residue final}
\end{equation}
 It can be also checked that the same residue is computed starting
from $(\ref{eq:ABphiphiB-2})$.

\subsection*{Axion emission}

Let us move on to the slightly more complicated example corresponding to the emission of two axions by a massive line. As we mentioned, the matter spectrum of the \12x12
double copy can be truncated to massive
vector fields once we consider
the emission of gravitons or dilatons, but not axions. On the other hand, via double copy we showed that the matter line can only produce axions in pairs.
An example of this is the four point amplitude for two axions:
\[
A_{4}^{\frac{1}{2}\otimes\frac{1}{2}}(W_{1}B_{3}B_{4}W_{2}^{*})=\frac{1}{2^{\left\lfloor d/2\right\rfloor -1}}K_{1324,1324}\left(A_{4,1324\,[\mu_{4}}^{\rm{QCD},[\mu_{3}}\bar{A}_{4,1324\,\nu_{4}]}^{\rm{QCD},\nu_{3}]}\right)\epsilon_{B_{3},\mu_{3}\nu_{3}}\epsilon_{B_{4}}^{\mu_{4}\nu_{4}}.
\]

Studying tree-level unitarity in this object leads to consider additional matter fields. For instance, consider  the channel $2p_{1}{\cdot}k_{3}\rightarrow0$ given by \eqref{eq:resp1k3 general}.
For two axion emissions,  the sum truncates at $|J|=5.$ The sum of the contributions
for $|J|=0$ and $|J|=1$ cancels out, therefore no Proca field will
propagate in this channel, as expected since $A_{3}^{\frac{1}{2}\otimes \frac{1}{2}}(W_{1}BW_{2}^{*})=0$. We can check that
the sum of the contributions for $|J|=2$ and $|J|=3$ equals the sum
of the contributions for $|J|=4$ and $|J|=5.$ Therefore, in this factorization
channel there is the propagation of particles associated to the structures $\{\gamma^{\mu_{1},\mu_{2}},\gamma^{\mu_{1}\mu_{2}\mu_{3}}\}$
or equivalently $\{\gamma^{\mu_{1}\mu_{2}\mu_{3}\mu_{4}},\gamma^{\mu_{1}\mu_{2}\mu_{3}\mu_{4}\mu_{5}}\}$. The propagation of these structures is what makes more
involved to write down a Lagrangian including the additional fields in general dimension. We leave this task for future work.
In $d=4$ however there is a simplification since the form $\gamma^{\mu_{1}\mu_{2}\mu_{3}\mu_{4}}$
can be dualized to a pseudoscalar, whereas the form $\gamma^{\mu_{1}\mu_{2}\mu_{3}\mu_{4}\mu_{5}}$
vanishes. The propagation of this pseudoscalar (as obtained in \cite{Johansson:2019dnu}) was pointed out
in the previous Appendix, as obtained from antisymmetrization of spinors in $d=4$. 
%%%%%%%%%%%%%%%%%%%%%%%%%%%%%%%%%%%%%%%%%%%%%%%%%%%%%%%%%%%%%%%%%%%%%%%%%%%%%%%%%%%%%%%%%%%%%%%%%%%%%%%%%%%%%%%%%%%%%%%%%%%%%%%%%%%%
%%%%%%%%%%%%%%%%%%%%%%%%%%%%%%%%%%%%%%%%%%%%%%%%%%%%%%%%%%%%%%%%%%%%%%%%%%%%%%%%%%%%%%%%%%%%%%%%%%%%%%%%%%%%%%%%%%%%%%%%%%%%%%%%%%%%

\section{ \texorpdfstring{$\mathcal{N}=4$}{N4} SUGRA in the form of Nicolai and Townsend}\label{ap:niktow}

In this Appendix we review the original construction of \cite{Nicolai1981}. 
There the axion pseudoscalar was dualized to a two-form at the level of the Lagrangian, i.e. off-shell. The starting point is the following bosonic action 

\begin{equation}
\mathcal{L}^{\mathcal{N=}4}=\sqrt{g}\left[R-2(\partial\phi)^{2}-2e^{4\phi}X_{\mu}X^{\mu}-e^{-2\phi}F_{\mu\nu}^{I}F_{I}^{\mu\nu}\right]+2\epsilon^{\mu\nu\rho\sigma}X_{\mu}A_{I\nu}F_{\rho\sigma}^{I}+\epsilon^{\mu\nu\rho\sigma}B_{\mu\nu}\partial_{\rho}X_{\sigma}.\label{eq:N4-1}
\end{equation}
Here $B_{\mu\nu}$ acts as a Lagrange multiplier imposing the condition
\begin{equation}
0=\frac{\delta\mathcal{L}}{\delta B_{\mu\nu}}=\epsilon^{\mu\nu\rho\sigma}\partial_{\rho}X_{\sigma}.
\end{equation}
We can solve such constraint locally by $X_{\mu}=\partial_{\mu}\chi$
(and globally provided certain topological conditions) and plug it
back on the action. Then, $\chi$ can be seen as a dynamical pseudoscalar carrying the degrees of freedom of the axion in four dimensions. The resulting Lagrangian then
reads

\begin{equation}\label{n4sug}
\left.\mathcal{L}^{\mathcal{N=}4}\right|_{X_{\mu}=\partial_{\mu}\chi}=\sqrt{g}\left[R-2(\partial\phi)^{2}-2e^{4\phi}(\partial\chi)^{2}-e^{-2\phi}F_{\mu\nu}^{I}F_{I}^{\mu\nu}-2\chi F_{\mu\nu}^{I}\star F_{I}^{\mu\nu}\right],
\end{equation}
where $\star F_{I}^{\mu\nu}=\frac{1}{2\sqrt{g}}\epsilon^{\mu\nu\rho\sigma}F_{\rho\sigma}^{I}$
is the Hodge dual. This is the standard form of the $\mathcal{N}=4$ Supergravity action given
in the Einstein frame, introduced in \cite{CREMMER197861,PhysRevD.15.2805}. A small paradox arises in that from the analysis of the main text we expect to find contact matter interactions between flavours. However, in this case we have six
flavours of photons which only interact through massless exchanges. The resolution of the paradox is that this is not a well posed statement, as we will see when integrating
out the scalar $\chi$, or more precisely, replacing it by the two
form $B_{\mu\nu}$. In order to do so we go back to \ref{eq:N4-1} and
solve the field equations of $X_{\mu}$ instead:

\begin{equation}\label{XU}
X^{\mu}=\frac{e^{-4\phi}}{2\sqrt{g}}\epsilon^{\mu\nu\rho\sigma}\left[A_{\nu}^{I}F_{I\rho\sigma}+\frac{1}{6}H_{\rho\sigma\nu}\right]\Rightarrow X_{\mu}=-\frac{e^{-4\phi}}{2}\sqrt{g}\epsilon_{\mu\nu\rho\sigma}\left[A^{I\nu}F_{I}^{\rho\sigma}+\frac{1}{6}H^{\rho\sigma\nu}\right],
\end{equation}
where $H=dB$. Inserting this back in (\ref{eq:N4-1}) gives, after
some algebra,
\begin{equation}\label{dual}
\sqrt{g}\left[R-2(\partial\phi)^{2}+3e^{-4\phi}(A_{I}^{\nu}F^{I\rho\sigma}+\frac{1}{6}H^{\nu\rho\sigma})(A_{J\nu}F_{\rho\sigma}^{J}+\frac{1}{6}H_{\nu\rho\sigma})-e^{-2\phi}F_{\mu\nu}^{I}F_{I}^{\mu\nu}\right]\,,
\end{equation}
which leads to the interaction tern $\sim A^2F^2$. Finally, note that the original Lagrangian \eqref{n4sug} is invariant under $\delta A_\mu ^I = -\partial_\mu \xi^I$ whereas the dualized one \eqref{dual} seems not to be. This is reconciled by imposing that $X^\mu$ in \eqref{XU} does not change under the $U(1)^6$ gauge transformations, which in turn can be achieved via

\begin{equation}
\delta B_{\mu \nu } = 2\xi_I F^I_{\mu \nu } \Rightarrow \delta H_{\mu \nu \rho} = 6\,\partial_{[\mu} \xi_I F^I_{\nu \rho] }\,.
\end{equation}
However, after restoring the coupling $\kappa$ as in the main text we find that $\delta B_{\mu \nu }= \mathcal{O}(\kappa)$ and hence the gauge symmetry does not shift the asymptotic axion states.

\section{Testing Amplitudes from CHY-like formulas}\label{ap:chy}

To construct the KLT product while taking different traces (Dirac
traces or Lorentz traces for dilatons) can be a cumbersome operation.
Hence in this section we will provide CHY-like formulas that automatically
implement double copy and at the same time avoid the computation of
QCD Feynman diagrams. For the case of the \12x12
theory we will more precisely use the connected prescription (as in \cite{He:2016dol} for the massless case) which was recently introduced for massive
particles in \cite{Cachazo:2018hqa,Geyer:2018xgb}  and unified in \cite{Schwarz:2019aat}.

\subsection{The  \texorpdfstring{\0x1}{01t} theory }\label{ap:01chy}

Based on the considerations of the main text it is direct to identify
the massless version of the \0x1 theory with the extended ``Einstein-Maxwell''
theory considered in the context of CHY in \cite{Cachazo:2014nsa}. Thus our conjecture is to assign the \textit{$d$-dimensional}
Lagrangian (\ref{eq:final01}) to such construction, even for the
massive case. As a warm-up for the \12x12 case
we give an overview on this construction.

In the CHY formulation \cite{Cachazo:2013gna} the amplitude is obtained by
solving the scattering equations $E_{i}=0$, $i=1,\ldots,n$, where

\begin{equation}
E_{i}:=\sum_{j\neq i}\frac{2P_{i}\cdot P_{j}}{\sigma_{ij}}\,,\quad\sigma_{ij}=\sigma_{i}-\sigma_{j}\label{eq:se}
\end{equation}
These equations feature an ${\rm SL(2,\mathbb{C})}$ redundancy due
to the fact that $\sum_{j\neq i}P_{i}\cdot P_{j}=0$ which requires
the momenta to be massless $P_{i}^{2}=0$, and in fact only $(n-3)$
of the $E_{i}$'s are independent. The gravitational and YM amplitudes
are given by

\begin{equation}\label{eq:chyf}  
\begin{split}
  A_{n}^{{\rm gr}}&=\int\frac{\prod_{i=1}^{n}d\sigma_{i}}{{\rm Vol}({\rm SL}(2,\mathbb{C}))}\prod_{i=1}^{n}{}'\delta(E_{i}){\rm Pf}'\Psi_{n}{\rm Pf}'\tilde{\Psi}_{n},\\ A_{n}^{{\rm YM}}(\alpha)&=\int\frac{\prod_{i=1}^{n}d\sigma_{i}}{{\rm Vol}({\rm SL}(2,\mathbb{C}))}\prod_{i=1}^{n} {}'\delta(E_{i}){\rm Pf}'\Psi_{n}{\rm PT(\alpha)}
\end{split}
\end{equation}
where the different ingredients are detailed in \cite{Cachazo:2013hca,Cachazo:2013iea}. Here
we just need to recall that the delta functions $\delta(E_{i})$ in
fact localize the integration to $(n-3)!$ solutions, hence making
it effectively a sum over Jacobians weighted by the integrands. For
$A_{n}^{{\rm YM}}(\alpha)$ the color ordering is encoded in the integrand

\begin{equation}
{\rm PT(\alpha)}:=\frac{1}{(\sigma_{\alpha_{1}}-\sigma_{\alpha_{2}})\cdots(\sigma_{\alpha_{n}}-\sigma_{\alpha_{1}})},
\end{equation}
whereas the polarization dependence is encoded in the reduced Pfaffian
of the matrix $\Psi$. The double copy construction (\ref{eq:massiveklt})
is already implemented in (\ref{eq:chyf}): It corresponds to the
replacement of one PT color actor by a second copy of the polarization
factor ${\rm Pf}'\tilde{\Psi}_{n}$. As observed in \cite{Cachazo:2014xea} we can
directly compactify the polarization vectors in ${\rm Pf}'\tilde{\Psi}_{n}$
as in (\ref{eq:compeps}) obtain the special YMS theory and the Einstein-Maxwell
theory.

It was observed in \cite{Naculich:2015zha} that the massive compactification
can be implemented in the CHY formalism to include up to three massive
interacting species.\footnote{For a recent generalization of this procedure see  \cite{Mizera:2019gea}.}
In our case we are solely interested in different species interacting
only through massless exchanges. The corresponding massive scattering
equations can be computed by setting the momenta as in (\ref{eq:masscom}).

As a preparation for next section let us give a simple example on
how the scattering equations can be naturally adapted for massive
particles. For two massive-lines of different species $a,b$ and no
external massless fields, the only independent equation in (\ref{eq:se})
reads

\begin{equation}
E_{1}=q^{2}+\frac{s-m_{a}^{2}-m_{b}^{2}}{1-\sigma}=0\Longrightarrow\sigma=1+\frac{s-m_{a}^{2}-m_{b}^{2}}{q^{2}}
\end{equation}
where $q=p_{1}+p_{2}$ and $s=(p_{1}+p_{3})^{2}$ and we have fixed
$(\sigma_{1},\ldots,\sigma_{4})=(1,0,\sigma,a)$, with $a\to\infty$.
It is straightforward to compute the four-massive amplitude $M_{4}$,
we write

\begin{align}
M_{4}^{(0,1)} & =\int\frac{\prod_{i=1}^{4}d\sigma_{i}}{{\rm Vol}({\rm sl}(2,\mathbb{C}))}\prod_{i=1}^{4}{}'\delta(E_{i}){\rm Pf}'\Psi_{4}{\rm Pf}'\tilde{\Psi}_{4}^{C}, \\
 & =\frac{\sigma(\sigma-1)}{q^{2}}\lim_{a\to\infty}({\rm Pf}'\Psi_{4}a^{2})({\rm Pf}'\tilde{\Psi}_{4}^{C} a^2),\label{eq:4ptm4}
\end{align}
where the factor $\lim_{a\to\infty}({\rm Pf}'\Psi_{4}a^{2})$ is standard
\cite{Cachazo:2013iea}, whereas the second copy of this factor simplifies
(under (\ref{eq:compeps})) to

\begin{equation}
{\rm Pf}'\tilde{\Psi}_{4}^{C}={\rm Pf}\left(\begin{array}{cccccc}
0 & \frac{q^{2}}{\sigma-a} & 0 & 0 & 0 & 0\\
\frac{-q^{2}}{\sigma-a} & 0 & 0 & 0 & 0 & 0\\
0 & 0 & 0 & 1 & 0 & 0\\
0 & 0 & -1 & 0 & 0 & 0\\
0 & 0 & 0 & 0 & 0 & \frac{1}{\sigma-a}\\
0 & 0 & 0 & 0 & \frac{-1}{\sigma-a} & 0
\end{array}\right)=\frac{q^{2}}{(\sigma-a)^{2}}\to\frac{q^{2}}{a^{2}}.
\end{equation}
Hence we get to the compact form
\begin{equation}
M_{4}^{(0,1)}=\sigma(\sigma-1)\lim_{a\to\infty}({\rm Pf}'\Psi_{4}a^{2})=\frac{(s-m_{a}^{2}-m_{b}^{2})(q^{2}+s-m_{a}^{2}-m_{b}^{2})}{q^{2}}\times\lim_{a\to\infty}({\rm Pf}'\Psi_{4}a^{2}).
\end{equation}
Of course, we have done nothing but to compactify a 4-graviton amplitude
to obtain (\ref{eq:4ptm4}). However, at higher points the fact that
we do not need to implement the KLT kernel explicitly turns out to
be very efficient for testing our Lagrangian numerically. By including
an external (fat) graviton, we can compute the 5-pt. amplitude $M_{5}$
in the compactified version of $\mathcal{N=}4$ SUGRA, whose classical
limit matches with the amplitude considered in Section \ref{sec:sasb01}.

\subsection{The  \texorpdfstring{\12x12 }{1212t}theory}

Recently a rational map formalism has been introduced for 6D $(1,1)$
SYM Theory \cite{Cachazo:2018hqa,Geyer:2018xgb, Schwarz:2019aat}. This can be understood
as an extension of CHY that naturally produces superamplitudes in
six dimensions \cite{Heydeman:2017yww}, analogous to the Witten-RSV formalism
in four dimensions \cite{Witten:2003nn,Roiban:2004vt,Roiban:2004yf}.
Under dimensional reduction this theory generates the massive amplitudes
of $d=4$ SYM in the Coulomb branch \cite{Huang:2011um}, which color-kinematics duality was first studied in \cite{Chiodaroli:2015rdg}. On the other hand, the double copy 
is naturally incorporated into the rational map formula \cite{Cachazo:2018hqa}, in an analogous way to the original CHY construction \cite{Cachazo:2013hca}.

As for one matter line the Coulomb branch amplitudes coincide with
QCD, we can easily construct the \12x12 amplitudes
in this framework. Presumably, we can also construct the amplitudes with two massive
lines by applying the projections of \cite{He:2016dol,Dixon:2010ik}, although for this case
we employ the BCJ prescription as in Section 4. On the other hand, the downside of the Witten-RSV picture is that
we will have to restrict these checks to $d=4$ dimensions. In fact
the amplitudes are naturally produced in the massive spinor-helicity
formalism of \cite{Arkani-Hamed:2017jhn}, which we outlined already in appendix \ref{4 d double copy}.

For a review of the formalism see \cite{Schwarz:2019aat}, which has also explained
the equivalence of the two original formulations in \cite{Cachazo:2018hqa,Geyer:2018xgb}.
The massive SYM amplitudes can be written explicitly as the localized
integral
\begin{equation}
\begin{split}
\mathcal{A}_{n}^{{\rm SYM}}(\alpha)\delta^{4}(\sum_{i}p_{i})\delta(\sum m_{i})\delta(\sum\tilde{m}_{i}) & =\int d\mu \prod_{i=1}^{n}\delta^{2}\left(\sum_{j\neq i}\frac{\langle u_{j}u_{i}\rangle}{\sigma_{ji}}\lambda_{j}^{1}{-}\lambda_{i}^{1}v_{i}{-}\lambda_{i}^{1}\right)\\
 & \times\delta^{2}\left(\sum_{j\neq i}\frac{\langle u_{j}u_{i}\rangle}{\sigma_{ji}}\tilde{\lambda}_{j}^{1}{-}\lambda_{i}^{1}v_{i}{-}\lambda_{i}^{1}\right){\rm det}'(H){\rm PT(\alpha)}J_{F}
\end{split}
\end{equation}
where we have written  the measure $d\mu =\frac{\prod_{i}d\sigma_{i}dv_{i}d^{2}u_{i}^{a}}{{\rm vol}({\rm SL}(2,\mathbb{C})_{\sigma}\times{\rm SL}(2,\mathbb{C})_{u})}$,  and $\tilde{p}_{i}=\lambda_{i}^{a}\tilde{\lambda}_{i}^{b}\epsilon_{ab}$,
omitting spinor indices ($a,b$ are little-group indices). These satisfy
$\langle\lambda_{i}^{1}\lambda_{i}^{2}\rangle=m_{i}$ and $[\tilde{\lambda}_{i}^{1}\tilde{\lambda}_{i}^{2}]=\tilde{m}_{i}$,
so we can obtain massless particles (gluons) by taking $\lambda_{i}^{2}=\tilde{\lambda}_{i}^{1}=0$.
The equations localizing the integration are named \textit{Polarized Scattering
Equations} \cite{Geyer:2014fka,Geyer:2018xgb} and imply the (massive) scattering equations of the previous
subsection. In the following we will take the volume form to be

\begin{equation}
{\rm vol}({\rm SL}(2,\mathbb{C})_{\sigma}\times{\rm SL}(2,\mathbb{C})_{u})=d\sigma_{1}d\sigma_{2}d\sigma_{3}d^{2}u_{1}^{a}du_{3}^{1}\times\sigma_{12}\sigma_{23}\sigma_{31}u_{1}^{1}\langle u_{1}u_{3}\rangle
\end{equation}
We will also need to define the matrix 
\begin{equation}
H_{ij}=\begin{cases}
\frac{\langle\lambda_{i}^{1}\lambda_{j}^{1}\rangle-[\tilde{\lambda}_{i}\tilde{\lambda}_{j}]}{\sigma_{ij}} & i\neq j,\\
-\frac{1}{u_{i}^{2}}\sum_{k\neq i}\frac{u_{k}^{2}\langle\lambda_{i}^{1}\lambda_{k}^{1}\rangle-u_{k}^{2}[\tilde{\lambda}_{i}\tilde{\lambda}_{k}]}{\sigma_{ik}} & i=j,
\end{cases}
\end{equation}
and ${\rm det}'(H)=\frac{\det(H_{12,12})}{\langle u_{1}u_{2}\rangle^{2}}$
where $H_{12,12}$ corresponds to $H$ with the first and second rows
and columns deleted. The remaining object in the integrand is $J_{F}$,
which depends on the external fields we are considering. For one matter
line of massive fermions emitting gluons it is obtained via the Grassman
integration
\begin{equation}
\begin{split}
J_{F}^{s=\frac{1}{2}}{=}J_{F}(\psi^{a_{1}}g_{2}^{\pm}\cdots g_{n-1}^{\pm}\bar{\psi}^{b_{1}}){=}\int\prod_{i}d^{4}\eta_{i}^{aI}\times(\eta_{1}^{1,2}\eta_{1}^{2,2}\eta_{1}^{a_{1},1}\eta_{2}^{b_{1},1})\prod_{r\in+}\eta_{r}^{2,1}\eta_{r}^{2,2}\prod_{s\in-}\eta_{s}^{1,1}\eta_{s}^{1,2}\\
{\times}\prod_{i{=}1}^{n}\delta^{2}\left(\sum_{j\neq i}\frac{\langle u_{j}u_{i}\rangle}{\sigma_{ji}}\eta_{j}^{1,I}{-}\eta_{i}^{1,I}v_{i}{-}\eta_{i}^{1,I}\right)
\end{split}
\end{equation}
where $r$($s)$ range over the positive (negative) helicity gluons.
Implementation of this Jacobian is relatively direct using the Mathematica
package MatrixEDC. \footnote{The package is available at  \url{http://library.wolfram.com/infocenter/MathSource/683}.}
The advantage is that we can now implement double copy of this object
directly, instead of writing the KLT expansion and projecting into
the states. It is obtained by the replacement of the integrand

\begin{equation}
\boxed{{\rm det}'(H){\rm PT(\alpha)}J_{F}^{s=\frac{1}{2}}\to{\rm det}'(H)^{2}J_{F}^{s=1}},
\end{equation}
where
\begin{equation}
\begin{split}
J_{F}^{s=1}{=}J_{F}(W^{a_{1}a_{2}}H_{2}{\cdots} H_{n-1}\bar{W}^{b_{1}b_{2}}){=}\int\prod_{i}d^{4}\eta_{i}^{aI}d^{4}\tilde{\eta}_{i}^{aI}(\eta_{1}^{1,2}\eta_{1}^{2,2}\eta_{1}^{a_{1},1}\eta_{2}^{b_{1},1})(\tilde{\eta}_{1}^{1,2}\tilde{\eta}_{1}^{2,2}\tilde{\eta}_{1}^{a_{1},1}\tilde{\eta}_{2}^{b_{1},1})\\
\prod_{t}\left[\eta_{j}^{2,1}\eta_{j}^{2,2}\tilde{\eta}_{j}^{1,1}\tilde{\eta}_{j}^{1,2}+\tilde{\eta}_{j}^{2,1}\tilde{\eta}_{j}^{2,2}\eta_{j}^{1,1}\eta_{j}^{1,2}\right]\prod_{r\in+}\eta_{r}^{2,1}\eta_{r}^{2,2}\tilde{\eta}_{r}^{2,1}\tilde{\eta}_{r}^{2,2}\prod_{s\in-}\eta_{s}^{1,1}\eta_{s}^{1,2}\tilde{\eta}_{s}^{1,1}\tilde{\eta}_{s}^{1,2}\\
\prod_{i=1}^{n}\delta^{2}\left(\sum_{j\neq i}\frac{\langle u_{j}u_{i}\rangle}{\sigma_{ji}}\tilde{\eta}_{j}^{1,I}{-}\tilde{\eta}_{i}^{1,I}v_{i}{-}\tilde{\eta}_{i}^{1,I}\right)\prod_{i=1}^{n}\delta^{2}\left(\sum_{j\neq i}\frac{\langle u_{j}u_{i}\rangle}{\sigma_{ji}}\eta_{j}^{1,I}{-}\eta_{i}^{1,I}v_{i}{-}\eta_{i}^{1,I}\right)
\end{split}
\end{equation}
where $r$(s) range over the positive (negative) helicity gravitons
and $t$ ranges over the dilaton states. Despite the various ingredients
in the formula, the implementation in Mathematica is relatively fast.
For instance, we are mostly interested in the pure dilaton case as
it enables us to check the exponentials in (\ref{eq:1212dc-1}): For
$n=5$, that is $A_{5}^{(\frac{1}{2},\frac{1}{2})}(W_{1}\phi_{2}\phi_{3}\phi_{4}\bar{W}_{5})$,
the computation of $J_{F}^{s=1}$ is readily automated and takes about
15 minutes to perform.
\bibliographystyle{JHEP}
\bibliography{references}

% Please avoid comments such as "For a review'', "For some examples",
% "and references therein" or move them in the text. In general,
% please leave only references in the bibliography and move all
% accessory text in footnotes.

% Also, please have only one work for each \bibitem.

\end{document}